\documentclass[pdflatex,sn-standardnature]{sn-jnl}
\usepackage{adjustbox}
\usepackage[left]{lineno}

\usepackage{amsmath}
\usepackage[utf8]{inputenc}
\usepackage[varg]{txfonts}
\usepackage{natbib}
\usepackage{soul} 

\newcommand*\arcsec{\ensuremath{^{\prime\prime}}}

\definecolor{linkcolor}{rgb}{0.41960784, 0.40784314, 0.39607843}

\newcommand{\lbe}{$\lambda_{\rm e}$}
\newcommand{\lbp}{$\lambda_{\rm p}$}
\newcommand{\pgp}{$P_{\rm GP}$}

\newcommand{\refjnl}[1]{{#1}}

\def\aj{\refjnl{Astron. J.}}              
             
\def\apj{\refjnl{Astrophys. J.}}                 
\def\apjl{\refjnl{Astrophys. J.}}                
\def\apjs{\refjnl{Astrophys. J. Suppl.}}               
           
\def\apss{\refjnl{Astrophys. Space Sci.}}             
\def\aap{\refjnl{Astron. Astrophys.}}

\def\mnras{\refjnl{Mon. Not. R. Astron. Soc.}}

\def\psj{\refjnl{PSJ}}               
\def\pasp{\refjnl{Proc. Acad. Sci. Pacific}}               
\def\pasj{\refjnl{Pub. Astron. Soc. Japan}}             
\def\pnas{\refjnl{Proc. Nat. Acad. Sci. USA}}

\def\nat{\refjnl{Nature}}

\def\procspie{\refjnl{Proc. SPIE}}
\def\icarus{\refjnl{Icarus}}

\begin{document}

\title[The HD~110067 planetary system]{A resonant sextuplet of sub-Neptunes transiting the bright star HD~110067}

\author{
R. Luque$^{1,\ast}$, 
H. P. Osborn$^{2,3,\dagger}$, 
A. Leleu$^{4,2,\dagger}$, 
E. Pallé$^{5,6,\dagger}$,
A. Bonfanti$^{7}$, 
O. Barragán$^{8}$, 
T. G. Wilson$^{9,10,11}$, 
C. Broeg$^{2,12}$, 
A. Collier Cameron$^{9}$, 
M. Lendl$^{4}$
% ...
\\ \\
\footnotesize{
$^{1}$ Department of Astronomy \& Astrophysics, University of Chicago, Chicago, IL 60637, USA \\
$^{2}$ Space Research and Planetary Sciences, Physics Institute, University of Bern, Gesellschaftsstrasse 6, 3012 Bern, Switzerland \\
$^{3}$ Department of Physics and Kavli Institute for Astrophysics and Space Research, Massachusetts Institute of Technology, Cambridge, MA 02139, USA \\
$^{4}$ Observatoire Astronomique de l'Université de Genève, Chemin Pegasi 51, 1290 Versoix, Switzerland \\
$^{5}$ Instituto de Astrofisica de Canarias, Via Lactea s/n, 38200 La Laguna, Tenerife, Spain \\
$^{6}$ Departamento de Astrofisica, Universidad de La Laguna, Astrofísico Francisco Sanchez s/n, 38206 La Laguna, Tenerife, Spain \\
$^{7}$ Space Research Institute, Austrian Academy of Sciences, Schmiedlstrasse 6, A-8042 Graz, Austria \\
$^{8}$ Sub-department of Astrophysics, Department of Physics, University
of Oxford, Oxford, OX1 3RH, UK \\
$^{9}$ Centre for Exoplanet Science, SUPA School of Physics and Astronomy, University of St Andrews, North Haugh, St Andrews KY16 9SS, UK \\
$^{10}$ Department of Physics, University of Warwick, Gibbet Hill Road, Coventry CV4 7AL, UK \\
$^{11}$ Centre for Exoplanets and Habitability, University of Warwick, Coventry, CV4 7AL, UK \\
$^{12}$ Center for Space and Habitability, University of Bern, Gesellschaftsstrasse 6, 3012 Bern, Switzerland \\
$^{13}$ Astrophysics Group, Lennard Jones Building, Keele University, Staffordshire, ST5 5BG, United Kingdom \\
$^{14}$ Dipartimento di Fisica, Universita degli Studi di Torino, via Pietro Giuria 1, I-10125, Torino, Italy \\
$^{15}$ Cavendish Laboratory, JJ Thomson Avenue, Cambridge CB3 0HE, UK \\
$^{16}$ Institute of Astronomy, Faculty of Physics, Astronomy and Informatics, Nicolaus Copernicus University, Grudzicadzka 5, 87-100 Torun, Poland \\
$^{17}$ NASA Ames Research Center, Moffett Field, CA 94035, USA \\
$^{18}$ Research Institute for Advanced Computer Science, Universities Space Research Association, Washington, DC 20024, USA \\
$^{19}$ SETI Institute, Mountain View, CA 94043, USA \\
$^{20}$ Institut de Ciencies de l'Espai (ICE, CSIC), Campus UAB, Can Magrans s/n, 08193 Bellaterra, Spain \\
$^{21}$ Institut d’Estudis Espacials de Catalunya (IEEC), Gran Capità 2-4, 08034 Barcelona, Spain \\
$^{22}$ INAF, Osservatorio Astrofisico di Torino, Via Osservatorio, 20, I-10025 Pino Torinese To, Italy \\
$^{23}$ Instituto de Astrof\'isica e Ci\^encias do Espa\c{c}o, Universidade do Porto, CAUP, Rua das Estrelas, 4150-762 Porto, Portugal \\
$^{24}$ Departamento de F\'isica e Astronomia, Faculdade de Ci\^encias, Universidade do Porto, Rua do Campo Alegre, 4169-007 Porto, Portugal \\
$^{25}$ Mullard Space Science Laboratory, University College London, Holmbury St Mary, Dorking, Surrey, RH5 6NT, UK \\
$^{26}$ Instituto de Astrofísica de Andalucía (IAA-CSIC), Glorieta de la Astronomía s/n, 18008 Granada, Spain \\
$^{27}$ European Space Agency (ESA), European Space Research and Technology Centre (ESTEC), Keplerlaan 1, 2201 AZ Noordwijk, The Netherlands \\
$^{28}$ Admatis, 5. Kandó Kálmán Street, 3534 Miskolc, Hungary \\
$^{29}$ Depto. de Astrofisica, Centro de Astrobiologia (CSIC-INTA), ESAC campus, 28692 Villanueva de la Cañada (Madrid), Spain \\
$^{30}$ Instituto de Astrofisica e Ciencias do Espaco, Universidade do Porto, CAUP, Rua das Estrelas, 4150-762 Porto, Portugal \\
$^{31}$ Departamento de Fisica e Astronomia, Faculdade de Ciencias, Universidade do Porto, Rua do Campo Alegre, 4169-007 Porto, Portugal \\
$^{32}$ Université Grenoble Alpes, CNRS, IPAG, 38000 Grenoble, France \\
$^{33}$ INAF, Osservatorio Astronomico di Padova, Vicolo dell'Osservatorio 5, 35122 Padova, Italy \\
$^{34}$ Department of Astronomy, California Institute of Technology, 1200 E. California Blvd, Pasadena, CA 91106 \\
$^{35}$ Department of Astronomy, Stockholm University, AlbaNova University Center, 10691 Stockholm, Sweden \\
$^{36}$ Institute of Planetary Research, German Aerospace Center (DLR), Rutherfordstrasse 2, 12489 Berlin, Germany \\
$^{37}$ Universidad Nacional Autónoma de México, Instituto de Astronomía, AP 70-264, Ciudad de México,  04510, México \\
$^{38}$ Department of Physics and Astronomy, Georgia State University, Atlanta, GA 30302-4106, USA \\
$^{39}$ RECONS Institute, Chambersburg, PA 17201, USA \\
$^{40}$ Center for Astrophysics ${\rm \mid}$ Harvard {\rm \&} Smithsonian, 60 Garden Street, Cambridge, MA 02138, USA \\
$^{41}$ Université de Paris Cité, Institut de physique du globe de Paris, CNRS, 1 Rue Jussieu, F-75005 Paris, France \\
$^{42}$ McDonald Observatory and Center for Planetary Systems Habitability, The University of Texas, Austin, Texas, USA \\
$^{43}$ Department of Physics and Astronomy, University of  Kansas, Lawrence, KS, USA \\
$^{44}$ Division of Geological and Planetary Sciences, 1200 E California Boulevard, Pasadena, CA 91125, USA \\
$^{45}$ Centre for Mathematical Sciences, Lund University, Box 118, 221 00 Lund, Sweden \\
$^{46}$ Aix Marseille Univ, CNRS, CNES, LAM, 38 rue Frédéric Joliot-Curie, 13388 Marseille, France \\
$^{47}$ Astrobiology Research Unit, Université de Liège, Allée du 6 Août 19C, B-4000 Liège, Belgium \\
$^{48}$ Space sciences, Technologies and Astrophysics Research (STAR) Institute, Université de Liège, Allée du 6 Août 19C, 4000 Liège, Belgium \\
$^{49}$ Centre Vie dans l’Univers, Faculté des sciences, Université de Genève, Quai Ernest-Ansermet 30, 1211 Genève 4, Switzerland \\
$^{50}$ Space Telescope Science Institute, 3700 San Martin Drive, Baltimore, MD, 21218, USA \\
$^{51}$ Leiden Observatory, University of Leiden, PO Box 9513, 2300 RA Leiden, The Netherlands \\
$^{52}$ Department of Space, Earth and Environment, Chalmers University of Technology, Onsala Space Observatory, 439 92 Onsala, Sweden \\
$^{53}$ Komaba Institute for Science, The University of Tokyo, 3-8-1 Komaba, Meguro, Tokyo 153-8902, Japan \\
$^{54}$ Th\:uringer Landessternwarte Tautenburg, Sternwarte 5, D-07778 Tautenburg, Germany \\
$^{55}$ Department of Astrophysics, University of Vienna, Türkenschanzstrasse 17, 1180 Vienna, Austria \\
$^{56}$ Department of Multi-Disciplinary Sciences, Graduate School of Arts and Sciences, The University of Tokyo, 3-8-1 Komaba, Meguro, Tokyo 153-8902, Japan \\
$^{57}$ Konkoly Observatory, HUN-REN Research Centre for Astronomy and Earth Sciences, 1121 Budapest, Konkoly Thege Miklós út 15-17, Hungary \\
$^{58}$ ELTE E\"otv\"os Loránd University, Institute of Physics, Pázmány Péter sétány 1/A, 1117 Budapest, Hungary \\
$^{59}$ Lund Observatory, Division of Astrophysics, Department of Physics, Lund University, Box 43, SE-221 00 Lund, Sweden \\
$^{60}$ IMCCE, UMR8028 CNRS, Observatoire de Paris, PSL Univ., Sorbonne Univ., 77 av. Denfert-Rochereau, 75014 Paris, France \\
$^{61}$ Institut d'astrophysique de Paris, UMR7095 CNRS, Université Pierre \& Marie Curie, 98bis blvd. Arago, 75014 Paris, France \\
$^{62}$ Astrobiology Center, 2-21-1 Osawa, Mitaka, Tokyo 181-8588, Japan \\
$^{63}$ National Astronomical Observatory of Japan, 2-21-1 Osawa, Mitaka, Tokyo 181-8588, Japan \\
$^{64}$ Department of Astronomical Science, The Graduated University for Advanced Studies, SOKENDAI, 2-21-1, Osawa, Mitaka, Tokyo, 181-8588, Japan \\
$^{65}$ U.S. Naval Observatory, Washington, D.C. 20392, USA \\
$^{66}$ Max Planck Institute for Astronomy, Heidelberg, Germany \\
$^{67}$ Instituto de Astronomía, Universidad Católica del Norte, Angamos 0610, 1270709, Antofagasta, Chile \\
$^{68}$ INAF, Osservatorio Astrofisico di Catania, Via S. Sofia 78, 95123 Catania, Italy \\
$^{69}$ Institute of Optical Sensor Systems, German Aerospace Center (DLR), Rutherfordstrasse 2, 12489 Berlin, Germany \\
$^{70}$ Dipartimento di Fisica e Astronomia "Galileo Galilei", Universita degli Studi di Padova, Vicolo dell'Osservatorio 3, 35122 Padova, Italy \\
$^{71}$ ETH Zurich, Department of Physics, Wolfgang-Pauli-Strasse 2, CH-8093 Zurich, Switzerland \\
$^{72}$ Landessternwarte, Zentrum für Astronomie der Universität Heidelberg, D-69117 Heidelberg, Germany \\
$^{73}$ Zentrum für Astronomie und Astrophysik, Technische Universität Berlin, Hardenbergstr. 36, D-10623 Berlin, Germany \\
$^{74}$ Institut fuer Geologische Wissenschaften, Freie Universitaet Berlin, Maltheserstrasse 74-100,12249 Berlin, Germany \\
$^{75}$ Astronomy Department and Van Vleck Observatory, Wesleyan University, Middletown, CT 06459, USA \\
$^{76}$ Instituto de Astronomía, Universidad Nacional Autónoma de México, Apdo. Postal 877, 22860 Ensenada, B.C., Mexico \\
$^{77}$ Department of Astronomy, University of Maryland, College Park, MD 20742, USA \\
$^{78}$ NASA Goddard Space Flight Center, Greenbelt, MD 20771, USA \\
$^{79}$ Department of Earth, Atmospheric and Planetary Sciences, Massachusetts Institute of Technology, Cambridge, MA 02139, USA \\
$^{80}$ Department of Aeronautics and Astronautics, MIT, 77 Massachusetts Avenue, Cambridge, MA 02139, USA \\
$^{81}$ ELTE E\"otv\"os Loránd University, Gothard Astrophysical Observatory, 9700 Szombathely, Szent Imre h. u. 112, Hungary \\
$^{82}$ HUN-REN-ELTE Exoplanet Research Group, 9700 Szombathely, Szent Imre h. u. 112, Hungary \\
$^{83}$ German Aerospace Center (DLR), Institute of Optical Sensor Systems, Rutherfordstraße 2, 12489 Berlin \\
$^{84}$ Institute of Astronomy, University of Cambridge, Madingley Road, Cambridge, CB3 0HA, UK \\
$^{85}$ Department of Astrophysical Sciences, Princeton University, Princeton, NJ 08544, USA
}
\\ \\ 
$^\ast$Corresponding author. E-mail: rluque@uchicago.edu.
\\
$^\dagger$These authors contributed equally to this work.
}

\maketitle
\newpage



\textbf{
Planets with radii between that of the Earth and Neptune (hereafter referred to as ``sub-Neptunes'') are found in close-in orbits around more than half of all Sun-like stars \cite{Howard2012,Fressin13}. Yet, their composition, formation, and evolution remain poorly understood \cite{Bean2021}. 
The study of multi-planetary systems offers an opportunity to investigate the outcomes of planet formation and evolution while controlling for initial conditions and environment. Those in resonance (with their orbital periods related by a ratio of small integers) are particularly valuable because they imply a system architecture practically unchanged since its birth. 
Here, we present the observations of six transiting planets around the bright nearby star HD~110067. We find that the planets follow a chain of resonant orbits. A dynamical study of the innermost planet triplet allowed the prediction and later confirmation of the orbits of the rest of the planets in the system. 
The six planets are found to be sub-Neptunes with radii ranging from 1.94 to 2.85 $R_{\oplus}$. Three of the planets have measured masses, yielding low bulk densities that suggest the presence of large hydrogen-dominated atmospheres.
}
\medskip

\begin{figure}[t]
    \centering
    \includegraphics[width=0.99\columnwidth]{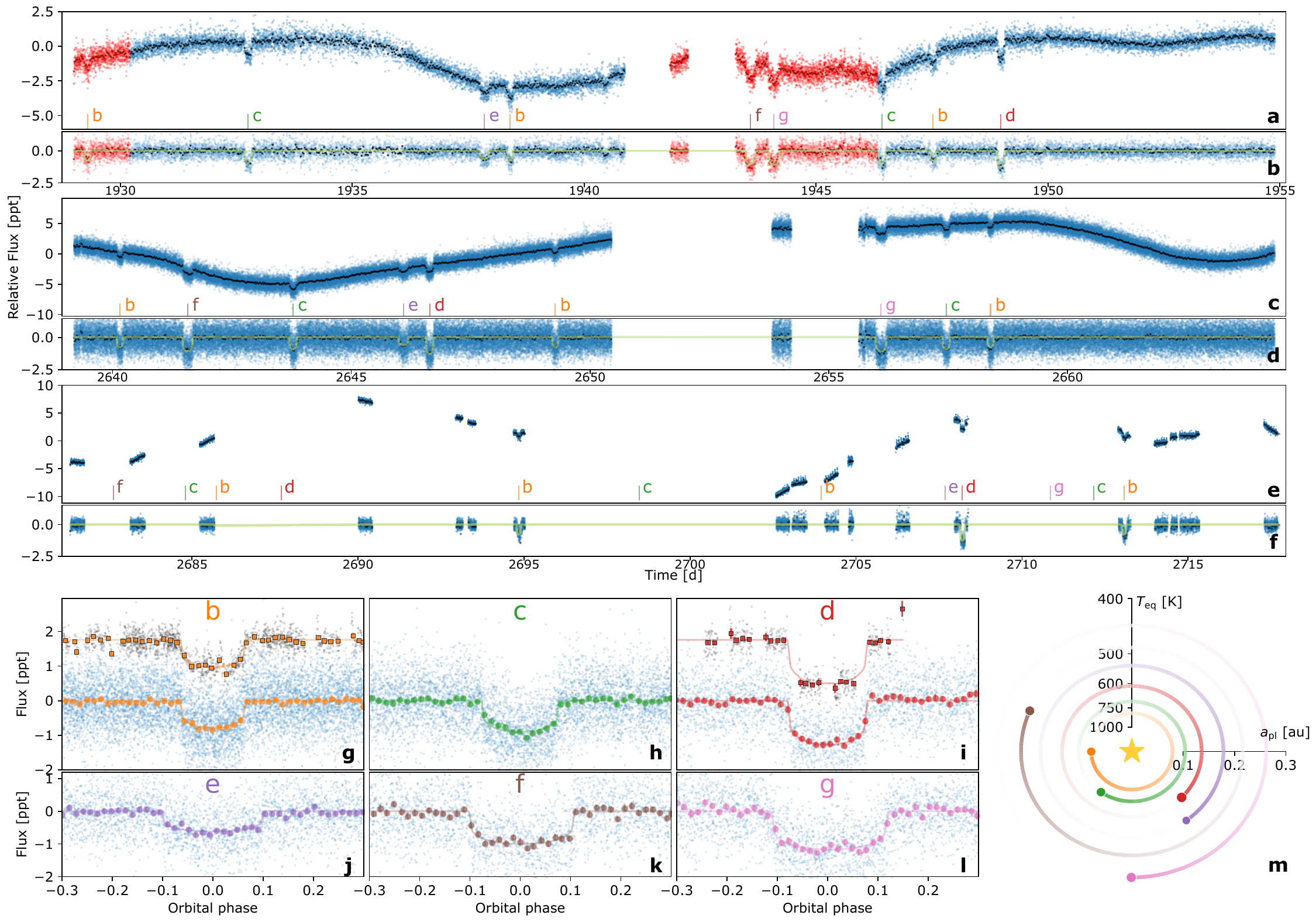}
    \caption{\textbf{Space-based photometry from TESS and CHEOPS of HD~110067}. Transits identified for each planet after our analyses are each associated with a different color. \textbf{a-f,} Photometric time series of TESS Sector 23 (\textbf{a,b}, 2-minute cadence), TESS Sector 49 (\textbf{c,d}, 20-second cadence), and CHEOPS (\textbf{e,f}). Red points in \textbf{a} show the reprocessed data affected by scattered light and high levels of sky background. Time units are in TESS Julian Date $\mathrm{(TJD \equiv BJD-2457000)}$, where BJD is the Barycentric Julian Date in units of days.
    Detrended light curves and the final model are shown in \textbf{b,d,f}. \textbf{g-l,} TESS phase-folded transits of each planet in the system. The transit model and binned photometry are color-coded following the same convention. For HD~110067~b (\textbf{g}) and d (\textbf{i}), CHEOPS photometry is shown atop the TESS photometry with an arbitrary offset for clarity. \textbf{m,} top-down view of the planetary system. Axes show the distance from the central star as a function of the semi-major axis and equilibrium temperature of the planet.} 
    \label{fig:phot}
\end{figure}

HD~110067 (TIC 347332255) is a bright K0-type star in the constellation of Coma Berenices with mass and radius of approximately 80 percent of the Sun's. The Transiting Exoplanet Survey Satellite (TESS) monitored HD~110067 as part of its observations of Sector 23 \cite{Ricker2015}. The data, processed by the TESS Science Processing Operations Center (SPOC) \cite{SPOC}, exhibited several dips that could be associated with transiting planets. SPOC reported two candidates: one with an orbital period of 5.642\,days based on three dips with apparently similar depth and duration, and a second one with an unconstrained orbital period based on a single event. TESS re-observed the star two years later in Sector 49, revealing nine further transits incompatible with the previously announced candidates (Fig.~\ref{fig:phot}). 

\medskip
Combining TESS Sectors 23 and 49 we could associate a fraction of the transits with two new planet candidates: HD~110067~b, a planet with an orbital period of 9.114\,days, and HD~110067~c, a planet with an orbital period of 13.673\,days. For the remaining unidentified transit events (two in Sector 23 and four in Sector 49), we attempted to match them by modeling them individually using a purely shape-based transit model agnostic to the orbital period (Methods and Extended Data Fig.\ref{fig:ind_transits}). Then, we compared them in duration-depth space, allowing us to identify two ``duo-transits'' (planets seen to transit only once in each of the two widely-separated sectors) and two single-transits (solitary transit events seen only in Sector 49). Duo-transits have orbital periods limited to a finite number of harmonics or aliases which are constrained at the short-period end by the extent of continuous photometry observed before/after transit and at the long end by the distance between transits. Targeted observations with the CHaracterising ExOPlanets Satellite (CHEOPS) \cite{CHEOPS} allowed us to rule out many of these aliases and confirm a third planet in the system, HD~110067~d, with an orbital period of 20.519\,days (Fig.~\ref{fig:phot}).

\medskip
The orbital periods of planets HD~110067~b, c, and d (9.114, 13.673, and 20.519 days, respectively) have ratios very close to 3/2 ($P_c/P_b=1.5003 $ and $P_d/P_c=1.5007$). Mean-motion resonances (MMRs) are orbital configurations where the period ratio of a pair of planets is oscillating near a rational number of the form $(k+q)/k$, where $k$ and $q$ are integers. First-order MMRs (where $q=1$) are the most common among planetary systems, as well as resonant chains. For the two innermost pairs of planets (bc and cd), we define two resonant angles $\phi_1= 2 \lambda_b -3\lambda_c + \varpi_c$ and  $\phi_2= 2 \lambda_c -3\lambda_d + \varpi_c$, where $\lambda$ is the mean longitude of the planets and $\varpi$ its longitude of periastron. Since $d\lambda_x/dt=2\pi/P_x$, and given the aforementioned period ratios between $b$, $c$, and $d$, a generalized Laplace relation links the triplet of planets via $2/P_b-5/P_c+3/P_d \approx 0$ \cite{Sinclair1975MNRAS.171...59S,Morbidelli2002mcma.book.....M,Papaloizou2015}.  This relation ensures that the associated Laplace angle $ \Psi_{bcd} = \phi_1-\phi_2 = 2\lambda_b-5\lambda_c+3\lambda_d $, evolves slowly ($d\Psi_{bcd}/dt \approx 0$), indicating that the three planets might indeed be trapped in a Laplace chain of 3/2 resonances. 

\medskip
We explored the possibility that the remaining ``unmatched'' dips correspond to planets that continue the first-order generalized Laplace resonant chain of HD~110067~b, c, and d. Using a Laplace relation similar to the one above, we can predict the possible orbital period of the planets, in a similar fashion as the discoveries of TOI-178~f and TRAPPIST-1~h \cite{Leleu2021A&A...649A..26L,Luger2017NatAs...1E.129L}. We assume that the next planet in the system must continue the generalized Laplace chain of first-order resonances. Considering the most common first-order MMRs (2/1, 3/2, 4/3, 5/4, 6/5), the only continuation of the chain that matches the spare duo-transit supports the presence of a fourth planet in the system, HD~110067~e, with an orbital period of 30.7931\,days (in a 3/2 MMR with planet d).

\medskip
We are then left with two mono-transits in TESS Sector 49, which we assume that they correspond to two separate planets f and g that continue the chain. With single transits, we cannot use the same method as above and must introduce a novel argument. In nature, all known three-body resonant chains are close to an equilibrium, i.e. an equilibrium value of their angles $\Psi$ \cite{Gozdziewski2016MNRAS.455L.104G,Leleu2021A&A...649A..26L,Agol2021,Dai2023AJ....165...33D} (Extended Data Fig.~\ref{fig:Laplace_amp}). Assuming low eccentricities, only the period and a single transit of each planet are required to estimate the value of $\Psi$. We can therefore try out different periods for planets f and g, then use their single transits to estimate $\Psi_{def}$ and $\Psi_{efg}$, and see if it lands close to an equilibrium of the chain. Here, we try the same set of first-order MMRs (2/1, 3/2, 4/3, 5/4, 6/5) between e and f and between f and g (50 combinations in total, see Methods). Out of those, the only combination not excluded by existing data that can be close to an equilibrium of the chain is the one where $P_f/P_e=4/3$ and $P_g/P_f=4/3$, yielding the three outer generalized Laplace angles at less than 20 degrees from the closest equilibrium (Methods, Extended Data Table~4, and Extended Data Fig.~\ref{fig:Cminimize}). 

\medskip
According to this prediction, planets HD~110067~f and g would have orbital periods of 41.0575 and 54.7433 days, respectively. If so, both planets would have transited during TESS Sector 23 observations, but during a time when the effect of scattered light and the relative contributions of the Earth and Moon to the image backgrounds were highly significant. These frames are typically discarded, but a detailed reprocessing of the TESS Sector 23 observations triggered by this prediction showed two additional transit events at 1943.6\,TJD and 1944.1\,TJD matching exactly our model based on the system's resonant dynamics (Fig.~\ref{fig:phot}). Additionally, a ground-based multi-instrument photometric campaign to catch a predicted transit of HD~110067~f (Methods and Extended Data Fig.~\ref{fig:groundcampaign}) recovered a statistically significant detection ($\Delta$WAIC = 9.5) with a consistent depth and duration at the expected time, confirming its orbital period. 

\begin{figure}[t]
    \centering
    \includegraphics[width=0.99\columnwidth]{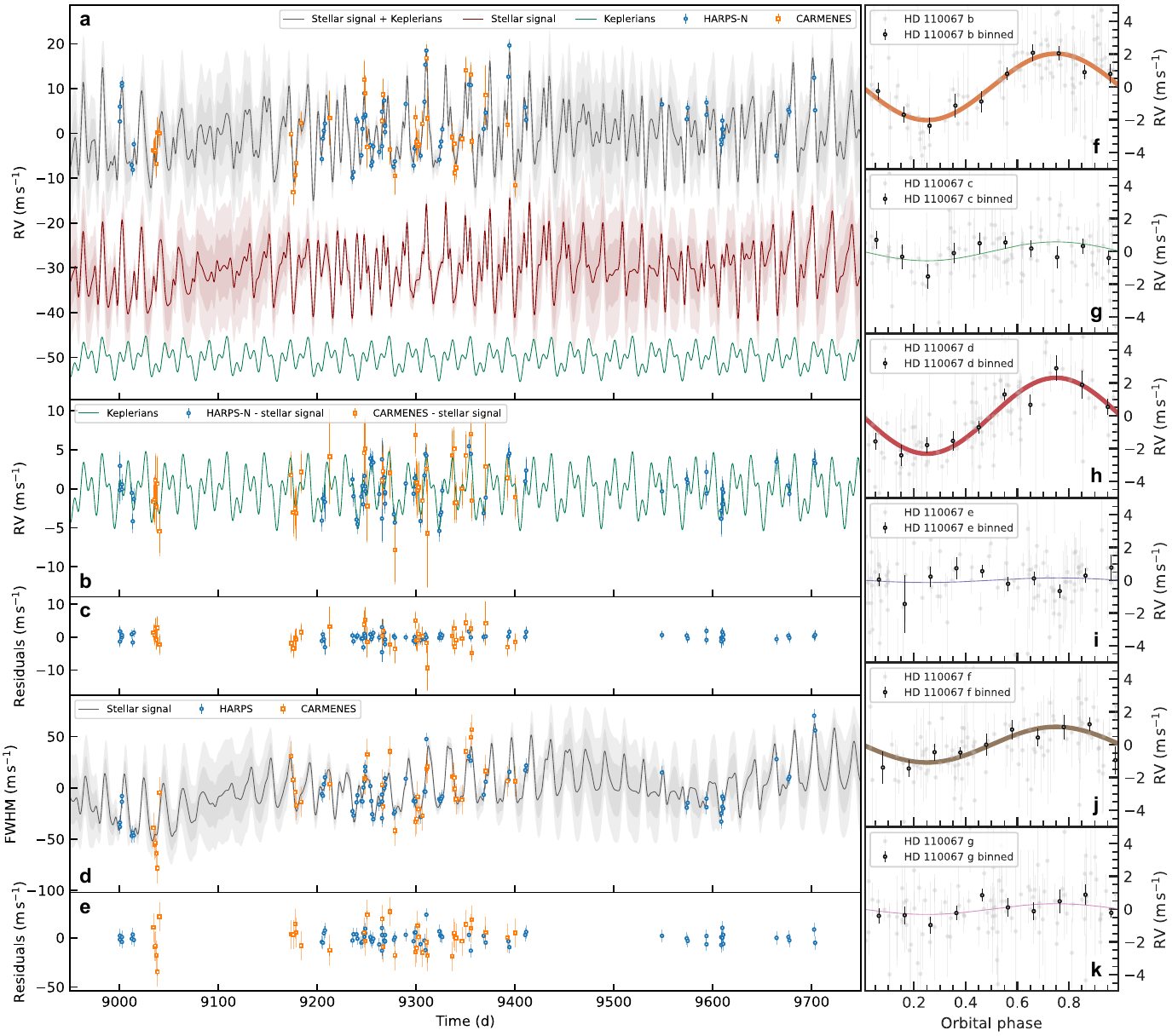}
    \caption{\sep \textbf{Radial velocity data from CARMENES and HARPS-N of HD~110067.} Time units are the same as Fig.~\ref{fig:phot}. \textbf{a-e,} radial velocity (RV) and full-width at half-maximum (FWHM) time-series after being corrected by inferred offsets. Each panel shows: RV data together with full, stellar, and planetary inferred models (\textbf{a}); RV data with the stellar model subtracted (\textbf{b}); RV residuals (\textbf{c}); FWHM data together with the inferred stellar model (\textbf{d}); and FWHM residuals (\textbf{e}). HARPS-N (blue) and CARMENES (orange) measurements are shown with solid circles with 1$\sigma$ error bars with a semi-transparent error bar extension accounting for the inferred jitter. The solid lines show the inferred full model coming from our multi-dimensional Gaussian process model (Methods), lightly shaded areas showing the 1 and 2$\sigma$ credibility intervals of the corresponding model. For the RV time series (\textbf{a}) we also show the inferred stellar (red) and planetary (green) recovered signals with an offset for better clarity. \textbf{f-k,} phase-folded RV signals for all the planets following the subtraction of the systemic velocities, stellar signal, and other planets. Nominal RV observations are shown as light gray points. Solid points show data binned to a tenth of the orbital phase. The inferred model is shown with a solid line following the color-coding of Fig.~\ref{fig:phot}. The planets with mass measurement uncertainties smaller than 3$\sigma$ (\textbf{f,h,j}) are marked with thicker lines. }
    \label{fig:rvs}
\end{figure}

\medskip
Also, we collected high-precision radial velocities of the star with the CARMENES \cite{CARMENES20} and HARPS-N \cite{HARPN} instruments. The dominant signals in the data correspond to the star's magnetic activity rather than the planetary companions but, by applying state-of-the-art analyses to model stellar activity (Methods), we could independently confirm the detection of HD~110067~f, measuring an orbital period and phase (using agnostic priors on both quantities) that matched the transits. We used the radial velocities of the system to measure precise masses for three of the planets (HD~110067~b, d, and f) and place upper limits on the remaining ones (Fig.~\ref{fig:rvs}). Although transit timing variations caused by the planets' mutual gravitational interactions are expected in resonant systems \cite{holman2005}, our photometric analysis did not measure any significant deviation from the linear ephemeris (below 5\,min for the inner triplet), likely due to the low number of individual transits for each planet. Further monitoring of the system will enable an independent measurement of the planetary masses using this technique.

\begin{figure}[t]
    \centering
    \includegraphics[width=0.99\columnwidth]{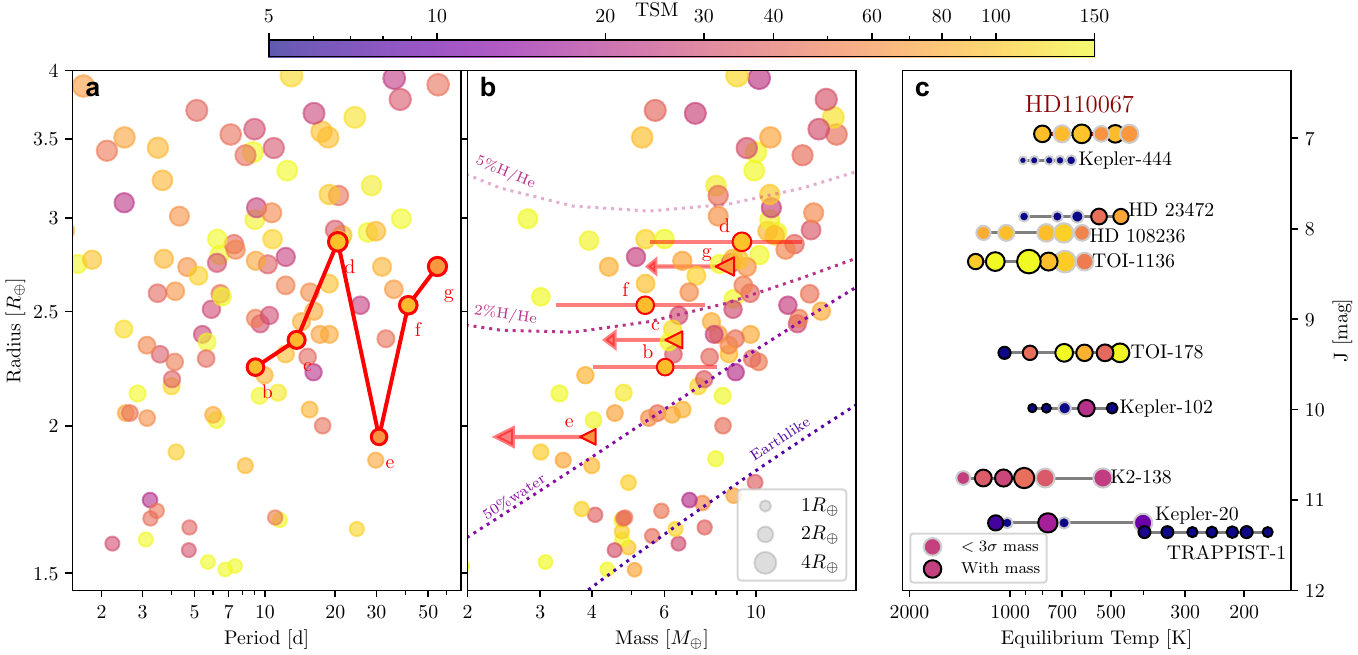}
    \caption{\textbf{Properties of the HD~110067 system compared to the known sub-Neptune-sized planet population. } \textbf{a,} period-radius diagram. \textbf{b,} mass-radius diagram. Mass-radius relationships for different planet compositions are taken from ref.~\cite{Zeng2019}. The 3$\sigma$ mass upper limits are shown for planets HD~110067~c, e, and g; 1$\sigma$ error bars for the rest. \textbf{c,} host star brightness in $J$-band magnitude as a function of planet equilibrium temperature showing systems with five or more transiting planets. Lower magnitudes indicate brighter stars. In all plots, point size is proportional to the planet radius while the color represents a proxy of the expected atmospheric scale height in transmission spectroscopy using JWST following the metric of ref.~\cite{Kempton2018PASP..130k4401K}. Where mass is not known, we use the empirical mass-radius relation of ref.~\cite{Otegi2020} to compute this metric. Planet population properties were retrieved from the NASA Exoplanet Archive on May 2023.} 
    \label{fig:tsm_mr}
\end{figure}

\medskip
The HD~110067 planetary system is thus comprised of at least six transiting planets orbiting in a chain of first-order MMRs (3/2---3/2---3/2---4/3---4/3). The planets have radii ranging between 1.94 and 2.85\,times the radius of the Earth, orbital periods between 9 and 55\,days, and equilibrium temperatures between 440 and 800\,K (Fig.~\ref{fig:tsm_mr}). Thanks to the constraints on the planetary masses and their location above the radius valley \cite{Fulton17,2018MNRAS.479.4786V}, our internal composition modeling concludes that all the planets in the system (with the exception perhaps of planet e, which remains undetected in the radial velocity data) must possess large hydrogen-dominated atmospheres to explain their relatively low bulk densities (see Supplementary). A summary of the most relevant properties of the system is presented in Table~1. We note that for both planets e and g their orbital period measurement relies on a prediction based on the dynamical properties of the system, but an independent third transit observation confirming each orbit has not been obtained yet. Given the low mutual inclination of the system (below 1\,deg), additional transiting planets may yet be found at periods longer than 70\,days, which would correspond to orbits within or beyond the habitable zone of the star \cite{Kasting1993Icar..101..108K,Kopparapu14}. 

\medskip
From an observational point of view, HD~110067 is the brightest star found to host more than four transiting exoplanets. The current delicate configuration of the planetary orbits in HD~110067 rules out any violent event over the billion-year history of the system \cite{Izidoro2021}, making it a rare ``fossil'' \cite{Fabrycky2014} to study migration mechanisms and the properties of its protoplanetary disk in a pristine environment. The combination of host star brightness and the inferred presence of extended atmospheres in the majority of its planets makes HD~110067 the most favorable multi-planetary sub-Neptune system to be observed in transmission spectroscopy with JWST (Fig.~\ref{fig:tsm_mr}). HD~110067 offers a chance to gain insight into the nature of sub-Neptune planets and where, how, and under what conditions resonant chains form and survive.

\clearpage
\newpage

\begin{table}[t]
    \footnotesize 
    \caption{\textbf{Summary of the planetary parameters of the HD~110067 system.}}
    \label{tab:derivedparams}
    \begin{tabular}{lrrr} 
        \hline
        \hline
        \noalign{\smallskip}
        Parameter$^{(a)}$ & HD~110067~b & HD~110067~c  & HD~110067~d  \\
        \noalign{\smallskip}
        \hline
        \noalign{\smallskip}
        \multicolumn{4}{c}{\it Fit parameters} \\[0.1cm]
        \noalign{\smallskip}
        $P$ (d)                          & $9.113678 \pm 1\times10^{-5}$       & $13.673694\pm2.4\times10^{-5}$   & $20.519617\pm4\times10^{-5}$    \\[0.1 cm]
        $t_0$$^{(b)}$ (d)                      & $2640.15797\pm0.00036$        & $2657.45704\pm0.0007$    & $2708.20282\pm0.0008$     \\[0.1 cm]
        $R_{\rm p}/R_\star$                 & $0.0256\pm0.0002$           & $0.0278\pm0.0003$       & $0.0332\pm0.0003$     \\[0.1 cm]
        $b = (a/R_\star)\cos i_{\rm p}$     & $0.355\pm0.033$             & $0.155\pm0.078$         & $0.488\pm0.023$      \\[0.1 cm]
        $a/R_\star$                         & $21.66\pm0.24$               & $28.37\pm0.30$            & $37.21\pm0.40$      \\[0.1 cm]
        $K$ (m\,s$^{-1}$)                   & $2.03^{+0.62}_{-0.65}$        & $< 1.55$                  & $2.32^{+0.89}_{-0.88}$    \\[0.1 cm]
        \noalign{\smallskip}
        \multicolumn{4}{c}{\it Derived physical parameters} \\[0.1cm]
        \noalign{\smallskip}
        $M_{\rm p}$ ($M_\oplus$)            & $5.69_{-1.82}^{+1.78}$    & $< 6.3$                  & $8.52_{-3.25}^{+3.31}$     \\[0.1 cm]
        $R_{\rm p}$ ($R_\oplus$)            & $2.200 \pm 0.030$         & $2.388 \pm 0.036$         & $2.852 \pm 0.039$    \\[0.1 cm]
        $a_{\rm p}$ (au)                    & $0.0793 \pm 0.00096$      & $0.1039 \pm 0.0013$      & $0.1362 \pm 0.0017$     \\[0.1 cm]
        $T_\textnormal{eq}$ (K)$^{(c)}$     & $800 \pm 10$              & $699 \pm 9$              & $602 \pm 8$     \\[0.1 cm]
        $t_T$ (d)$^{(d)}$                & $0.12895\pm0.00078$           & $0.15435\pm0.00121$       & $0.15907\pm0.00158$    \\[0.1 cm]
        $i$ (deg)  & $89.061\pm0.099$   & $89.687\pm0.163$  & $89.248\pm0.046$  \\[0.1 cm]
        \noalign{\smallskip}
        \hline
        \hline
        \hline
        \noalign{\smallskip}
         & $\dagger$HD~110067~e & HD~110067~f & $\dagger$HD~110067~g \\
        \noalign{\smallskip}
        \hline
        \noalign{\smallskip}
        \multicolumn{4}{c}{\it Derived fit parameters} \\[0.1cm]
        \noalign{\smallskip}
        $P$ (d)                          & $30.793091\pm1.2\times10^{-5}$       & $41.05854\pm1\times10^{-4}$        & $54.76992\pm2\times10^{-4}$ \\[0.1 cm]
        $t_0$ (d)                               & $2646.0919\pm0.0011$      & $2641.5763\pm0.001$      & $2656.0921\pm0.002$ \\[0.1 cm]
        $R_{\rm p}/R_\star$                 & $0.0226\pm0.0004$           & $0.03026\pm0.00039$           & $0.0303\pm0.00051$     \\[0.1 cm]
        $b = (a/R_\star)\cos i_{\rm p}$     & $0.113\pm0.074$              & $0.337\pm0.043$             & $0.338\pm0.087$    \\[0.1 cm]
        $a/R_\star$                         & $48.77\pm0.52$                & $59.08\pm0.52$                & $71.59\pm0.78$    \\[0.1 cm]
        $K$ (m\,s$^{-1}$)                   & $< 0.80$                      & $1.09^{+0.40}_{-0.42}$        & $< 1.30$    \\[0.1 cm]
        \noalign{\smallskip}
        \multicolumn{4}{c}{\it Derived physical parameters} \\[0.1cm]
        \noalign{\smallskip}
        $M_{\rm p}$ ($M_\oplus$)            & $< 3.9$                  & $5.04_{-1.94}^{+1.89}$    & $<8.4$  \\[0.1 cm]
        $R_{\rm p}$ ($R_\oplus$)            & $1.940 \pm 0.040$         & $2.601 \pm 0.042$         & $2.607 \pm 0.052$ \\[0.1 cm]
        $a_{\rm p}$ (au)                    & $0.1785 \pm 0.0022$       & $0.2163 \pm 0.0026$      & $0.2621 \pm 0.0032$  \\[0.1 cm]
        $T_\textnormal{eq}$ (K)             & $533 \pm 7$               & $489 \pm 6$               & $440 \pm 6$  \\[0.1 cm]
        $t_T$ (d)                        & $0.20265\pm0.00244$           & $0.2152\pm0.0022$           & $0.2361\pm0.0052$    \\[0.1 cm]
        $i$ (deg)  & $89.867\pm0.089$   & $89.673\pm0.046$  & $89.729\pm0.073$ \\[0.1 cm]
        \noalign{\smallskip}
        \hline
    \end{tabular}
     \\ \\
      \textbf{Footnotes.}  
      \textit{(a)} Error bars denote the $68\%$ posterior credibility intervals. Upper limits correspond to $99.7\%$ posterior credibility intervals.
      \textit{(b)} Mid-transit epoch is measured using the full baseline of transits observed with TESS, CHEOPS, and ground-based facilities. Units are expressed in terms of the TESS Julian Date, TJD = BJD$-$2457000, where BJD is the Barycentric Julian Date in units of days. Reference time system is TDB (Barycentric Dynamical Time). 
      \textit{(c)} Equilibrium temperatures were calculated assuming zero Bond albedo and perfect energy redistribution.
      \textit{(d)} Transit duration $t_T$ is measured from the first to the last contact. \\
$\dagger$ The orbital period for planets e and g have been measured from their assigned duo-transits placing a prior based on our prediction from the resonant chain analysis. 
\end{table}

\clearpage
\newpage



\section*{Methods}

\subsection{Data}

\subsubsection{TESS photometry} \label{subsec:tess}

HD~110067 was observed in sectors S23 and S49 (2020~March~18 to 2020~April~16 and 2022~February~26 to 2022~March~26, respectively) of the Transiting Exoplanet Survey Satellite (TESS) \cite{Ricker2015}. The star was included on the \textit{TESS} Candidate Target List (CTL) \cite{Stassun2018AJ....156..102S} and therefore the target was observed at 120\,s cadence in both sectors. The target images are processed by the TESS Science Processing Operations Center (SPOC) pipeline at NASA Ames \cite{SPOC}, which calibrates the pixels, performs simple aperture photometry (SAP), flags poor-quality data, and removes systematic trends to create the so-called ``Presearch Data Conditioning'' light curve (PDCSAP) \cite{Stumpe2012, Stumpe2014, 2012PASP..124.1000S}. Finally, the SPOC pipeline runs a wavelet-based transiting planet search for periodic exoplanets \cite{2002ApJ...575..493J,2010SPIE.7740E..0DJ,2020TPSkdph} which in the case of the S23 data revealed two threshold-crossing events (TCEs) --- i.e. candidate planets --- which passed Data Validation checks \cite{Twicken:DVdiagnostics2018,Li:DVmodelFit2019}. Manual vetting of these two candidates resulted in the assignment of two TESS Object of Interest (TOI), namely TOI-1835.01 and TOI-1835.02 \cite{2021ApJS..254...39G}.

\medskip
The first TCE was alerted by SPOC as TOI-1835.01 with a period of 5.641\,d. Although two of the three transits associated with this ephemeris appeared to be of similar depth and duration, the third did not appear to be associated with a clear transit. However, this could have been due to proximity to a systematic dip caused by a momentum dump. A second TCE, TOI-1835.02, was alerted as a single transit at $1948.98$\,TJD TESS Julian Date $\mathrm{(TJD \equiv BJD-2457000)}$, where BJD is the Barycentric Julian Date in units of days, as opposed to the period of 11.107\,d proposed by SPOC due to an apparent discrepancy in the transit depth between the two transits purportedly linked by the transiting planet search. On the other hand, a peak in the background flux ruled out the planetary origin of an apparent transit feature at TJD=1940.5. Both the depth and shape of this feature were strongly dependent on the detrending method used in the lightcurve, which is indicative of its spurious nature.

\medskip
HD~110067 was later reobserved by TESS in S49. In order to confirm the TCEs and plan immediate follow-up, we downloaded the TESS Image CAlibrator Full Frame Images (TICA FFI) \cite{Fausnaugh2020} only a week after downlink for each of the first and then second orbits. We computed a light curve from the 10-minute cadence TICA FFI cut-outs using SAP and clipped regions of high brightness due to the Earth and Moon, as well as parts of the light curve affected by systematics such as momentum dumps. The first orbit alone revealed at least 5 new clear transit features. These all appeared to have varying depths and durations, and none were compatible with the 5.64\,d period implied by TOI-1835.01. The second orbit showed three more clear transit events, making a total of eight in S49 and five in S23. A 20-second cadence target pixel file was later made available for HD~110067, which resulted in higher-precision photometry.

\medskip
Manual vetting of both S23 and S49 PDCSAP light curves revealed unexpectedly large systematic uncertainties (Gini's mean difference, i.e., the average point-to-point absolute difference, of 215\,ppm in 3-hour bins) which were absent from the uncorrected SAP flux. This has been seen in many bright stars which possess stellar variability \cite[e.g.,][]{2021AJ....162...54H, Osborn2022HIP8618}. In order to better correct these systematics, we performed a custom extraction of the TESS light curves for both sectors using the quaternion detrending technique against spacecraft motion developed by \cite{Vanderburg2019ApJ...881L..19V}. This involved fitting a model consisting of a linear combination of a basis spline (with breakpoints spaced every 1.5\,d to model long-timescale stellar or instrumental variability) and decorrelation with parameters linked to systematic flux changes, namely the means and standard deviations of the spacecraft quaternion time series (and the squared time series) within each exposure. Using a linear least squares technique (matrix inversion), we solved for the best-fit coefficients of our free parameters while iteratively excluding $3\sigma$ outliers from the fit until convergence was reached. After calculating the best-fit model for each aperture's systematics, we then subtracted it from the uncorrected light curve and identified the aperture that produced the light curve with the lowest photometric scatter. The final light curves used in our subsequent analyses (Gini's mean difference of 130\,ppm in 3-hour bins) are shown in Fig.~\ref{fig:phot}.

\medskip
Finally, as discussed above, a large portion of the data was missing from the PDCSAP light curves due to high levels of scattered light and sky background from the Earth and Moon. 
The dates affected coincide with the potential transit events of planets f and g based on our dynamical model prediction. Therefore, in order to recover data affected by scattered light, we performed a custom extraction of the TESS light curves for both sectors using a pixel level decorrelation (PLD) method \cite{Deming_2009,Luger2016AJ....152..100L, exoplanet:luger18} implemented in the \texttt{PLDCorrector} class of the community Python package \texttt{lightkurve} \cite{lightkurve}. This method employs (i) a spline polynomial fit to describe stellar variability, (ii) Principal Component Analysis (PCA) eigenmodes to model the background light, and (iii) the PLD technique to account for pointing and mechanical effects. Before applying the \texttt{PLDCorrector}, we add the background flux and errors estimated by the TESS SPOC pipeline back onto the SAP light curve. Flux level, fraction, and crowding adjustments are then applied to the corrected light curve. To automatically optimize the selection of parameter values for the PLD corrector, we evaluate the resulting light curve using the Savitsky-Golay Combined Differential Photometric Precision (sgCDPP) proxy algorithm \citep{Gilliland_2011,Van_Cleve_2016} implemented in \texttt{lightkurve}. For a grid of PLD corrector parameter values, we calculate the harmonic mean of these quantities and select the corrected light curve that minimizes it. We use this data for the cadences missing in the quaternion-detrended light curve in our final analyses (marked with a different color in Fig.~\ref{fig:phot}).

\subsubsection{CHEOPS photometry}

The CHaracterising ExOPlanets Satellite (CHEOPS) mission is a European Space Agency small-class mission dedicated to studying bright, nearby exoplanet host stars for the purpose of making high-precision photometric observations of transiting planets \cite{CHEOPS}. We collected 19 separate visits of HD~110067 with CHEOPS between 2022~April~11 and 2022~May~17 under Guaranteed Time Observing programs ID-048 and ID-031. The goal of these observations is (1) to confirm the true orbital period of single- and duo-transiting planet candidates and (2) to improve the planetary radius precision and ephemeris of confirmed planets. This has been done for large planets producing deep eclipses from the ground \cite{Schanche2022,Ulmer-Moll2022}, and for small planets from space \cite{Osborn2021,Osborn2022HIP8618,Tuson2023}. An observing log summarizing the duration of each visit, its average observing efficiency (considering the gaps produced by Earth occultations or passages over the South Atlantic Anomaly along the spacecraft's low-Earth orbit), and photometric precision are presented in Extended Data Table~1.

\medskip
To provide the highest quality photometric precision, we opted to perform custom photometric extraction of the CHEOPS imagettes using point-spread-function (PSF) photometry as implemented by the \texttt{PIPE} package \cite{Szabo2021A&A...654A.159S,Morris2021A&A...653A.173M}. For bright targets such as HD~110067, light curves generated with \texttt{PIPE} exhibit lower median absolute differences than those generated by the CHEOPS Data Reduction Pipeline \cite{Hoyer2020A&A...635A..24H}.  The shorter cadence of the CHEOPS imagettes allows a higher cadence light curve, and PSF detrending is also better at removing trends due to systematic factors and background stars. As various PSF models have already been generated and vary as a function of stellar temperature, we opted to use a PSF generated using the star HD~189733, with a similar spectral type of HD~110067. In order to preserve inter-visit flux differences, we normalized the entire CHEOPS data together instead of individually. This revealed clear visit-to-visit flux differences due to stellar rotation with an amplitude larger than that of TESS (as stellar activity is typically more pronounced at bluer bandpasses). The final light curves used in our subsequent analyses are shown in Fig.~\ref{fig:phot} and Fig.~S1 in the Supplementary.

\subsubsection{Ground-based photometric campaign}

We carried out a campaign on the night of May 23rd, 2022 to attempt to confirm the 41.05-day period orbit of HD~110067~f as predicted by our resonance chain analysis. Photometric observations were taken using 14 telescopes using seven different filters, which observed from various locations to continuously cover a temporal baseline of more than 11 hours (between 2022-05-23UT22:52:55 and 2022-05-24UT10:01:33). This window is long enough to catch the 5-hour transit expected from 02:52 to 07:12UT. However, no single location was able to cover both ingress and egress. A summary of the observations is shown in Extended Data Table~2. Details from each individual observation are shown below. Extended Data Fig.~\ref{fig:groundcampaign} shows the data and best-fit models as discussed in Sect.~\ref{sec:groundphot}.

\paragraph{Teide Observatory}
We observed HD~110067 on May 23rd, 2022 using the MuSCAT2 instrument installed at the 1.5-m Telescopio Carlos Sánchez (TCS) located at the Teide Observatory, Spain \cite{2019JATIS...5a5001N}. The images were taken simultaneously in $g$, $r$, $i$, and $z_s$ filters with the telescope heavily defocused and with short exposure times of 3 to 5 sec, depending on the band, to avoid saturation. Relative light curves for each band and instrument of HD~110067 were extracted by aperture photometry using a custom pipeline \cite{Parviainen2020A&A...633A..28P} with optimal aperture radii of $8.\!\!^{\prime\prime}1$ to $11.\!\!^{\prime\prime}3$ depending on the band. Note that there was a technical problem on the dome of the TCS between BJD-2459723 = 0.488 and 0.526; we discarded the data taken during this period.

\medskip
We also observed HD~110067 on May 23rd, 2022 with one of the 1-m telescopes from Las Cumbres Observatory (LCO) global network located at the Teide Observatory, Spain \cite{Brown2013}. The observations were obtained through Director's Discretionary Time program 2022A-005 (PI: Wilson). We collected 181 frames with an exposure time of 20\,s, covering 2.5\,h, using the $4096\times4096$\,pix SINISTRO camera. The images were calibrated by the standard LCO {\tt BANZAI} pipeline \cite{McCully2018}. Differential photometric data were extracted using {\tt AstroImageJ} (AIJ) \citep{Collins:2017}.

\paragraph{Paranal Observatory}
We observed HD~110067 on May 23rd, 2022 using the Next Generation Transit Survey (NGTS) facility located at ESO's Paranal Observatory in Chile \cite{Wheatley2018}. NGTS consists of twelve 20-cm, f/2.8 telescopes with Andor cameras and red-sensitive (600--900\,nm) deep-depletion e2v CCDs. Nine NGTS telescopes observed from 23:14 to 04:35UT, covering a predicted transit ingress of HD~110067~f, and spanning an airmass range of 1.7--2.5. Two telescopes started observing two hours late due to a technical issue. All nine telescopes were defocused to avoid saturating the bright target star during the 10-second exposures. The NGTS camera shutters were not functional and so were kept open during the entire observing block. That caused the stars to streak during the 1.5-second readout sequences but without any apparent detrimental effect on the photometry. Observing without using the shutters is now the standard operation mode of NGTS. We performed standard differential aperture photometry, using large aperture radii of 6.5--8.0 pixels, and carefully selecting comparison stars to avoid those that exhibited variability. The light curve of each telescope was normalized individually and no detrending was performed. 

\paragraph{F. L. Whipple Observatory}
We observed HD~110067 on May 24th, 2022 using the \textit{Tierras} instrument installed at the refurbished 1.3-m telescope located at the F.\,L.\,Whipple Observatory atop Mount Hopkins, Arizona, United States. The instrument is designed to regularly achieve a photometric precision of 250\,ppm on a time scale of both 10\,min and a complete observing season. The design choices that permit this precision include a four-lens focal reducer and field-flattener that increase the field-of-view of the telescope, a custom narrow bandpass filter centered around 863.5\,nm to minimize precipitable water vapor errors, and a fully automated mode of operation \cite{Tierras}. A total of 1262 4-second exposures were gathered with \textit{Tierras} for HD~110067. Astrometric calibrations were done in real-time during data gathering and were stored in WCS headers in the FITS files. The FITS files were then passed through the \textit{Tierras} image reduction pipeline to perform bias corrections and image stitching (the CCD chip is read out through separate amplifiers). AIJ was used for photometric extraction. These observations were gathered shortly after \textit{Tierras} started science operations, and the data were not flat-fielded since knowledge of the flat-field was incomplete at the time. The RMS of the 15-minute binned data is 323\,ppm. The photometric precision on this target is ultimately limited by scintillation, as the target was observed down to an airmass of 2.37. The observations were mildly affected by cirrus.

\paragraph{San Pedro Mártir Observatory}
We observed HD~110067 on May 24th, 2022 with the 1-m SAINT-EX telescope at the Observatorio Astron\'omico Nacional de la Sierra de San Pedro M\'artir in Baja California, Mexico \cite{2020A&A...642A..49D}. SAINT-EX is equipped with a deep-depleted and back-illuminated Andor IKON CCD and a filter wheel. The observations were defocused and acquired in the ``zcut'' filter, a custom filter optimized to reduce the systematic uncertainties in the light curves of red stars due to precipitable water vapor, with an exposure time of 10\,s. The data were reduced with AIJ using the standard corrections for bias, flat-fielding, and dark current. AIJ was also utilized to do the aperture photometry of the time series, producing the light curves and relevant meta-data. The observations were mildly affected by high-altitude cirrus.

\paragraph{Haleakala Observatory}
We observed HD~110067 on May 24th, 2022 using the MuSCAT3 instrument mounted on the 2-m Faulkes Telescope North (FTN) at Haleakala Observatory on Maui, Hawaii, United States \cite{2020SPIE11447E..5KN}. The images were taken simultaneously in $g$, $r$, $i$, and $z_s$ filters with the telescope heavily defocused and with short exposure times of 3 to 5 sec, depending on the band, to avoid saturation. Relative light curves for each band and instrument were extracted by aperture photometry using a custom pipeline \cite{2011PASJ...63..287F} with optimal aperture radii of $8.\!\!^{\prime\prime}1$ to $11.\!\!^{\prime\prime}3$ depending on the band. There was a guiding issue on the FTN around BJD-2459723 = 0.795, which caused a large shift of the stellar positions on the detectors; we treated the MuSCAT3 data as two independent datasets separated by that time.

\subsubsection{High-resolution imaging}

As part of our standard process for validating transiting exoplanets, and to assess the possible contamination of bound or unbound companions on the derived planetary radii \cite{ciardi2015}, we observed HD~110067 with near-infrared (NIR) adaptive optics (AO) imaging at Palomar Observatory and with optical speckle imaging at Gemini North. Gaia DR3 is also used to provide additional constraints on the presence of undetected stellar companions and wide companions.  No close-in ($\lesssim 1\arcsec$) stellar companions were detected by either the NIR adaptive optics or optical speckle imaging.
    
\paragraph{Palomar Observatory}
The Palomar Observatory observations of HD~110067 were made with the PHARO instrument \cite{hayward2001} behind the natural guide star AO system P3K \cite{dekany2013} on 2020~Jan~08 in a standard 5-point quincunx dither pattern with steps of 5\arcsec\ in the narrow-band Br-$\gamma$ filter. Each dither position was observed three times, offset in position from each other by 0.5\arcsec\ for a total of 15 frames; with an integration time of 1.4 seconds per frame, respectively for total on-source times of 21 seconds. PHARO has a pixel scale of $0.025\arcsec$ per pixel for a total field of view of $\sim25\arcsec$. The sensitivities of the final combined AO image were determined by injecting simulated sources azimuthally around the primary target every $20^\circ $ at separations of integer multiples of the central source's FWHM \cite{furlan2017}. The Palomar data have a sensitivity $\Delta {\rm mag} = 2$ at 0.1\arcsec\ and $\Delta {\rm mag} = 9$ at 1\arcsec; the final sensitivity curve is shown in Fig.~S2 of the Supplementary.

\paragraph{Gemini Observatory}
We observed HD~110067 with the `Alopeke speckle imaging camera at Gemini North on 2020~June~10 \cite{Scott2021FrASS...8..138S}. We obtained five sets of 1000 frames, each frame having an integration time of 60~ms, obtaining images in each of the instrument's two bands (centered at 562~nm and 832~nm). The observations were reduced using our standard software pipeline \cite{Howell(2011)} and reached a 5$\sigma$ sensitivity of $\Delta {\rm mag} = 7$ (blue channel) and $\Delta {\rm mag} = 6.8$ (red channel) at separations of 0.5\arcsec. The reconstructed speckle images show no evidence of additional nearby point sources. The final sensitivity curve is shown in Fig.~S2 of the Supplementary.
	
\paragraph{Gaia Space Observatory}
In addition to the high-resolution imaging, we have utilized Gaia to identify any wide stellar companions that may be bound members of the system \cite{mugrauer2020,mugrauer2021}. There are no additional widely separated companions identified by Gaia that have the same distance and proper motion as HD~110067. Additionally, the Gaia DR3 astrometry provides additional information on the possibility of inner companions that may have gone undetected by either Gaia or the high-resolution imaging data. The Gaia Renormalised Unit Weight Error (RUWE) is a metric, similar to a reduced chi-square, where values that are $\lesssim 1.4$ indicate that the Gaia astrometric solution is consistent with the star being single whereas RUWE values $\gtrsim 1.4$ may indicate an astrometric excess noise, possibly caused the presence of an unseen companion \cite[e.g., ][]{ziegler2020}.  HD~110067 has a Gaia EDR3 RUWE value of 0.94 indicating that the astrometric fit is consistent with a single-star model.

\subsubsection{Radial velocity monitoring}

\paragraph{Calar Alto Observatory}
We observed HD~110067 using the CARMENES instrument \cite{CARMENES20} installed at the 3.5-m telescope of Calar Alto Observatory in Almería, Spain, between 3 July 2020 and 4 July 2021. We collected 39 high-resolution spectra under the observing programs F20-3.5-011 (PI: Nowak) and H20-3.5-013 (PI: Luque). Radial velocities and additional spectral indicators were derived using \texttt{raccoon} \cite{Lafarga2020} and \texttt{serval} \cite{SERVAL}. While the mean internal precision of the template matching \texttt{serval} RVs is $3.1\,\mathrm{m\,s^{-1}}$, the precision of the cross-correlation method \texttt{raccoon} RVs is $2.9\,\mathrm{m\,s^{-1}}$, so we used the latter in our analyses.

\paragraph{Roque de los Muchachos Observatory}
We observed HD~110067 with the HARPS-N spectrograph mounted at the 3.6\,m Telescopio Nazionale Galileo \cite{HARPN} of Roque de los Muchachos observatory in La Palma, Spain, between 30 May 2020 and 4 May 2022. We collected 72 high-resolution spectra under the observing programs CAT19A\_162 (PI: Nowak), CAT21A\_119 (PI: Nowak) and ITP19\_1 (PI: Pall\'e) that were used to measure the photospheric properties of the star and precise radial velocities. Radial velocities and additional spectral indicators were derived using an online version of the DRS pipeline \cite{2014SPIE.9147E..8CC}, the YABI tool, and \texttt{serval} \cite{SERVAL}. Both the YABI- and \texttt{serval}-derived radial velocities have a median internal precision of $1.0\,\mathrm{m\,s^{-1}}$, but we used the YABI ones (based on the cross-correlation method) in our final analyses for consistency with the CARMENES dataset.

\subsection{Stellar parameters}

\subsubsection{Photospheric parameters and abundances}

To properly characterize the planetary system around HD~110067, we first conduct a series of analyses to determine the properties of the host star. We derive the stellar spectral parameters by applying the widely used \texttt{ARES+MOOG} tools to our co-added HARPS-N spectra \cite{Santos2013,Sousa2014,Sousa2021}. \texttt{ARES} \cite{Sousa2007,Sousa2015} measures the equivalent widths of iron lines in the spectrum that are converted into stellar atmospheric parameters using the \texttt{MOOG} radiative transfer code \cite{Sneden1973} applied to Kurucz model atmospheres \cite{Kurucz1993}. In Extended Data Table~3 we report the effective temperature $T_{\mathrm{eff}}$, surface gravity $\log g$, and metallicity {[Fe/H]}, obtained upon convergence of ionization and excitation equilibria within this method. Additionally, we measure the stellar $v \sin i$ from the HARPS-N spectra using \texttt{ZASPE} \cite{ZASPE}.

\medskip
We further study the photospheric parameters by conducting a classical curve-of-growth analysis on our co-added HARPS-N spectrum using our aforementioned spectral parameters in order to obtain {[Mg/H]} and {[Si/H]} abundances for HD~110067. Utilizing the \texttt{ARES+MOOG} framework detailed above, we obtain the equivalent widths \cite{Sousa2015} for these elements, which are converted to abundances assuming local thermodynamic equilibrium \cite{Sneden1973,Kurucz1993}. The specific details of this analysis are beyond the scope of this paper and can be found in refs.\cite{Adibekyan2012,Adibekyan2015}. We report the stellar abundances in Extended Data Table~3.

\subsubsection{Physical parameters}

Using our spectral parameters and the \texttt{ATLAS} \cite{Kurucz1993,Castelli2003} and \texttt{PHOENIX} \cite{Allard2014} catalogs, we build spectral energy distributions of HD~110067 that we compare to optical and infrared broadband photometry of the star (see Extended Data Table~3) to derive the stellar angular diameter and effective temperature via the infrared flux method \cite{Blackwell1977}. This is conducted in an MCMC approach \cite{Schanche2020,Wilson2022} within which we convert the angular diameter to the stellar radius using the {\it Gaia} EDR3 offset-corrected parallax \cite{Lindegren2021} with model uncertainties accounted for using a Bayesian modeling averaging. We report the stellar radius $R_\star$ in Extended Data Table~3.

\medskip
Last, we complete our stellar characterization by determining the mass and age of HD~110067. We constrain two sets of stellar evolutionary models with help of our derived values for $T_{\mathrm{eff}}$, $\log g$, and $R_\star$ \cite{Bonfanti2021}. On the one hand, we use an isochrone placement algorithm \cite{Bonfanti2015,Bonfanti2016} and interpolate over pre-computed grids of \texttt{PARSEC\,v1.2S} \cite{Marigo2017} isochrones. On the other hand, we use the Code Liègeois d’Évolution Stellaire \cite{Scuflaire2008} combined with a Levenberg-Marquadt minimization scheme \cite{Salmon2021} to optimize the best-fitting evolutionary track. The results from the two methods are combined to determine the mass and age of the star that is reported in Extended Data Table~3.

\medskip
The [Fe/H] and age of HD~110067 indicate that this star could belong either to the galactic thick disk stellar population or be an older member of the galactic thin disk. The values of [Mg/H] and [Si/H], being within $1\sigma$ of [Fe/H], show that the star is not enhanced in $\alpha$-capture elements and are indicative of a typical thin disk chemical composition. We determined the kinematic properties of HD~110067 by using the Gaia EDR3 astrometry to compute the Local Standard of Rest space velocities of this star following ref.\cite{2006MNRAS.367.1329R}. From these velocities, we compute that the probability of kinematic membership in the galactic thin disk is $0.9911 \pm 0.0029$. Thus, we conclude that HD~110067 is on the older, more metal-poor end of the distribution of the galactic thin disk stellar population.

\subsection{Analysis}

\subsubsection{Space-based photometry modeling}

We performed simultaneous modeling of the space-based photometry. We used the quaternion-detrended TESS data combined with the PLD-detrended data for the missing S23 gaps, and the \texttt{PIPE}-detrended CHEOPS data for the three visits containing transits. We built transit models for the six planets with \texttt{exoplanet} \cite{exoplanet:exoplanet}. Due to its nature as a rotating telescope on a near-Earth orbit, even PSF-detrended CHEOPS photometry can include systematic trends. However, these typically correlate with other measurements, for example, roll angle, background, or contaminant flux. In order to not bias the transit model and to better propagate uncertainties on the derived parameters, we performed CHEOPS decorrelation alongside our photometric transit modeling. We first fitted each CHEOPS transit individually alongside multiple possible decorrelation factors, allowing us to assess which decorrelation factors are most useful. This also enabled us to test whether such decorrelation is shared among all CHEOPS visits or individual to a single light curve. From this analysis, we included the following parameters in the linear correlation: position centroids, the second harmonic of the cosine of the roll-angle, $\cos{2\Phi}$, the change in telescope temperature, and quadratic trends with the x-y centroids. CHEOPS data have also been known to contain flux trends that vary stochastically as a function of roll angle over shorter frequencies \cite[see e.g.][]{Delrez2021}. These are not well removed using simple trigonometric functions, hence we also modeled a flexible spline shared between all visits to model shorter-timescale variation. To incorporate stellar variability, a floating mean and flux trend were also fitted to each CHEOPS visit, as well as an individual jitter term. 

\medskip
Informative priors were used on limb darkening parameters using the theoretical quadratic limb darkening parameters for TESS \cite{Claret2018A&A...618A..20C} and CHEOPS \cite{Claret2021RNAAS...5...13C}, with uncertainties inflated to 0.1 in all cases to guard against systematic offsets. The impact parameter and radius ratio are fitted from a broad uniform and log-normal prior, respectively, while the period and mid-transit epoch are fitted using broad normal priors from the transits identified and modeled above. Stellar parameters from Extended Data Table~3 were used as inputs to the model with Gaussian priors. Orbits were assumed circular in all cases which is a good approximation for planetary systems with multiple transiting planets \cite{VanEylenAlbrecht2015,Xie2016,HaddenLithwick2017}. The prior and posterior distributions of each parameter in the model are shown in Table~S1 of the Supplementary.

\subsubsection{Properties of the unmatched transits}

Our first modeling of the TESS space-based photometry was able to account for a total of 5 transits of planet b (two in TESS S23 and three in TESS S49) and 4 transits of planet c (two each in TESS S23 and S49). However, this analysis left six ``unmatched'' transits in the original TESS light curves. In order to pair the transits, we fitted each transit individually using a purely shape-based transit model agnostic to the orbital period using \texttt{MonoTools} \cite{MonoTools}. From this analysis, we then compared each transit in duration-depth space, allowing us to clearly see that both transits from S23 shared unique regions of this parameter space with two more transits seen in S49 (duo-transits), while the two longest-duration transits seen only in S49 were solitary (mono-transits). Extended Data Fig.~\ref{fig:ind_transits} shows this result. 

\medskip
We then modeled both duo- and single-transits using \texttt{MonoTools} fitting. This allows long-period planets to be modeled in a way that the transit model is agnostic of the orbital period with the implied period distributions being manipulated using priors. This technique works for transits with single- or duo-transits. In the case of two transit events separated by a long gap, the planetary transit is fitted leaving the orbital period open, and the implied transit shape is used to calculate the probability for each of the possible period aliases. For single transits, potential orbital period windows are computed. In both cases, the period probability distribution comes from a combination of a simple period prior (longer period planets are geometrically disfavored) \cite{KippingSingleTrans}, an eccentricity prior (eccentric orbits are disfavored in multi-transiting systems) \cite{van2019orbital}, and a stability prior using the orbits of other planets in the system (orbit-crossing is disallowed) \cite[further details in][]{OsbornTOI2076,Osborn2022HIP8618}. The resulting marginalized period predictions for planets HD~110067~d, e, f, and g are shown in Fig.~S3, 
with posterior values of $ 21.6^{+2.9}_{-1.6} $, $ 29.9^{+4.6}_{-3.3}$, $40.1^{+7.1}_{-5.1}$, and $47.0 \pm 8.0$\,days, respectively.

\subsubsection{Continuing the resonant chain}

In this section, we expand the analyses that led to the prediction of the orbits of planets HD~110067~e, f, and g based on the generalized Laplace resonant configuration of the three inner transiting planets in the system. We assume that all events mentioned in the previous section are transits that belong to planets that continue the resonant chain. 
 
\medskip
For transiting systems, generalized three-body Laplace angles can be estimated in 0th order in eccentricity, defined as $\Psi_{e=0}$, from the times of mid-transit and the orbital period of the planets \cite[see e.g.,][]{Mills2016}. This estimation differs from the actual generalized three-body Laplace angle proportionally to the eccentricities \cite[eq. 15 of][]{SiFa2021}. Interestingly, for known systems with a chain of three-body resonances, all $\Psi_{e=0}$ lie close to an equilibrium of the chain, as seen in Extended Data Fig.~\ref{fig:Laplace_amp}. The largest distance is $\sim 43$\,degrees for the inner triplet of K2-138 \cite{Lopez2019}. For HD~110067, the estimated angle $\Psi_{e=0,bcd}$ is also at about $\sim 44$\,degrees from its theorized 180-degree equilibrium. Through the study of transit timing variations over several years, one can get constraints on the underlying generalized three-body Laplace angles. In known cases, one can see that these angles oscillate with amplitudes of a few tens of degrees at most around their equilibrium value, see Fig.~2 of \cite{Gozdziewski2016MNRAS.455L.104G} for Kepler-60, Fig.~25 of \cite{Agol2021} for TRAPPIST-1. 

\medskip
As shown above, the two events at 2646.088\,TJD and 1937.851\,TJD have fully consistent shapes. Among the probable periods computed with \texttt{MonoTools} (Fig.~S3), 
$P_e=30.7931$\,days is the only one that continues the resonant chain, with $P_e/P_d=1.5007$, landing inside the common 3:2 MMR (see Fig.~S4 in the Supplementary). 
We compute the observed value of the associated generalized three-body Laplace angle $\Psi_{e=0,cde}=169.995$\,deg, which is at only $10$\,deg from the expected 180-degree equilibrium. We hence predict a period of $30.7931$\,days for planet HD~110067~e if it is in the resonant chain.

\medskip
For the remaining two mono-transits, we try a set of first-order MMRs (2/1, 3/2, 4/3, 5/4, 6/5) between planet \#4 and \#5 and the same between planet \#5 and \#6 (hence 25 combinations). Each of these combinations has to be tested assuming that the transit at TJD=2641.5778 belongs to the 5th planet, and TJD=2656.0944 belongs to the 6th planet (case A), and vice-versa (case B). Fortunately, many of these 50 possibilities are excluded by existing data. We end up with 4 possibilities for case A and 9 for case B. As seen in Fig.~\ref{fig:Laplace_amp}, all known chains of Laplace resonances have either their estimated generalized three-body Laplace angle $\Psi_{e=0}$, or their actual generalized three-body Laplace angle $\Psi$ close to an equilibrium of the chain. We will hence favor the configurations that are closest to an equilibrium of the chain. For each case, the distance of each estimated angle to its closest equilibrium $\Delta\Psi=\lvert \Psi_{e=0}-\Psi_{eq} \rvert$ is given in Extended Data Table~4. The case A2, with $P_f/P_e=4/3$ and $P_g/P_f=4/3$, comes out as a favorite, with the three outer generalized three-body Laplace angles at less than 20\,deg from the closest equilibrium. In addition, one can note that $4/3$ MMRs are relatively common in resonant chains (see Fig.~S4 in the Supplementary).

\medskip
For completeness, we study the role that the eccentricity of the orbit plays in the prediction. To estimate the generalized three-body Laplace angle 
\begin{equation}
\begin{aligned}
 \Psi = l \lambda_1 - (l+m) \lambda_2 + m \lambda_3 , \\
\end{aligned}
\label{eq:LaplaceAngle}
\end{equation}
at a given epoch, we estimate the value of the $\lambda_j$ as follow.
Transits occur when the true longitude of the planet is equal to $l_0=-\pi/2$. At first order in the eccentricity, 
\begin{equation}
\begin{aligned}
 \lambda_0=l_0-2e\sin (l_0-\varpi) = -\frac{\pi}{2} + 2 e\cos(\varpi) \, .
\end{aligned}
\label{eq:lambdafl}
\end{equation}
We then assume the planet to be in a circular, unperturbed orbit to compute the value of its mean longitude at the time of transit $t_0$, $\lambda_0=-\pi/2$. We hence obtain
\begin{equation}
\begin{aligned}
 \lambda (t)= -  \frac{\pi}{2} + \frac{(t-t_0)}{P} 2\pi\, .
\end{aligned}
\label{eq:lambdaft}
\end{equation}
The error on $\Psi$ made by assuming zero eccentricity is hence, at first order \cite{SiFa2021}:
\begin{equation}
\begin{aligned}
 \lvert \Psi - \Psi_{e=0} \rvert = \lvert 2 l e_1\cos \varpi_1  - 2 (l+m)  e_2\cos \varpi_2 + 2 m  e_3\cos \varpi_3 \rvert.
\end{aligned}
\label{eq:LaplaceDist}
\end{equation}
This error can thus be substantial (several tens of degrees) if the eccentricities are of the order of several parts per hundred, as it is the case for Kepler-223 \cite{Mills2016}. Therefore, we study if a given combination of the eccentricities and longitudes of periastron can make $\Psi_{e=0}$ closer to the equilibrium than $\Psi$ actually is, or vice-versa. We check this for the cases presented in Extended Data Table~4. 

\medskip
Each case sets the orbital period of the planets and their mid-transit time. We estimate the planetary masses using the mass-radius relation from \cite{Otegi2020}. Then, varying the remaining parameters $k_i=e_i \cos \varpi_i$ and $h_i=e_i \sin \varpi_i$, we minimize the cost function
\begin{equation}
\begin{aligned}
 C=\mathcal{{A}} (\Psi_{bcd}) + \mathcal{{A}} (\Psi_{cde}) + \mathcal{{A}} (\Psi_{efg}) + \mathcal{{A}} (\Psi_{fgh}) 
\end{aligned}
\label{eq:cost}
\end{equation}
over 200 years, where $\mathcal{{A}} (\Psi)=2\pi$ if $\Psi$ circulates, and $\mathcal{{A}} (\Psi)$ is the peak-to-peak amplitude of libration of $\Psi$ otherwise. For each case, 40 MCMC runs are conducted to minimize $C$, using \texttt{REBOUND} \cite{Rebound} for the $N$-body integration and \texttt{samsam} \cite{samsam} for the MCMC. For each run, the $k_i$ and $h_i$ parameters are randomly initialized in the $[-0.05,0.05]$ range, which are also their boundaries during the MCMC runs. This allows eccentricities that are comparable to those of Kepler-60 \citep{RIVERS3} and Kepler-223 \citep{Mills2016}, which are other known chains for which the inner planets are far enough from the star to not have their eccentricities damped by tides. 

\medskip
The best solution of each fit is shown in Extended Data Fig.~\ref{fig:Cminimize}. Case A2 is the only one for which the best solutions consistently have a peak-to-peak amplitude of the generalized three-body Laplace angles below 50\,degrees on average across the four angles. In all other cases, we were not able to find values of the $k_i$ and $k_j$ parameters below $85$\,deg of amplitude on average, with the exception of case A0 for which an average of $\approx 66$\,deg was reached. The best solutions found across all MCMC runs for the A2 and A0 cases integrated for 1000 years of evolution are shown in Fig.~S5 of the Supplementary. 
This analysis shows that the A2 case remains the one with the highest potential of being close to an equilibrium, while showing that all other cases cannot have an amplitude of libration smaller than $66$\,deg on average across their generalized three-body Laplace angles, regardless of the values of the $k_i$ and $k_j$ parameters. The case A2, with $P_f/P_e=4/3$ and $P_g/P_f=4/3$, is hence our prediction for the outer architecture of the HD~110067 system.

\subsubsection{Confirming the predictions}

\paragraph{Recovering the missing cadences of TESS S23 observations}

Based on the dynamical analysis presented above, the likely orbital periods associated with the two mono-transits observed in TESS S49 are approximately 41.05 and 54.74\,days, respectively. According to this prediction, both planets transited their host star during TESS S23 observations, but at a time when the photometry was highly affected by scattered light and sky background contamination. The Earth was a significant source of scattered light at the beginning of both S23 orbits (TJD=1928.09 and TJD=1941.83) and the Moon was a significant source of scattered light for a few days after the beginning of the second orbit (between 1942 and 1947\,TJD). The cadences affected were flagged by SPOC, thus not leaving enough valid data to derive cotrending basis vectors and missing in the PDCSAP light curve.

\medskip
Our custom extraction using the PLD method from Sect.~\ref{subsec:tess} was able to recover the missing data, showing two mono-transits at 1943.6 and 1944.1\,TJD (Fig.~\ref{fig:phot} and also Fig.~S6 in the Supplementary). 
Using \texttt{MonoTools}, we confirmed that the transits were consistent in duration-depth space with the two mono-transits from TESS S49 and separated by an integer number of orbits that matched the orbital periods predicted in our dynamical analysis for planets f and g. With this data reduction, all six planets in the HD~110067 system have been detected in transit at least twice, allowing a precise orbital period determination if we impose priors based on the hypothesis that all planets are trapped in a chain of first-order MMRs. Additionally, we recovered an additional transit of planet b at the beginning of TESS S23 observations.

\paragraph{Modeling of the ground-based photometric campaign} \label{sec:groundphot}

Targeted observations of HD~110067 were carried out on the night of May 23rd, 2022 to attempt to confirm the 41.05-day orbit of HD~110067\,f as predicted via our resonance chain analysis. In order to reveal whether a transit was present in the combined dataset, we built a combined photometric model using all ground-based observations. In order to remove spurious systematic trends in a way that does not bias any transit fit, we opted to perform simultaneous linear decorrelation of each photometric data set using the various meta-data time series available. In all cases, for example, we included an airmass term in the decorrelation as well as a measure of the FWHM width. We also included two position-centroid terms (for MuSCAT-2 and -3), information on comparison star total counts and FWHM width (for LCO, Tierras, and SAINT-EX), and interpolated color time series derived from the relative shift in flux across bands in the MuSCAT-2 and -3 filters \cite[as used in ][]{Osborn2021}. In all cases, the metadata were normalized to a time series with $\mu=0$, $\sigma=1$ and modeled using a single scaling parameter with a normal prior of $\mu=0$, $\sigma=0.5$. Quadratic limb darkening parameters were also constrained using normal priors dictated by theoretical limb darkening parameters as computed for each of the nine passbands using \texttt{LDTK} \cite{ldtk} and with inflated uncertainties following the methodology of the space-based photometric analysis. Each of the four time series observed by MuSCAT-3 was split into two around an observing gap that occurred due to the star passing close to the zenith at 2022-05-24UT06:57. Individual time series were used for each of the nine NGTS telescopes, which were decorrelated independently. The final result is 24 individual photometric time series. An offset was also applied to each light curve, as well as a single global slope parameter to include the possibility of stellar activity.

\medskip
The transit parameters were constrained based on those found in a fit of the TESS S49 mono-transit. The predicted period used was $4/3 \times P_e = 41.051\pm0.1$\,d, with the uncertainty implying a divergence from the perfect integer period of $2.4\times10^{-3}$ --- larger than those values found for the inner three planets. We limited the period to $41.0\pm0.2$\,days to ensure a transit fit that could be explored with the temporal baseline of the photometry. Due to the non-continuous nature of the photometry, the probability density function of the observed transit time is likely to be asymmetric and could potentially have multiple minima. Therefore, analyses using classical sampling techniques (Markov Chain Monte Carlo, Hamiltonian Monte Carlo, etc.) may not reveal the full picture. In order to initially test this, we kept all other parameters equal but split the range of periods covered by the time series into 36 bins across the anticipated period range ($40.8<P<41.2$) and fitted a constrained model for each. This would allow us to see the variation in the goodness-of-the-fit as a function of transit epoch. The transit model was built with \texttt{exoplanet} \cite{exoplanet:exoplanet} and optimized using \texttt{pymc3} \cite{exoplanet:pymc3}, specifically with the \texttt{pymc3-ext} sampling which enabled correlated parameters for each time series to be grouped together, speeding up the computation. In order to assess whether or not a transit model was justified over a flat model, we used the Watanabe-Akaike information criterion (WAIC) \cite{watanabe2010asymptotic,vehtari2017practical} as implemented in \texttt{arviz} \cite{arviz}.

\medskip
Our results show a preference for a transit at the expected period $P=41.04\pm0.01$\,days, with a $\Delta$WAIC of 9.5 over a transit-free model, as can be seen in Extended Data Fig.~\ref{fig:groundcampaign}. The majority of instruments showed a weak preference for a peak at $P\sim41.05$d, with the exception of LCO (which observed no in-transit data) and SAINT-EX (which is the most affected by cirrus). This is only equivalent to moderate evidence for a $\sim41.051$d period of HD110067\,f. The lower two panels of Extended Data Fig.~\ref{fig:groundcampaign} show that both models (with or without transit) fit reasonably well. This is in part because systematic effects dominate over astrophysical signals for transits with depths below 1000\,ppm, especially when the target star is observed at a low airmass. A further peak in WAIC is seen at $\sim40.9$\,d, but this hypothetical transit is covered only by the initial 1\,h of MuSCAT-2 photometry and does not fit our predicted period, hence we consider it spurious. This campaign shows that such transits are at the very limit of what is possible with ground-based observations. However, observations during a more favorable observing season and without technical issues such as the meridian flip of MuSCAT-3 (which unluckily coincided with the expected egress) may have constrained better the presence of a transit.

\paragraph{Modeling of the radial velocity data}

We carried out an initial frequency-based exploration of the CARMENES and HARPS-N spectroscopic data sets to see which significant signals are present and those related to stellar activity using periodograms \cite{GLS}. Figures~S7 and S8 in the Supplementary 
show that the dominant signal in the generalized Lomb-Scargle periodograms of both the radial velocities and main activity indicators (CCF-FWHM, differential line width, Mount Wilon's S-index, H$\alpha$ emission) is attributable to the rotational period of the star, measured photometrically to be approximately 20\,days using TESS and CHEOPS data. As it is well known, stellar activity induces spurious radial velocity signals \cite[e.g.,][]{saar97,hatzes02,meunier10,dumusque14}, which should be properly removed to unveil induced Keplerian motions in the star. We followed two independent approaches to model the data and minimize the impact of stellar activity effects on the detection and mass determination of the planets in the system.

\medskip
\textsc{Method I: SN-fit and breakpoint algorithm} \\
Spurious radial velocity signals induced by stellar activity come from the line shape variations in stellar spectra. Those can be quantified through the full-width-at-half-maximum (FWHM) and the asymmetry of the cross-correlation function (CCF) computed from the spectra \cite[e.g.,][]{queloz01,boisse09,dumusque16}. Following \cite{simola19}, we first fit Skew Normal (SN) functions to the CCFs available from HARPS-N and CARMENES. An SN function is not only characterized by a location and a scale parameter (which are the counterparts of the mean and standard deviation of a Gaussian), but it has a further free parameter that expresses its skewness (hereafter denoted with $\gamma$). For each observation, through the SN-fit we were able to retrieve the stellar radial velocity ($\overline{RV}$, quantified through the SN median), the $\mathrm{FWHM_{SN}}$, the contrast $A$, and the asymmetry $\gamma$. The errors $\sigma_{\mathrm{RV}}$ of the $\overline{RV}$ measurements were inferred using a bootstrap approach. Denoting with $f_{CCF}$ the flux of a CCF data point, each point was perturbed by sampling values from a Normal distribution whose standard deviation is equal to $\sqrt{f_{\mathrm{CCF}}}$, since the errors affecting the CCF data points are expected to be Poissonian.

\medskip
After that, we applied the breakpoint (\emph{bp}) method \citep{simola22} to both the HARPS-N and CARMENES RV time series. The algorithm has been designed to detect those locations along the RV time series where the correlation changes against the vector [$\mathrm{FWHM_{SN}}$, $A$, $\gamma$] are statistically significant. The goal is to then detrend the RV time series by applying a piece-wise interpolation to each segment found by the \emph{bp} algorithm rather than performing an overall correction to the whole time series. In this way we are able to better correct for the contamination of stellar variability as shown by ref.\cite{simola22} and ref.\cite{Bonfanti2023arXiv230210664B}. Finally, we jointly analyzed the RV time series using the \texttt{MCMCI} code \cite{bonfanti20}, where we switched off the interaction with stellar evolutionary models to speed up the computations. We set up the detrending function on each piece-wise stationary segment found by the \emph{bp} algorithm as a polynomial of the following form
\begin{multline}
RV_{\star} = \beta_0 + \displaystyle\sum_{k=1}^{k_t}{\beta_{k,t} t^k} + \displaystyle\sum_{k=1}^{k_F}{\beta_{k,F} \mathrm{FWHM_{SN}}^k} + \displaystyle\sum_{k=1}^{k_A}{\beta_{k,A} A^k} + \\
+ \displaystyle\sum_{k=1}^{k_{\gamma}}{\beta_{k,\gamma} \gamma^k} + \displaystyle\sum_{k=1}^{k_R}{\beta_{k,R} \log{R'_{\mathrm{HK}}}^k},
\label{eq:RVactivity}    
\end{multline}
where ($k_t$, $k_F$, $k_A$, $k_{\gamma}$, $k_R$) is the vector of the polynomial orders whose optimal value has been established by launching several MCMC preliminary runs and selecting that combination which produces the minimum Bayesian Information Criterion (BIC) \cite{schwarz78}.

\medskip
After performing a longer \texttt{MCMCI} run made of 4 independent runs (300\,000 steps each), which successfully converged as checked through the Gelman-Rubin test \citep{gelman1992}, we retrieved the posterior distributions of the system parameters. Their median values along with their error bars at the 1$\sigma$ level are reported in Tables~S2 and S3 of the Supplementary. 
The full RV time series and the phase-folded RVs of those planets whose detection is above the 3$\sigma$ level (planets d and f) are shown in Fig.~S9 of the Supplementary. 

\medskip
\textsc{Method II: multi-dimensional GP} \\
On the other hand, we also perform a multidimensional Gaussian Process (GP) approach to characterize the stellar and planetary signals in our RV time series as in \cite{Rajpaul2015,pyaneti2}. This approach has been proven useful to disentangle stellar and planetary signals in multi-planet systems \cite[e.g.,][]{Barragan2022,Zicher2022}.
We create $N-$dimensional GP models, including $N$ time-series $\mathcal{A}_i$,  as
\begin{equation}
\begin{matrix}
 \mathcal{A}_1 =  A_{1} G(t) + B_{1} \dot{G}(t) \\
\vdots \\
\mathcal{A}_N =  A_{N} G(t) + B_{N} \dot{G}(t), \\
\end{matrix}
\label{eq:gps}
\end{equation}
\noindent
where the variables $A_{1}$, $B_{1}$, $\cdots$, $A_{N}$, $B_{N}$, are free parameters which relate the individual time-series to $G(t)$ and $\dot{G}(t)$. In this approach, $G(t)$ is assumed to be a latent (unobserved) variable that represents the projected area of the visible stellar disk that is covered by active regions as a function of time. 

\medskip
We model the stellar signal using a GP whose covariance between two times $t_i$ and $t_j$ is given by
\begin{equation}
   \gamma_{{\rm QP},i,j} = \exp 
    \left[
    - \frac{\sin^2[\pi(t_i - t_j)/P_{\rm GP}]}{2 \lambda_{\rm P}^2}
    - \frac{(t_i - t_j)^2}{2\lambda_{\rm e}^2}
    \right],
    \label{eq:gamma}
\end{equation}
\noindent
where $\gamma_{{\rm QP},i,j}$ is the Quasi-Periodic (QP) kernel, whose hyperparameters are, \pgp, the GP characteristic period, \lbp, the inverse of the harmonic complexity, and \lbe, the long term evolution timescale. 

\medskip
We perform a two-dimensional GP model between the RVs and FWHM. We note that these quantities are equivalent in the HARPS-N and CARMENES data. The multidimensional covariance matrix was created using the kernel given in Eq.~\eqref{eq:gamma} and its derivatives \cite{Rajpaul2015,pyaneti2}.  We assume that RVs can be described as $\mathcal{A}_i =  A_{i} G(t) + B_{i} \dot{G}(t)$, while the FWHM time series is described as $\mathcal{A}_i =  A_{i} G(t)$. The planetary signals were included in the model as the mean function of the RV time series. We use $N$ Keplerian signals (where $N$ is the number of planetary signals); each one of them depends on the time of minimum conjunction $t_0$, orbital period $P$, and Doppler semi-amplitude, $K$. All orbits are fixed to be circular, so the eccentricity and angle of periastron are fixed. For the FWHM the mean function was treated as an offset, noting that we include a different offset per instrument. We also include a jitter term per time series and per instrument to account for unaccounted systematic errors.

\medskip
We perform MCMC samplings of the parameter space using the code \texttt{pyaneti} \cite{pyaneti,pyaneti2}. We sample the parameter space with 250 walkers and create the posterior distributions with the last 5000 iterations of converged chains with a thin factor of 10. This leads to posterior distributions of 125\,000 points for each sampled parameter. Figure~\ref{fig:rvs} 
shows the spectroscopic time series resulting from this joint analysis. Median values along with their 1$\sigma$ uncertainties are reported in Table~S4 of the Supplementary.

\medskip
Modeling techniques employing GP are particularly subject to overfitting giving their flexibility to reproduce the data \cite[see e.g.,][]{Cale21,Blunt23}. To test the robustness of our GP model, we carried out a cross-validation analysis. We repeat the two-dimensional GP model described in this section but applied only to the HARPS-N data. Then, we create a predictive model with the inferred parameters and overlay the CARMENES data with median offsets subtracted (similar to a training/evaluation set for machine learning algorithms). Figure~S10 
shows this analysis, zoomed in to the 2021 observing campaign.
The plot shows that the RV data is in agreement with the predictive model, suggesting that our assumption that the stellar signal imprinted in the RVs can be described with a two-dimensional GP is valid for the time span of our observations. For the CCF FWHM CARMENES data, the correlation with the RV measurements is not as strong as for the HARPS-N one, thus the prediction is less accurate in this case.

\medskip
Both the two methods clearly detect planets HD~110067~d and f, \textsc{Method II} also detects planet b, however, the detection levels slightly differ. Recalling that the two different techniques are based on different RV extraction methodologies and on a different treatment of stellar activity, on the one hand, the slight output tension suggests that the RV data alone do not strongly constrain all the six Keplerian signals. We tested additional stellar mitigation approaches, such as sinusoid-fitting at the stellar rotation period and its harmonics or GP decorrelation as a function of time only, but they all turned unsuccessful at constraining the masses of any of the planets (only the time-dependent GP model could recover the signal of planet f, but with a much larger uncertainty, $K_{\rm f} = 2.0\pm1.0\,\mathrm{m\,s^{-1}}$). On the other hand, the RV semi-amplitudes inferred from the two methods shown in the manuscript are compatible within $\sim$\,1.5\,$\sigma$, with the statistical tension $\Delta_{\mathrm{I-II}}$ below $\sim$\,1$\sigma$ for planets b, c, d, and g. Extended Data Fig.~\ref{fig:Kcomparison} displays the pairs of posterior density functions for the RV semi-amplitudes of each planet for comparison.

\medskip
For HD~110067~f, after imposing a Gaussian prior centered around 41.05\,days in both \textsc{Method I} and \textsc{Method II} RV models, we indeed recover a significant RV signal with a detection level of $\sim$\,3$\sigma$. The planet transits only twice in the TESS data and the ground-based photometric campaign hints at moderate evidence for a planetary transit compatible with this value. Therefore, to secure an independent detection of planet f from spectroscopy, we performed an RV-only analysis imposing a uniform unbounded prior (between 30 to 54 days) to the orbital period. The MCMC converged and detected a clear Keplerian signal with a period $P_{f,\mathrm{uni}}=40.2\pm0.2$ d. Thus, the RV data independently suggest the presence of a putative planet having a period close to 41.05 d. The tension at the $\sim$\,4$\sigma$ level with the predictions from the resonant chain model and the transit observations prove that the current RV data set cannot fully constrain the entire architecture of the planetary system.

\medskip
We finally checked whether there are further Keplerian signals within the RV time series. In particular, given that both planets e and g were not detected via our previous RV analyses where model-dependent values of the orbital periods were imposed as priors, we investigated the presence of potential planetary signals at different periods that could be attributed to planet e or g. To this end, we performed an MCMC run, where we modeled planets b and c (which are clearly confirmed by the transit events) along with planets d and f (which are clearly detected also in the RV time series). As a result, we produced the Generalized Lomb Scargle periodograms \cite{GLS} of the residuals, obtained after subtracting the Keplerian signals of all four planets from the activity-cleaned time series (Fig.~S11). 
The high false alarm probability level of the highest peak in both the HARPS-N (18\%) and CARMENES (10\%) residuals suggests that there are no signals left in the RV data that could be associated with other planets or a misidentification of the orbital periods of planets e and g.

\subsubsection{Final model}
We computed a final model of the photometric and spectroscopic data sets of the HD~110067 system. Based on the analyses above, neither the light curves nor the radial velocities are precise enough to constrain the eccentricity of the planets. Assuming circular orbits, the photometry and radial velocity thus only constrain jointly the period and phase of a given planet in the system. However, the transit data dominate the precision of these two quantities (by several orders of magnitude). Therefore, for our final model, we opt to perform an independent analysis of the photometry and radial velocity datasets, where priors inform the planet periods and phases in the radial velocity model based on the posterior distributions of the photometry-only fit. Besides, the large number of free parameters in each of the models makes it computationally expensive to run a joint fit, not to mention the complications for numerical samplers to explore the vast multi-dimensional parameter space. Table~1 shows the most relevant planetary parameters of the system based on the photometric fit from Table~S1, 
the radial velocity fit using \textsc{Method II} from Table~S4, 
and the stellar parameters from Extended Data Table~3. A corner plot with the posterior distribution of the fitted transit parameters is shown in Fig.~S12. 
The resulting best-fit models and corresponding credibility bands are presented in Fig.~\ref{fig:phot} for the TESS and CHEOPS photometry and Fig.~\ref{fig:rvs} for the radial velocities.

\subsubsection{Planetary internal structures}

Using a Bayesian analysis \cite{Dorn2015,Dorn2017}, we computed the possible internal structures of the six planets of the system, using the results provided in Tables~1, Extended Data Table~3 and the planetary masses from \textsc{Method I} and \textsc{Method II}. The forward model used to compute the likelihood is based on a four-layer structure: a central core (iron and sulfur), a silicate mantle (containing Si, Mg, and Fe), a water layer, and a gas layer (H and He). The equation of state (EOS) of water is the one of ref.\cite{Haldemann2020},  the core EOS is the one of ref.\cite{Hakim2018}, and we use for EOS of ref.\cite{Sotin2007} for the silicate mantle.  The thickness of the gas envelope, which depends on the planetary age, mass, etc., is derived from ref.\cite{Lopez2014}. Note that the influence of the gas layer on the innermost planet (compression and thermal effect) is not included in our model, as the mass of gas layer for the six planets is small (see below). The planetary Si/Mg/Fe molar ratio in all planets is assumed to be equal to the stellar one. The prior distribution of the mass fractions of the three innermost layers (core, mantle, and water layer) is assumed to be uniform on the simplex --- the surface defined by the sum of the three mass fractions equal to one. In addition, the mass fraction of the water layer is assumed to be 50\% at most \cite{Thiabaud2014,Marboeuf2014}, and for the gas mass, we use a uniform log prior.

\medskip
The results from this analysis are shown in Extended Data Fig.~\ref{fig:gasfrac_box}. Our model shows that the gas mass content of all planets is of the order of $10^{-3} M_\oplus$ to $10^{-1} M_\oplus$ (median value, see Table~S5), 
with the notable exception of HD~110067~e (median value of $\sim 10^{-7} M_\oplus$ using the masses from \textsc{Method I}, $\sim 10^{-3} M_\oplus$ using \textsc{Method II}). The apparent lack of an atmosphere of planet~e (located just outside of planet~d, which is the most gas-rich of the system, according to the internal structure models) is puzzling. If confirmed by future better determination of its density, the origin of the peculiar internal structure of planet e will have to be understood in the context of the very fragile architecture of the whole HD~110067 system. On the other hand, the water fraction for all planets is essentially unconstrained, due to the still large uncertainty in the planetary masses. However, according to simulations of combined planetary formation and evolution, independently of the accretion mechanism (planetesimal- or pebble-based) all the planets in the system have masses and radii consistent with a formation beyond the ice line \cite{Venturini2020A&A...643L...1V,Emsenhuber2021A&A...656A..70E,Izidoro2022ApJ...939L..19I}. Therefore, it is possible that even though the water content is unconstrained in our model, the planets' cores are rich in volatiles. JWST observations of some atmospheric trace gases (particularly ammonia, methane, and/or methanol) could be used as a proxy for the presence of a deep or shallow surface that could break the degeneracies from internal composition models using bulk density measurements alone \cite{Hu2021ApJ...921L...8H,Tsai2021ApJ...922L..27T}.


\clearpage
\newpage\bigskip

\begin{thebibliography}{100}
\expandafter\ifx\csname url\endcsname\relax
  \def\url#1{\burl{#1}}\fi
\expandafter\ifx\csname urlprefix\endcsname\relax\def\urlprefix{URL }\fi
\providecommand{\bibinfo}[2]{#2}
\providecommand{\eprint}[2][]{\url{#2}}
\providecommand{\doi}[1]{\url{https://doi.org/#1}}
\bibcommenthead

\bibitem{Howard2012}
\bibinfo{author}{{Howard}, A.~W.} \emph{et~al.}
\newblock \bibinfo{title}{{Planet Occurrence within 0.25 AU of Solar-type Stars
  from Kepler}}.
\newblock \emph{\bibinfo{journal}{\apjs}} \textbf{\bibinfo{volume}{201}}~(2),
  \bibinfo{pages}{15} (\bibinfo{year}{2012}).

\bibitem{Fressin13}
\bibinfo{author}{{Fressin}, F.} \emph{et~al.}
\newblock \bibinfo{title}{{The False Positive Rate of Kepler and the Occurrence
  of Planets}}.
\newblock \emph{\bibinfo{journal}{\apj}} \textbf{\bibinfo{volume}{766}}~(2),
  \bibinfo{pages}{81} (\bibinfo{year}{2013}).

\bibitem{Bean2021}
\bibinfo{author}{{Bean}, J.~L.}, \bibinfo{author}{{Raymond}, S.~N.} \&
  \bibinfo{author}{{Owen}, J.~E.}
\newblock \bibinfo{title}{{The Nature and Origins of Sub-Neptune Size
  Planets}}.
\newblock \emph{\bibinfo{journal}{J. Geophys. Res. Planets}}
  \textbf{\bibinfo{volume}{126}}~(1), \bibinfo{pages}{e06639}
  (\bibinfo{year}{2021}).

\bibitem{Ricker2015}
\bibinfo{author}{{Ricker}, G.~R.} \emph{et~al.}
\newblock \bibinfo{title}{{Transiting Exoplanet Survey Satellite (TESS)}}.
\newblock \emph{\bibinfo{journal}{Journal of Astronomical Telescopes,
  Instruments, and Systems}} \textbf{\bibinfo{volume}{1}}~(1),
  \bibinfo{pages}{014003} (\bibinfo{year}{2015}).

\bibitem{SPOC}
\bibinfo{author}{{Jenkins}, J.~M.} \emph{et~al.}
\newblock \bibinfo{editor}{Chiozzi, G.} \& \bibinfo{editor}{Guzman, J.~C.}
  (eds) \emph{\bibinfo{title}{{The TESS science processing operations
  center}}}.
\newblock (eds \bibinfo{editor}{Chiozzi, G.} \& \bibinfo{editor}{Guzman,
  J.~C.}) \emph{\bibinfo{booktitle}{Software and Cyberinfrastructure for
  Astronomy IV}}, Vol. \bibinfo{volume}{9913}, \bibinfo{pages}{99133E}.
  \bibinfo{organization}{International Society for Optics and Photonics}
  (\bibinfo{publisher}{SPIE}, \bibinfo{year}{2016}).

\bibitem{CHEOPS}
\bibinfo{author}{{Benz}, W.} \emph{et~al.}
\newblock \bibinfo{title}{{The CHEOPS mission}}.
\newblock \emph{\bibinfo{journal}{Exper. Astron.}}
  \textbf{\bibinfo{volume}{51}}~(1), \bibinfo{pages}{109--151}
  (\bibinfo{year}{2021}).

\bibitem{Sinclair1975MNRAS.171...59S}
\bibinfo{author}{{Sinclair}, A.~T.}
\newblock \bibinfo{title}{{The orbital resonance amongst the Galilean
  satellites of Jupiter.}}
\newblock \emph{\bibinfo{journal}{\mnras}} \textbf{\bibinfo{volume}{171}},
  \bibinfo{pages}{59--72} (\bibinfo{year}{1975}).

\bibitem{Morbidelli2002mcma.book.....M}
\bibinfo{author}{{Morbidelli}, A.}
\newblock \emph{\bibinfo{title}{{Modern celestial mechanics : aspects of solar
  system dynamics}}}  (\bibinfo{publisher}{Advances in Astronomy \&
  Astrophysics}, \bibinfo{year}{2002}).

\bibitem{Papaloizou2015}
\bibinfo{author}{{Papaloizou}, J. C.~B.}
\newblock \bibinfo{title}{{Three body resonances in close orbiting planetary
  systems: tidal dissipation and orbital evolution}}.
\newblock \emph{\bibinfo{journal}{Int. J. Astrobiology}}
  \textbf{\bibinfo{volume}{14}}~(2), \bibinfo{pages}{291--304}
  (\bibinfo{year}{2015}).

\bibitem{Leleu2021A&A...649A..26L}
\bibinfo{author}{{Leleu}, A.} \emph{et~al.}
\newblock \bibinfo{title}{{Six transiting planets and a chain of Laplace
  resonances in TOI-178}}.
\newblock \emph{\bibinfo{journal}{\aap}} \textbf{\bibinfo{volume}{649}},
  \bibinfo{pages}{A26} (\bibinfo{year}{2021}).

\bibitem{Luger2017NatAs...1E.129L}
\bibinfo{author}{{Luger}, R.} \emph{et~al.}
\newblock \bibinfo{title}{{A seven-planet resonant chain in TRAPPIST-1}}.
\newblock \emph{\bibinfo{journal}{Nat. Astron.}}
  \textbf{\bibinfo{volume}{1}}, \bibinfo{pages}{0129} (\bibinfo{year}{2017}).

\bibitem{Gozdziewski2016MNRAS.455L.104G}
\bibinfo{author}{{Go{\'z}dziewski}, K.}, \bibinfo{author}{{Migaszewski}, C.},
  \bibinfo{author}{{Panichi}, F.} \& \bibinfo{author}{{Szuszkiewicz}, E.}
\newblock \bibinfo{title}{{The Laplace resonance in the Kepler-60 planetary
  system}}.
\newblock \emph{\bibinfo{journal}{\mnras}} \textbf{\bibinfo{volume}{455}}~(1),
  \bibinfo{pages}{L104--L108} (\bibinfo{year}{2016}).

\bibitem{Agol2021}
\bibinfo{author}{{Agol}, E.} \emph{et~al.}
\newblock \bibinfo{title}{{Refining the Transit-timing and Photometric Analysis
  of TRAPPIST-1: Masses, Radii, Densities, Dynamics, and Ephemerides}}.
\newblock \emph{\bibinfo{journal}{\psj}} \textbf{\bibinfo{volume}{2}}~(1),
  \bibinfo{pages}{1} (\bibinfo{year}{2021}).

\bibitem{Dai2023AJ....165...33D}
\bibinfo{author}{{Dai}, F.} \emph{et~al.}
\newblock \bibinfo{title}{{TOI-1136 is a Young, Coplanar, Aligned Planetary
  System in a Pristine Resonant Chain}}.
\newblock \emph{\bibinfo{journal}{\aj}} \textbf{\bibinfo{volume}{165}}~(2),
  \bibinfo{pages}{33} (\bibinfo{year}{2023}).

\bibitem{CARMENES20}
\bibinfo{author}{{Quirrenbach}, A.} \emph{et~al.}
\newblock \bibinfo{editor}{Evans, C.~J.}, \bibinfo{editor}{Bryant, J.~J.} \&
  \bibinfo{editor}{Motohara, K.} (eds) \emph{\bibinfo{title}{{The CARMENES
  M-dwarf planet survey}}}.
\newblock (eds \bibinfo{editor}{Evans, C.~J.}, \bibinfo{editor}{Bryant, J.~J.}
  \& \bibinfo{editor}{Motohara, K.}) \emph{\bibinfo{booktitle}{Ground-based and
  Airborne Instrumentation for Astronomy VIII}}, Vol. \bibinfo{volume}{11447},
  \bibinfo{pages}{114473C}. \bibinfo{organization}{International Society for
  Optics and Photonics} (\bibinfo{publisher}{SPIE}, \bibinfo{year}{2020}).

\bibitem{HARPN}
\bibinfo{author}{{Cosentino}, R.} \emph{et~al.}
\newblock \bibinfo{editor}{McLean, I.~S.}, \bibinfo{editor}{Ramsay, S.~K.} \&
  \bibinfo{editor}{Takami, H.} (eds) \emph{\bibinfo{title}{{Harps-N: the new
  planet hunter at TNG}}}.
\newblock (eds \bibinfo{editor}{McLean, I.~S.}, \bibinfo{editor}{Ramsay, S.~K.}
  \& \bibinfo{editor}{Takami, H.}) \emph{\bibinfo{booktitle}{Ground-based and
  Airborne Instrumentation for Astronomy IV}}, Vol. \bibinfo{volume}{8446},
  \bibinfo{pages}{84461V}. \bibinfo{organization}{International Society for
  Optics and Photonics} (\bibinfo{publisher}{SPIE}, \bibinfo{year}{2012}).

\bibitem{holman2005}
\bibinfo{author}{{Holman}, M.~J.} \& \bibinfo{author}{{Murray}, N.~W.}
\newblock \bibinfo{title}{{The Use of Transit Timing to Detect Terrestrial-Mass
  Extrasolar Planets}}.
\newblock \emph{\bibinfo{journal}{Science}}
  \textbf{\bibinfo{volume}{307}}~(5713), \bibinfo{pages}{1288--1291}
  (\bibinfo{year}{2005}).

\bibitem{Fulton17}
\bibinfo{author}{{Fulton}, B.~J.} \emph{et~al.}
\newblock \bibinfo{title}{{The California-Kepler Survey. III. A Gap in the
  Radius Distribution of Small Planets}}.
\newblock \emph{\bibinfo{journal}{\aj}} \textbf{\bibinfo{volume}{154}},
  \bibinfo{pages}{109} (\bibinfo{year}{2017}).

\bibitem{2018MNRAS.479.4786V}
\bibinfo{author}{{Van Eylen}, V.} \emph{et~al.}
\newblock \bibinfo{title}{{An asteroseismic view of the radius valley: stripped
  cores, not born rocky}}.
\newblock \emph{\bibinfo{journal}{\mnras}} \textbf{\bibinfo{volume}{479}},
  \bibinfo{pages}{4786--4795} (\bibinfo{year}{2018}).

\bibitem{Kasting1993Icar..101..108K}
\bibinfo{author}{{Kasting}, J.~F.}, \bibinfo{author}{{Whitmire}, D.~P.} \&
  \bibinfo{author}{{Reynolds}, R.~T.}
\newblock \bibinfo{title}{{Habitable Zones around Main Sequence Stars}}.
\newblock \emph{\bibinfo{journal}{\icarus}} \textbf{\bibinfo{volume}{101}}~(1),
  \bibinfo{pages}{108--128} (\bibinfo{year}{1993}).

\bibitem{Kopparapu14}
\bibinfo{author}{{Kopparapu}, R.~K.} \emph{et~al.}
\newblock \bibinfo{title}{{Habitable Zones around Main-sequence Stars:
  Dependence on Planetary Mass}}.
\newblock \emph{\bibinfo{journal}{\apjl}} \textbf{\bibinfo{volume}{787}},
  \bibinfo{pages}{L29} (\bibinfo{year}{2014}).

\bibitem{Izidoro2021}
\bibinfo{author}{{Izidoro}, A.} \emph{et~al.}
\newblock \bibinfo{title}{{Formation of planetary systems by pebble accretion
  and migration. Hot super-Earth systems from breaking compact resonant
  chains}}.
\newblock \emph{\bibinfo{journal}{\aap}} \textbf{\bibinfo{volume}{650}},
  \bibinfo{pages}{A152} (\bibinfo{year}{2021}).

\bibitem{Fabrycky2014}
\bibinfo{author}{{Fabrycky}, D.~C.} \emph{et~al.}
\newblock \bibinfo{title}{{Architecture of Kepler's Multi-transiting Systems.
  II. New Investigations with Twice as Many Candidates}}.
\newblock \emph{\bibinfo{journal}{\apj}} \textbf{\bibinfo{volume}{790}}~(2),
  \bibinfo{pages}{146} (\bibinfo{year}{2014}).

\bibitem{Zeng2019}
\bibinfo{author}{{Zeng}, L.} \emph{et~al.}
\newblock \bibinfo{title}{{Growth model interpretation of planet size
  distribution}}.
\newblock \emph{\bibinfo{journal}{\pnas}} \textbf{\bibinfo{volume}{116}}~(20),
  \bibinfo{pages}{9723--9728} (\bibinfo{year}{2019}).

\bibitem{Kempton2018PASP..130k4401K}
\bibinfo{author}{{Kempton}, E. M.~R.} \emph{et~al.}
\newblock \bibinfo{title}{{A Framework for Prioritizing the TESS Planetary
  Candidates Most Amenable to Atmospheric Characterization}}.
\newblock \emph{\bibinfo{journal}{\pasp}} \textbf{\bibinfo{volume}{130}}~(993),
  \bibinfo{pages}{114401} (\bibinfo{year}{2018}).

\bibitem{Otegi2020}
\bibinfo{author}{{Otegi}, J.~F.}, \bibinfo{author}{{Bouchy}, F.} \&
  \bibinfo{author}{{Helled}, R.}
\newblock \bibinfo{title}{{Revisited mass-radius relations for exoplanets below
  120 M$_{{\ensuremath{\oplus}}}$}}.
\newblock \emph{\bibinfo{journal}{\aap}} \textbf{\bibinfo{volume}{634}},
  \bibinfo{pages}{A43} (\bibinfo{year}{2020}).

\end{thebibliography}

\begin{thebibliography}{100}
\expandafter\ifx\csname url\endcsname\relax
  \def\url#1{\burl{#1}}\fi
\expandafter\ifx\csname urlprefix\endcsname\relax\def\urlprefix{URL }\fi
\providecommand{\bibinfo}[2]{#2}
\providecommand{\eprint}[2][]{\url{#2}}
\providecommand{\doi}[1]{\url{https://doi.org/#1}}
\bibcommenthead

\makeatletter
\addtocounter{NAT@ctr}{26}
\makeatother

\bibitem{Stassun2018AJ....156..102S}
\bibinfo{author}{{Stassun}, K.~G.} \emph{et~al.}
\newblock \bibinfo{title}{{The TESS Input Catalog and Candidate Target List}}.
\newblock \emph{\bibinfo{journal}{\aj}} \textbf{\bibinfo{volume}{156}}~(3),
  \bibinfo{pages}{102} (\bibinfo{year}{2018}).

\bibitem{Stumpe2012}
\bibinfo{author}{{Stumpe}, M.~C.} \emph{et~al.}
\newblock \bibinfo{title}{{Kepler Presearch Data Conditioning
  I{\textemdash}Architecture and Algorithms for Error Correction in Kepler
  Light Curves}}.
\newblock \emph{\bibinfo{journal}{\pasp}} \textbf{\bibinfo{volume}{124}}~(919),
  \bibinfo{pages}{985} (\bibinfo{year}{2012}).

\bibitem{Stumpe2014}
\bibinfo{author}{{Stumpe}, M.~C.} \emph{et~al.}
\newblock \bibinfo{title}{{Multiscale Systematic Error Correction via
  Wavelet-Based Bandsplitting in Kepler Data}}.
\newblock \emph{\bibinfo{journal}{\pasp}} \textbf{\bibinfo{volume}{126}},
  \bibinfo{pages}{100} (\bibinfo{year}{2014}).

\bibitem{2012PASP..124.1000S}
\bibinfo{author}{{Smith}, J.~C.} \emph{et~al.}
\newblock \bibinfo{title}{{Kepler Presearch Data Conditioning II - A Bayesian
  Approach to Systematic Error Correction}}.
\newblock \emph{\bibinfo{journal}{\pasp}} \textbf{\bibinfo{volume}{124}},
  \bibinfo{pages}{1000} (\bibinfo{year}{2012}).

\bibitem{2002ApJ...575..493J}
\bibinfo{author}{{Jenkins}, J.~M.}
\newblock \bibinfo{title}{{The Impact of Solar-like Variability on the
  Detectability of Transiting Terrestrial Planets}}.
\newblock \emph{\bibinfo{journal}{\apj}} \textbf{\bibinfo{volume}{575}},
  \bibinfo{pages}{493--505} (\bibinfo{year}{2002}).

\bibitem{2010SPIE.7740E..0DJ}
\bibinfo{author}{{Jenkins}, J.~M.} \emph{et~al.}
\newblock \bibinfo{editor}{{Radziwill}, N.~M.} \& \bibinfo{editor}{{Bridger},
  A.} (eds) \emph{\bibinfo{title}{{Transiting planet search in the Kepler
  pipeline}}}.
\newblock (eds \bibinfo{editor}{{Radziwill}, N.~M.} \&
  \bibinfo{editor}{{Bridger}, A.}) \emph{\bibinfo{booktitle}{Software and
  Cyberinfrastructure for Astronomy}}, Vol. \bibinfo{volume}{7740} of
  \emph{\bibinfo{series}{Society of Photo-Optical Instrumentation Engineers
  (SPIE) Conference Series}}, \bibinfo{pages}{77400D} (\bibinfo{year}{2010}).

\bibitem{2020TPSkdph}
\bibinfo{author}{{Jenkins}, J.~M.} \emph{et~al.}
\newblock \bibinfo{title}{{Kepler Data Processing Handbook: Transiting Planet
  Search}}.
\newblock \bibinfo{howpublished}{Kepler Science Document KSCI-19081-003}
  (\bibinfo{year}{2020}).

\bibitem{Twicken:DVdiagnostics2018}
\bibinfo{author}{{Twicken}, J.~D.} \emph{et~al.}
\newblock \bibinfo{title}{{Kepler Data Validation I - Architecture, Diagnostic
  Tests, and Data Products for Vetting Transiting Planet Candidates}}.
\newblock \emph{\bibinfo{journal}{\pasp}} \textbf{\bibinfo{volume}{130}}~(6),
  \bibinfo{pages}{064502} (\bibinfo{year}{2018}).

\bibitem{Li:DVmodelFit2019}
\bibinfo{author}{{Li}, J.} \emph{et~al.}
\newblock \bibinfo{title}{{Kepler Data Validation II-Transit Model Fitting and
  Multiple-planet Search}}.
\newblock \emph{\bibinfo{journal}{\pasp}} \textbf{\bibinfo{volume}{131}}~(996),
  \bibinfo{pages}{024506} (\bibinfo{year}{2019}).

\bibitem{2021ApJS..254...39G}
\bibinfo{author}{{Guerrero}, N.~M.} \emph{et~al.}
\newblock \bibinfo{title}{{The TESS Objects of Interest Catalog from the TESS
  Prime Mission}}.
\newblock \emph{\bibinfo{journal}{\apjs}} \textbf{\bibinfo{volume}{254}}~(2),
  \bibinfo{pages}{39} (\bibinfo{year}{2021}).

\bibitem{Fausnaugh2020}
\bibinfo{author}{{Fausnaugh}, M.~M.}, \bibinfo{author}{{Burke}, C.~J.},
  \bibinfo{author}{{Ricker}, G.~R.} \& \bibinfo{author}{{Vanderspek}, R.}
\newblock \bibinfo{title}{{Calibrated Full-frame Images for the TESS Quick Look
  Pipeline}}.
\newblock \emph{\bibinfo{journal}{Res. Not. Amer. Astron. Soc.}} \textbf{\bibinfo{volume}{4}}~(12), \bibinfo{pages}{251}
  (\bibinfo{year}{2020}).

\bibitem{2021AJ....162...54H}
\bibinfo{author}{{Hedges}, C.} \emph{et~al.}
\newblock \bibinfo{title}{{TOI-2076 and TOI-1807: Two Young, Comoving Planetary
  Systems within 50 pc Identified by TESS that are Ideal Candidates for Further
  Follow Up}}.
\newblock \emph{\bibinfo{journal}{\aj}} \textbf{\bibinfo{volume}{162}}~(2),
  \bibinfo{pages}{54} (\bibinfo{year}{2021}).

\bibitem{Osborn2022HIP8618}
\bibinfo{author}{{Osborn}, H.} \emph{et~al.}
\newblock \bibinfo{title}{{Two Warm Neptunes transiting HIP 9618 revealed by
  TESS \& Cheops}}.
\newblock \emph{\bibinfo{journal}{\mnras}} \textbf{\bibinfo{volume}{523}}~(2),
  \bibinfo{pages}{3069--3089} (\bibinfo{year}{2023}).

\bibitem{Vanderburg2019ApJ...881L..19V}
\bibinfo{author}{{Vanderburg}, A.} \emph{et~al.}
\newblock \bibinfo{title}{{TESS Spots a Compact System of Super-Earths around
  the Naked-eye Star HR 858}}.
\newblock \emph{\bibinfo{journal}{\apjl}} \textbf{\bibinfo{volume}{881}}~(1),
  \bibinfo{pages}{L19} (\bibinfo{year}{2019}).

\bibitem{Deming_2009}
\bibinfo{author}{{Deming}, D.} \emph{et~al.}
\newblock \bibinfo{title}{{Spitzer Secondary Eclipses of the Dense,
  Modestly-irradiated, Giant Exoplanet HAT-P-20b Using Pixel-level
  Decorrelation}}.
\newblock \emph{\bibinfo{journal}{\apj}} \textbf{\bibinfo{volume}{805}}~(2),
  \bibinfo{pages}{132} (\bibinfo{year}{2015}).

\bibitem{Luger2016AJ....152..100L}
\bibinfo{author}{{Luger}, R.} \emph{et~al.}
\newblock \bibinfo{title}{{EVEREST: Pixel Level Decorrelation of K2 Light
  Curves}}.
\newblock \emph{\bibinfo{journal}{\aj}} \textbf{\bibinfo{volume}{152}}~(4),
  \bibinfo{pages}{100} (\bibinfo{year}{2016}).

\bibitem{exoplanet:luger18}
\bibinfo{author}{{Luger}, R.} \emph{et~al.}
\newblock \bibinfo{title}{{starry: Analytic Occultation Light Curves}}.
\newblock \emph{\bibinfo{journal}{\aj}} \textbf{\bibinfo{volume}{157}},
  \bibinfo{pages}{64} (\bibinfo{year}{2019}).

\bibitem{lightkurve}
\bibinfo{author}{{Lightkurve Collaboration}} \emph{et~al.}
\newblock \bibinfo{title}{{Lightkurve: Kepler and TESS time series analysis in
  Python}}.
\newblock \bibinfo{howpublished}{Astrophysics Source Code Library}
  (\bibinfo{year}{2018}).

\bibitem{Gilliland_2011}
\bibinfo{author}{{Gilliland}, R.~L.} \emph{et~al.}
\newblock \bibinfo{title}{{Kepler Mission Stellar and Instrument Noise
  Properties}}.
\newblock \emph{\bibinfo{journal}{\apjs}} \textbf{\bibinfo{volume}{197}}~(1),
  \bibinfo{pages}{6} (\bibinfo{year}{2011}).

\bibitem{Van_Cleve_2016}
\bibinfo{author}{{Van Cleve}, J.~E.} \emph{et~al.}
\newblock \bibinfo{title}{{That's How We Roll: The NASA K2 Mission Science
  Products and Their Performance Metrics}}.
\newblock \emph{\bibinfo{journal}{\pasp}} \textbf{\bibinfo{volume}{128}}~(965),
  \bibinfo{pages}{075002} (\bibinfo{year}{2016}).

\bibitem{Schanche2022}
\bibinfo{author}{{Schanche}, N.} \emph{et~al.}
\newblock \bibinfo{title}{{TOI-2257 b: A highly eccentric long-period
  sub-Neptune transiting a nearby M dwarf}}.
\newblock \emph{\bibinfo{journal}{\aap}} \textbf{\bibinfo{volume}{657}},
  \bibinfo{pages}{A45} (\bibinfo{year}{2022}).

\bibitem{Ulmer-Moll2022}
\bibinfo{author}{{Ulmer-Moll}, S.} \emph{et~al.}
\newblock \bibinfo{title}{{Two long-period transiting exoplanets on eccentric
  orbits: NGTS-20 b (TOI-5152 b) and TOI-5153 b}}.
\newblock \emph{\bibinfo{journal}{\aap}} \textbf{\bibinfo{volume}{666}},
  \bibinfo{pages}{A46} (\bibinfo{year}{2022}).

\bibitem{Osborn2021}
\bibinfo{author}{{Osborn}, A.} \emph{et~al.}
\newblock \bibinfo{title}{{TOI-431/HIP 26013: a super-Earth and a sub-Neptune
  transiting a bright, early K dwarf, with a third RV planet}}.
\newblock \emph{\bibinfo{journal}{\mnras}} \textbf{\bibinfo{volume}{507}}~(2),
  \bibinfo{pages}{2782--2803} (\bibinfo{year}{2021}).

\bibitem{Tuson2023}
\bibinfo{author}{{Tuson}, A.} \emph{et~al.}
\newblock \bibinfo{title}{{TESS and CHEOPS discover two warm sub-Neptunes transiting the bright K-dwarf HD 15906}}.
\newblock \emph{\bibinfo{journal}{\mnras}} \textbf{\bibinfo{volume}{523}}~(2),
  \bibinfo{pages}{3090--3118} (\bibinfo{year}{2023}).

\bibitem{Szabo2021A&A...654A.159S}
\bibinfo{author}{{Szab{\'o}}, G.~M.} \emph{et~al.}
\newblock \bibinfo{title}{{The changing face of AU Mic b: stellar spots,
  spin-orbit commensurability, and transit timing variations as seen by CHEOPS
  and TESS}}.
\newblock \emph{\bibinfo{journal}{\aap}} \textbf{\bibinfo{volume}{654}},
  \bibinfo{pages}{A159} (\bibinfo{year}{2021}).

\bibitem{Morris2021A&A...653A.173M}
\bibinfo{author}{{Morris}, B.~M.} \emph{et~al.}
\newblock \bibinfo{title}{{CHEOPS precision phase curve of the Super-Earth 55
  Cancri e}}.
\newblock \emph{\bibinfo{journal}{\aap}} \textbf{\bibinfo{volume}{653}},
  \bibinfo{pages}{A173} (\bibinfo{year}{2021}).

\bibitem{Hoyer2020A&A...635A..24H}
\bibinfo{author}{{Hoyer}, S.} \emph{et~al.}
\newblock \bibinfo{title}{{Expected performances of the Characterising
  Exoplanet Satellite (CHEOPS). III. Data reduction pipeline: architecture and
  simulated performances}}.
\newblock \emph{\bibinfo{journal}{\aap}} \textbf{\bibinfo{volume}{635}},
  \bibinfo{pages}{A24} (\bibinfo{year}{2020}).

\bibitem{2019JATIS...5a5001N}
\bibinfo{author}{{Narita}, N.} \emph{et~al.}
\newblock \bibinfo{title}{{MuSCAT2: four-color simultaneous camera for the
  1.52-m Telescopio Carlos S{\'a}nchez}}.
\newblock \emph{\bibinfo{journal}{J. Astron. Telesc.
  Instrum. Syst.}} \textbf{\bibinfo{volume}{5}}~(1),
  \bibinfo{pages}{015001} (\bibinfo{year}{2019}).

\bibitem{Parviainen2020A&A...633A..28P}
\bibinfo{author}{{Parviainen}, H.} \emph{et~al.}
\newblock \bibinfo{title}{{MuSCAT2 multicolour validation of TESS candidates:
  an ultra-short-period substellar object around an M dwarf}}.
\newblock \emph{\bibinfo{journal}{\aap}} \textbf{\bibinfo{volume}{633}},
  \bibinfo{pages}{A28} (\bibinfo{year}{2020}).

\bibitem{Brown2013}
\bibinfo{author}{{Brown}, T.~M.} \emph{et~al.}
\newblock \bibinfo{title}{{Las Cumbres Observatory Global Telescope Network}}.
\newblock \emph{\bibinfo{journal}{\pasp}} \textbf{\bibinfo{volume}{125}}~(931),
  \bibinfo{pages}{1031} (\bibinfo{year}{2013}).

\bibitem{McCully2018}
\bibinfo{author}{{McCully}, C.} \emph{et~al.}
\newblock \bibinfo{title}{{Real-time processing of the imaging data from the
  network of Las Cumbres Observatory Telescopes using BANZAI}}.
\newblock \emph{\bibinfo{journal}{\procspie}} \textbf{\bibinfo{volume}{10707}},
  \bibinfo{pages}{107070K} (\bibinfo{year}{2018}).

\bibitem{Collins:2017}
\bibinfo{author}{{Collins}, K.~A.}, \bibinfo{author}{{Kielkopf}, J.~F.},
  \bibinfo{author}{{Stassun}, K.~G.} \& \bibinfo{author}{{Hessman}, F.~V.}
\newblock \bibinfo{title}{{AstroImageJ: Image Processing and Photometric
  Extraction for Ultra-precise Astronomical Light Curves}}.
\newblock \emph{\bibinfo{journal}{\aj}} \textbf{\bibinfo{volume}{153}},
  \bibinfo{pages}{77} (\bibinfo{year}{2017}).

\bibitem{Wheatley2018}
\bibinfo{author}{{Wheatley}, P.~J.} \emph{et~al.}
\newblock \bibinfo{title}{{The Next Generation Transit Survey (NGTS)}}.
\newblock \emph{\bibinfo{journal}{\mnras}} \textbf{\bibinfo{volume}{475}}~(4),
  \bibinfo{pages}{4476--4493} (\bibinfo{year}{2018}).

\bibitem{Tierras}
\bibinfo{author}{{Garcia-Mejia}, J.} \emph{et~al.}
\newblock \bibinfo{editor}{{Marshall}, H.~K.}, \bibinfo{editor}{{Spyromilio},
  J.} \& \bibinfo{editor}{{Usuda}, T.} (eds) \emph{\bibinfo{title}{{The Tierras
  Observatory: An ultra-precise photometer to characterize nearby terrestrial
  exoplanets}}}.
\newblock (eds \bibinfo{editor}{{Marshall}, H.~K.},
  \bibinfo{editor}{{Spyromilio}, J.} \& \bibinfo{editor}{{Usuda}, T.})
  \emph{\bibinfo{booktitle}{Ground-based and Airborne Telescopes VIII}}, Vol.
  \bibinfo{volume}{11445} of \emph{\bibinfo{series}{Society of Photo-Optical
  Instrumentation Engineers (SPIE) Conference Series}},
  \bibinfo{pages}{114457R} (\bibinfo{year}{2020}).

\bibitem{2020A&A...642A..49D}
\bibinfo{author}{{Demory}, B.~O.} \emph{et~al.}
\newblock \bibinfo{title}{{A super-Earth and a sub-Neptune orbiting the bright,
  quiet M3 dwarf TOI-1266}}.
\newblock \emph{\bibinfo{journal}{\aap}} \textbf{\bibinfo{volume}{642}},
  \bibinfo{pages}{A49} (\bibinfo{year}{2020}).

\bibitem{2020SPIE11447E..5KN}
\bibinfo{author}{Narita, N.} \emph{et~al.}
\newblock \bibinfo{editor}{Evans, C.~J.}, \bibinfo{editor}{Bryant, J.~J.} \&
  \bibinfo{editor}{Motohara, K.} (eds) \emph{\bibinfo{title}{{MuSCAT3: a
  4-color simultaneous camera for the 2m Faulkes Telescope North}}}.
\newblock (eds \bibinfo{editor}{Evans, C.~J.}, \bibinfo{editor}{Bryant, J.~J.}
  \& \bibinfo{editor}{Motohara, K.}) \emph{\bibinfo{booktitle}{Ground-based and
  Airborne Instrumentation for Astronomy VIII}}, Vol. \bibinfo{volume}{11447},
  \bibinfo{pages}{114475K}. \bibinfo{organization}{International Society for
  Optics and Photonics} (\bibinfo{publisher}{SPIE}, \bibinfo{year}{2020}).

\bibitem{2011PASJ...63..287F}
\bibinfo{author}{{Fukui}, A.} \emph{et~al.}
\newblock \bibinfo{title}{{Measurements of Transit Timing Variations for
  WASP-5b}}.
\newblock \emph{\bibinfo{journal}{\pasj}} \textbf{\bibinfo{volume}{63}},
  \bibinfo{pages}{287--300} (\bibinfo{year}{2011}).

\bibitem{ciardi2015}
\bibinfo{author}{{Ciardi}, D.~R.}, \bibinfo{author}{{Beichman}, C.~A.},
  \bibinfo{author}{{Horch}, E.~P.} \& \bibinfo{author}{{Howell}, S.~B.}
\newblock \bibinfo{title}{{Understanding the Effects of Stellar Multiplicity on
  the Derived Planet Radii from Transit Surveys: Implications for Kepler, K2,
  and TESS}}.
\newblock \emph{\bibinfo{journal}{\apj}} \textbf{\bibinfo{volume}{805}}~(1),
  \bibinfo{pages}{16} (\bibinfo{year}{2015}).

\bibitem{hayward2001}
\bibinfo{author}{{Hayward}, T.~L.} \emph{et~al.}
\newblock \bibinfo{title}{{PHARO: A Near-Infrared Camera for the Palomar
  Adaptive Optics System}}.
\newblock \emph{\bibinfo{journal}{\pasp}} \textbf{\bibinfo{volume}{113}}~(779),
  \bibinfo{pages}{105--118} (\bibinfo{year}{2001}).

\bibitem{dekany2013}
\bibinfo{author}{{Dekany}, R.} \emph{et~al.}
\newblock \bibinfo{title}{{PALM-3000: Exoplanet Adaptive Optics for the 5 m
  Hale Telescope}}.
\newblock \emph{\bibinfo{journal}{\apj}} \textbf{\bibinfo{volume}{776}}~(2),
  \bibinfo{pages}{130} (\bibinfo{year}{2013}).

\bibitem{furlan2017}
\bibinfo{author}{{Furlan}, E.} \emph{et~al.}
\newblock \bibinfo{title}{{The Kepler Follow-up Observation Program. I. A
  Catalog of Companions to Kepler Stars from High-Resolution Imaging}}.
\newblock \emph{\bibinfo{journal}{\aj}} \textbf{\bibinfo{volume}{153}}~(2),
  \bibinfo{pages}{71} (\bibinfo{year}{2017}).

\bibitem{Scott2021FrASS...8..138S}
\bibinfo{author}{{Scott}, N.~J.} \emph{et~al.}
\newblock \bibinfo{title}{{Twin High-resolution, High-speed Imagers for the
  Gemini Telescopes: Instrument description and science verification results}}.
\newblock \emph{\bibinfo{journal}{Frontiers in Astronomy and Space Sciences}}
  \textbf{\bibinfo{volume}{8}}, \bibinfo{pages}{138} (\bibinfo{year}{2021}).

\bibitem{Howell(2011)}
\bibinfo{author}{{Howell}, S.~B.}, \bibinfo{author}{{Everett}, M.~E.},
  \bibinfo{author}{{Sherry}, W.}, \bibinfo{author}{{Horch}, E.} \&
  \bibinfo{author}{{Ciardi}, D.~R.}
\newblock \bibinfo{title}{{Speckle Camera Observations for the NASA Kepler
  Mission Follow-up Program}}.
\newblock \emph{\bibinfo{journal}{\aj}} \textbf{\bibinfo{volume}{142}}~(1),
  \bibinfo{pages}{19} (\bibinfo{year}{2011}).

\bibitem{mugrauer2020}
\bibinfo{author}{{Mugrauer}, M.} \& \bibinfo{author}{{Michel}, K.-U.}
\newblock \bibinfo{title}{{Gaia search for stellar companions of TESS Objects
  of Interest}}.
\newblock \emph{\bibinfo{journal}{Astronomische Nachrichten}}
  \textbf{\bibinfo{volume}{341}}~(10), \bibinfo{pages}{996--1030}
  (\bibinfo{year}{2020}).

\bibitem{mugrauer2021}
\bibinfo{author}{{Mugrauer}, M.} \& \bibinfo{author}{{Michel}, K.-U.}
\newblock \bibinfo{title}{{Gaia search for stellar companions of TESS Objects
  of Interest II}}.
\newblock \emph{\bibinfo{journal}{Astronomische Nachrichten}}
  \textbf{\bibinfo{volume}{342}}~(6), \bibinfo{pages}{840--864}
  (\bibinfo{year}{2021}).

\bibitem{ziegler2020}
\bibinfo{author}{{Ziegler}, C.} \emph{et~al.}
\newblock \bibinfo{title}{{SOAR TESS Survey. I. Sculpting of TESS Planetary
  Systems by Stellar Companions}}.
\newblock \emph{\bibinfo{journal}{\aj}} \textbf{\bibinfo{volume}{159}}~(1),
  \bibinfo{pages}{19} (\bibinfo{year}{2020}).

\bibitem{Lafarga2020}
\bibinfo{author}{{Lafarga}, M.} \emph{et~al.}
\newblock \bibinfo{title}{{The CARMENES search for exoplanets around M dwarfs.
  Radial velocities and activity indicators from cross-correlation functions
  with weighted binary masks}}.
\newblock \emph{\bibinfo{journal}{\aap}} \textbf{\bibinfo{volume}{636}},
  \bibinfo{pages}{A36} (\bibinfo{year}{2020}).

\bibitem{SERVAL}
\bibinfo{author}{{Zechmeister}, M.} \emph{et~al.}
\newblock \bibinfo{title}{{Spectrum radial velocity analyser (SERVAL).
  High-precision radial velocities and two alternative spectral indicators}}.
\newblock \emph{\bibinfo{journal}{\aap}} \textbf{\bibinfo{volume}{609}},
  \bibinfo{pages}{A12} (\bibinfo{year}{2018}).

\bibitem{2014SPIE.9147E..8CC}
\bibinfo{author}{{Cosentino}, R.} \emph{et~al.}
\newblock \bibinfo{editor}{Ramsay, S.~K.}, \bibinfo{editor}{McLean, I.~S.} \&
  \bibinfo{editor}{Takami, H.} (eds) \emph{\bibinfo{title}{{HARPS-N @ TNG, two
  year harvesting data: performances and results}}}.
\newblock (eds \bibinfo{editor}{Ramsay, S.~K.}, \bibinfo{editor}{McLean, I.~S.}
  \& \bibinfo{editor}{Takami, H.}) \emph{\bibinfo{booktitle}{Ground-based and
  Airborne Instrumentation for Astronomy V}}, Vol. \bibinfo{volume}{9147},
  \bibinfo{pages}{91478C}. \bibinfo{organization}{International Society for
  Optics and Photonics} (\bibinfo{publisher}{SPIE}, \bibinfo{year}{2014}).

\bibitem{Santos2013}
\bibinfo{author}{{Santos}, N.~C.} \emph{et~al.}
\newblock \bibinfo{title}{{SWEET-Cat: A catalogue of parameters for Stars With
  ExoplanETs. I. New atmospheric parameters and masses for 48 stars with
  planets}}.
\newblock \emph{\bibinfo{journal}{\aap}} \textbf{\bibinfo{volume}{556}},
  \bibinfo{pages}{A150} (\bibinfo{year}{2013}).

\bibitem{Sousa2014}
\bibinfo{author}{{Sousa}, S.~G.}
\newblock \emph{\bibinfo{title}{{ARES + MOOG: A Practical Overview of an
  Equivalent Width (EW) Method to Derive Stellar Parameters}}},
  \bibinfo{pages}{297--310} (\bibinfo{publisher}{Springer},
  \bibinfo{year}{2014}).

\bibitem{Sousa2021}
\bibinfo{author}{{Sousa}, S.~G.} \emph{et~al.}
\newblock \bibinfo{title}{{SWEET-Cat 2.0: The Cat just got SWEETer. Higher
  quality spectra and precise parallaxes from Gaia eDR3}}.
\newblock \emph{\bibinfo{journal}{\aap}} \textbf{\bibinfo{volume}{656}},
  \bibinfo{pages}{A53} (\bibinfo{year}{2021}).

\bibitem{Sousa2007}
\bibinfo{author}{{Sousa}, S.~G.}, \bibinfo{author}{{Santos}, N.~C.},
  \bibinfo{author}{{Israelian}, G.}, \bibinfo{author}{{Mayor}, M.} \&
  \bibinfo{author}{{Monteiro}, M.~J.~P.~F.~G.}
\newblock \bibinfo{title}{{A new code for automatic determination of equivalent
  widths: Automatic Routine for line Equivalent widths in stellar Spectra
  (ARES)}}.
\newblock \emph{\bibinfo{journal}{\aap}} \textbf{\bibinfo{volume}{469}}~(2),
  \bibinfo{pages}{783--791} (\bibinfo{year}{2007}).

\bibitem{Sousa2015}
\bibinfo{author}{{Sousa}, S.~G.}, \bibinfo{author}{{Santos}, N.~C.},
  \bibinfo{author}{{Adibekyan}, V.}, \bibinfo{author}{{Delgado-Mena}, E.} \&
  \bibinfo{author}{{Israelian}, G.}
\newblock \bibinfo{title}{{ARES v2: new features and improved performance}}.
\newblock \emph{\bibinfo{journal}{\aap}} \textbf{\bibinfo{volume}{577}},
  \bibinfo{pages}{A67} (\bibinfo{year}{2015}).

\bibitem{Sneden1973}
\bibinfo{author}{{Sneden}, C.~A.}
\newblock \emph{\bibinfo{title}{{Carbon and Nitrogen Abundances in Metal-Poor
  Stars.}}}
\newblock Ph.D. thesis, \bibinfo{school}{THE UNIVERSITY OF TEXAS AT AUSTIN.}
  (\bibinfo{year}{1973}).

\bibitem{Kurucz1993}
\bibinfo{author}{{Kurucz}, R.~L.}
\newblock \emph{\bibinfo{title}{{SYNTHE spectrum synthesis programs and line
  data}}}  (\bibinfo{publisher}{Astrophysics Source Code Library},
  \bibinfo{year}{1993}).

\bibitem{ZASPE}
\bibinfo{author}{{Brahm}, R.}, \bibinfo{author}{{Jord{\'a}n}, A.},
  \bibinfo{author}{{Hartman}, J.} \& \bibinfo{author}{{Bakos}, G.}
\newblock \bibinfo{title}{{ZASPE: A Code to Measure Stellar Atmospheric
  Parameters and their Covariance from Spectra}}.
\newblock \emph{\bibinfo{journal}{\mnras}} \textbf{\bibinfo{volume}{467}}~(1),
  \bibinfo{pages}{971--984} (\bibinfo{year}{2017}).

\bibitem{Adibekyan2012}
\bibinfo{author}{{Adibekyan}, V.~Z.} \emph{et~al.}
\newblock \bibinfo{title}{{Chemical abundances of 1111 FGK stars from the HARPS
  GTO planet search program. Galactic stellar populations and planets}}.
\newblock \emph{\bibinfo{journal}{\aap}} \textbf{\bibinfo{volume}{545}},
  \bibinfo{pages}{A32} (\bibinfo{year}{2012}).

\bibitem{Adibekyan2015}
\bibinfo{author}{{Adibekyan}, V.} \emph{et~al.}
\newblock \bibinfo{title}{{Identifying the best iron-peak and
  {\ensuremath{\alpha}}-capture elements for chemical tagging: The impact of
  the number of lines on measured scatter}}.
\newblock \emph{\bibinfo{journal}{\aap}} \textbf{\bibinfo{volume}{583}},
  \bibinfo{pages}{A94} (\bibinfo{year}{2015}).

\bibitem{Castelli2003}
\bibinfo{author}{{Castelli}, F.} \& \bibinfo{author}{{Kurucz}, R.~L.}
\newblock \bibinfo{editor}{{Piskunov}, N.}, \bibinfo{editor}{{Weiss}, W.~W.} \&
  \bibinfo{editor}{{Gray}, D.~F.} (eds) \emph{\bibinfo{title}{{New Grids of
  ATLAS9 Model Atmospheres}}}.
\newblock (eds \bibinfo{editor}{{Piskunov}, N.}, \bibinfo{editor}{{Weiss},
  W.~W.} \& \bibinfo{editor}{{Gray}, D.~F.})
  \emph{\bibinfo{booktitle}{Modelling of Stellar Atmospheres}}, Vol.
  \bibinfo{volume}{210} of \emph{\bibinfo{series}{IAU Symposium}},
  \bibinfo{pages}{A20} (\bibinfo{year}{2003}).

\bibitem{Allard2014}
\bibinfo{author}{{Allard}, F.}
\newblock \bibinfo{editor}{{Booth}, M.}, \bibinfo{editor}{{Matthews}, B.~C.} \&
  \bibinfo{editor}{{Graham}, J.~R.} (eds) \emph{\bibinfo{title}{{The BT-Settl
  Model Atmospheres for Stars, Brown Dwarfs and Planets}}}.
\newblock (eds \bibinfo{editor}{{Booth}, M.}, \bibinfo{editor}{{Matthews},
  B.~C.} \& \bibinfo{editor}{{Graham}, J.~R.})
  \emph{\bibinfo{booktitle}{Exploring the Formation and Evolution of Planetary
  Systems}}, Vol. \bibinfo{volume}{299}, \bibinfo{pages}{271--272}
  (\bibinfo{year}{2014}).

\bibitem{Blackwell1977}
\bibinfo{author}{{Blackwell}, D.~E.} \& \bibinfo{author}{{Shallis}, M.~J.}
\newblock \bibinfo{title}{{Stellar angular diameters from infrared photometry.
  Application to Arcturus and other stars; with effective temperatures.}}
\newblock \emph{\bibinfo{journal}{\mnras}} \textbf{\bibinfo{volume}{180}},
  \bibinfo{pages}{177--191} (\bibinfo{year}{1977}).

\bibitem{Schanche2020}
\bibinfo{author}{{Schanche}, N.} \emph{et~al.}
\newblock \bibinfo{title}{{WASP-186 and WASP-187: two hot Jupiters discovered
  by SuperWASP and SOPHIE with additional observations by TESS}}.
\newblock \emph{\bibinfo{journal}{\mnras}} \textbf{\bibinfo{volume}{499}}~(1),
  \bibinfo{pages}{428--440} (\bibinfo{year}{2020}).

\bibitem{Wilson2022}
\bibinfo{author}{{Wilson}, T.~G.} \emph{et~al.}
\newblock \bibinfo{title}{{A pair of sub-Neptunes transiting the bright K-dwarf
  TOI-1064 characterized with CHEOPS}}.
\newblock \emph{\bibinfo{journal}{\mnras}} \textbf{\bibinfo{volume}{511}}~(1),
  \bibinfo{pages}{1043--1071} (\bibinfo{year}{2022}).

\bibitem{Lindegren2021}
\bibinfo{author}{{Lindegren}, L.} \emph{et~al.}
\newblock \bibinfo{title}{{Gaia Early Data Release 3. Parallax bias versus
  magnitude, colour, and position}}.
\newblock \emph{\bibinfo{journal}{\aap}} \textbf{\bibinfo{volume}{649}},
  \bibinfo{pages}{A4} (\bibinfo{year}{2021}).

\bibitem{Bonfanti2021}
\bibinfo{author}{{Bonfanti}, A.} \emph{et~al.}
\newblock \bibinfo{title}{{CHEOPS observations of the HD 108236 planetary
  system: a fifth planet, improved ephemerides, and planetary radii}}.
\newblock \emph{\bibinfo{journal}{\aap}} \textbf{\bibinfo{volume}{646}},
  \bibinfo{pages}{A157} (\bibinfo{year}{2021}).

\bibitem{Bonfanti2015}
\bibinfo{author}{{Bonfanti}, A.}, \bibinfo{author}{{Ortolani}, S.},
  \bibinfo{author}{{Piotto}, G.} \& \bibinfo{author}{{Nascimbeni}, V.}
\newblock \bibinfo{title}{{Revising the ages of planet-hosting stars}}.
\newblock \emph{\bibinfo{journal}{\aap}} \textbf{\bibinfo{volume}{575}},
  \bibinfo{pages}{A18} (\bibinfo{year}{2015}).

\bibitem{Bonfanti2016}
\bibinfo{author}{{Bonfanti}, A.}, \bibinfo{author}{{Ortolani}, S.} \&
  \bibinfo{author}{{Nascimbeni}, V.}
\newblock \bibinfo{title}{{Age consistency between exoplanet hosts and field
  stars}}.
\newblock \emph{\bibinfo{journal}{\aap}} \textbf{\bibinfo{volume}{585}},
  \bibinfo{pages}{A5} (\bibinfo{year}{2016}).

\bibitem{Marigo2017}
\bibinfo{author}{{Marigo}, P.} \emph{et~al.}
\newblock \bibinfo{title}{{A New Generation of PARSEC-COLIBRI Stellar
  Isochrones Including the TP-AGB Phase}}.
\newblock \emph{\bibinfo{journal}{\apj}} \textbf{\bibinfo{volume}{835}},
  \bibinfo{pages}{77} (\bibinfo{year}{2017}).

\bibitem{Scuflaire2008}
\bibinfo{author}{{Scuflaire}, R.} \emph{et~al.}
\newblock \bibinfo{title}{{CL{\'E}S, Code Li{\'e}geois d'{\'E}volution
  Stellaire}}.
\newblock \emph{\bibinfo{journal}{\apss}} \textbf{\bibinfo{volume}{316}},
  \bibinfo{pages}{83--91} (\bibinfo{year}{2008}).
  .

\bibitem{Salmon2021}
\bibinfo{author}{{Salmon}, S.~J.~A.~J.}, \bibinfo{author}{{Van Grootel}, V.},
  \bibinfo{author}{{Buldgen}, G.}, \bibinfo{author}{{Dupret}, M.~A.} \&
  \bibinfo{author}{{Eggenberger}, P.}
\newblock \bibinfo{title}{{Reinvestigating {\ensuremath{\alpha}} Centauri AB in
  light of asteroseismic forward and inverse methods}}.
\newblock \emph{\bibinfo{journal}{\aap}} \textbf{\bibinfo{volume}{646}},
  \bibinfo{pages}{A7} (\bibinfo{year}{2021}).

\bibitem{2006MNRAS.367.1329R}
\bibinfo{author}{{Reddy}, B.~E.}, \bibinfo{author}{{Lambert}, D.~L.} \&
  \bibinfo{author}{{Allende Prieto}, C.}
\newblock \bibinfo{title}{{Elemental abundance survey of the Galactic thick
  disc}}.
\newblock \emph{\bibinfo{journal}{\mnras}} \textbf{\bibinfo{volume}{367}}~(4),
  \bibinfo{pages}{1329--1366} (\bibinfo{year}{2006}).

\bibitem{exoplanet:exoplanet}
\bibinfo{author}{Foreman-Mackey, D.} \emph{et~al.}
\newblock \bibinfo{title}{dfm/exoplanet: exoplanet v0.2.1}
  (\bibinfo{year}{2019}).

\bibitem{Delrez2021}
\bibinfo{author}{{Delrez}, L.} \emph{et~al.}
\newblock \bibinfo{title}{{Transit detection of the long-period volatile-rich
  super-Earth {\ensuremath{\nu}}$^{2}$ Lupi d with CHEOPS}}.
\newblock \emph{\bibinfo{journal}{Nat. Astron.}}
  \textbf{\bibinfo{volume}{5}}, \bibinfo{pages}{775--787}
  (\bibinfo{year}{2021}).

\bibitem{Claret2018A&A...618A..20C}
\bibinfo{author}{{Claret}, A.}
\newblock \bibinfo{title}{{A new method to compute limb-darkening coefficients
  for stellar atmosphere models with spherical symmetry: the space missions
  TESS, Kepler, CoRoT, and MOST}}.
\newblock \emph{\bibinfo{journal}{\aap}} \textbf{\bibinfo{volume}{618}},
  \bibinfo{pages}{A20} (\bibinfo{year}{2018}).

\bibitem{Claret2021RNAAS...5...13C}
\bibinfo{author}{{Claret}, A.}
\newblock \bibinfo{title}{{Limb and Gravity-darkening Coefficients for the
  Space Mission CHEOPS}}.
\newblock \emph{\bibinfo{journal}{Research Notes of the American Astronomical
  Society}} \textbf{\bibinfo{volume}{5}}~(1), \bibinfo{pages}{13}
  (\bibinfo{year}{2021}).

\bibitem{VanEylenAlbrecht2015}
\bibinfo{author}{{Van Eylen}, V.} \& \bibinfo{author}{{Albrecht}, S.}
\newblock \bibinfo{title}{{Eccentricity from Transit Photometry: Small Planets
  in Kepler Multi-planet Systems Have Low Eccentricities}}.
\newblock \emph{\bibinfo{journal}{\apj}} \textbf{\bibinfo{volume}{808}}~(2),
  \bibinfo{pages}{126} (\bibinfo{year}{2015}).

\bibitem{Xie2016}
\bibinfo{author}{{Xie}, J.-W.} \emph{et~al.}
\newblock \bibinfo{title}{{Exoplanet orbital eccentricities derived from
  LAMOST-Kepler analysis}}.
\newblock \emph{\bibinfo{journal}{\pnas}} \textbf{\bibinfo{volume}{113}}~(41),
  \bibinfo{pages}{11431--11435} (\bibinfo{year}{2016}).

\bibitem{HaddenLithwick2017}
\bibinfo{author}{{Hadden}, S.} \& \bibinfo{author}{{Lithwick}, Y.}
\newblock \bibinfo{title}{{Kepler Planet Masses and Eccentricities from TTV
  Analysis}}.
\newblock \emph{\bibinfo{journal}{\aj}} \textbf{\bibinfo{volume}{154}}~(1),
  \bibinfo{pages}{5} (\bibinfo{year}{2017}).

\bibitem{MonoTools}
\bibinfo{author}{{Osborn}, H.~P.}
\newblock \bibinfo{title}{{MonoTools: Planets of uncertain periods detector and
  modeler}}.
\newblock \bibinfo{howpublished}{Astrophysics Source Code Library, record
  ascl:2204.020} (\bibinfo{year}{2022}).

\bibitem{KippingSingleTrans}
\bibinfo{author}{{Kipping}, D.}
\newblock \bibinfo{title}{{The Orbital Period Prior for Single Transits}}.
\newblock \emph{\bibinfo{journal}{Research Notes of the American Astronomical
  Society}} \textbf{\bibinfo{volume}{2}}~(4), \bibinfo{pages}{223}
  (\bibinfo{year}{2018}).

\bibitem{van2019orbital}
\bibinfo{author}{Van~Eylen, V.} \emph{et~al.}
\newblock \bibinfo{title}{The orbital eccentricity of small planet systems}.
\newblock \emph{\bibinfo{journal}{The Astronomical Journal}}
  \textbf{\bibinfo{volume}{157}}~(2), \bibinfo{pages}{61}
  (\bibinfo{year}{2019}) .

\bibitem{OsbornTOI2076}
\bibinfo{author}{{Osborn}, H.~P.} \emph{et~al.}
\newblock \bibinfo{title}{{Uncovering the true periods of the young
  sub-Neptunes orbiting TOI-2076}}.
\newblock \emph{\bibinfo{journal}{\aap}} \textbf{\bibinfo{volume}{664}},
  \bibinfo{pages}{A156} (\bibinfo{year}{2022}).

\bibitem{Mills2016}
\bibinfo{author}{{Mills}, S.~M.} \emph{et~al.}
\newblock \bibinfo{title}{{A resonant chain of four transiting, sub-Neptune
  planets}}.
\newblock \emph{\bibinfo{journal}{\nat}} \textbf{\bibinfo{volume}{533}}~(7604),
  \bibinfo{pages}{509--512} (\bibinfo{year}{2016}).

\bibitem{SiFa2021}
\bibinfo{author}{{Siegel}, J.~C.} \& \bibinfo{author}{{Fabrycky}, D.}
\newblock \bibinfo{title}{{Resonant Chains of Exoplanets: Libration Centers for
  Three-body Angles}}.
\newblock \emph{\bibinfo{journal}{\aj}} \textbf{\bibinfo{volume}{161}}~(6),
  \bibinfo{pages}{290} (\bibinfo{year}{2021}).

\bibitem{Lopez2019}
\bibinfo{author}{{Lopez}, T.~A.} \emph{et~al.}
\newblock \bibinfo{title}{{Exoplanet characterisation in the longest known
  resonant chain: the K2-138 system seen by HARPS}}.
\newblock \emph{\bibinfo{journal}{\aap}} \textbf{\bibinfo{volume}{631}},
  \bibinfo{pages}{A90} (\bibinfo{year}{2019}).

\bibitem{Rebound}
\bibinfo{author}{{Rein}, H.} \& \bibinfo{author}{{Liu}, S.~F.}
\newblock \bibinfo{title}{{REBOUND: an open-source multi-purpose N-body code
  for collisional dynamics}}.
\newblock \emph{\bibinfo{journal}{\aap}} \textbf{\bibinfo{volume}{537}},
  \bibinfo{pages}{A128} (\bibinfo{year}{2012}).

\bibitem{samsam}
\bibinfo{author}{{Delisle}, J.-B.}
\newblock \bibinfo{title}{{samsam: Scaled Adaptive Metropolis SAMpler}}.
\newblock \bibinfo{howpublished}{Astrophysics Source Code Library, record
  ascl:2207.011} (\bibinfo{year}{2022}).

\bibitem{RIVERS3}
\bibinfo{author}{{Leleu}, A.} \emph{et~al.}
\newblock \bibinfo{title}{{Removing biases on the density of sub-Neptunes
  characterised via transit timing variations. Update on the mass-radius
  relationship of 34 Kepler planets}}.
\newblock \emph{\bibinfo{journal}{\aap}} \textbf{\bibinfo{volume}{669}},
  \bibinfo{pages}{A117} (\bibinfo{year}{2023}).

\bibitem{ldtk}
\bibinfo{author}{Parviainen, H.} \& \bibinfo{author}{Aigrain, S.}
\newblock \bibinfo{title}{{ldtk: Limb Darkening Toolkit}}.
\newblock \emph{\bibinfo{journal}{\mnras}} \textbf{\bibinfo{volume}{453}}~(4),
  \bibinfo{pages}{3821--3826} (\bibinfo{year}{2015}).

\bibitem{exoplanet:pymc3}
\bibinfo{author}{Salvatier, J.}, \bibinfo{author}{Wiecki, T.~V.} \&
  \bibinfo{author}{Fonnesbeck, C.}
\newblock \bibinfo{title}{Probabilistic programming in python using pymc3}.
\newblock \emph{\bibinfo{journal}{PeerJ Comput. Sci.}}
  \textbf{\bibinfo{volume}{2}}, \bibinfo{pages}{e55} (\bibinfo{year}{2016}) .

\bibitem{watanabe2010asymptotic}
\bibinfo{author}{Watanabe, S.} \& \bibinfo{author}{Opper, M.}
\newblock \bibinfo{title}{Asymptotic equivalence of bayes cross validation and
  widely applicable information criterion in singular learning theory.}
\newblock \emph{\bibinfo{journal}{J. Mach. Learn. Res.}}
  \textbf{\bibinfo{volume}{11}}~(12) (\bibinfo{year}{2010}) .

\bibitem{vehtari2017practical}
\bibinfo{author}{Vehtari, A.}, \bibinfo{author}{Gelman, A.} \&
  \bibinfo{author}{Gabry, J.}
\newblock \bibinfo{title}{Practical bayesian model evaluation using
  leave-one-out cross-validation and waic}.
\newblock \emph{\bibinfo{journal}{Stat. Comput.}}
  \textbf{\bibinfo{volume}{27}}, \bibinfo{pages}{1413--1432}
  (\bibinfo{year}{2017}).

\bibitem{arviz}
\bibinfo{author}{{ArviZ Developers}}.
\newblock \bibinfo{title}{{ArviZ: Exploratory analysis of Bayesian models}}.
\newblock \bibinfo{howpublished}{Astrophysics Source Code Library, record
  ascl:2004.012} (\bibinfo{year}{2020}).

\bibitem{GLS}
\bibinfo{author}{{Zechmeister}, M.} \& \bibinfo{author}{{K{\"u}rster}, M.}
\newblock \bibinfo{title}{{The generalised Lomb-Scargle periodogram. A new
  formalism for the floating-mean and Keplerian periodograms}}.
\newblock \emph{\bibinfo{journal}{\aap}} \textbf{\bibinfo{volume}{496}},
  \bibinfo{pages}{577--584} (\bibinfo{year}{2009}).

\bibitem{saar97}
\bibinfo{author}{{Saar}, S.~H.} \& \bibinfo{author}{{Donahue}, R.~A.}
\newblock \bibinfo{title}{{Activity-Related Radial Velocity Variation in Cool
  Stars}}.
\newblock \emph{\bibinfo{journal}{\apj}} \textbf{\bibinfo{volume}{485}}~(1),
  \bibinfo{pages}{319--327} (\bibinfo{year}{1997}).

\bibitem{hatzes02}
\bibinfo{author}{{Hatzes}, A.~P.}
\newblock \bibinfo{title}{{Starspots and exoplanets}}.
\newblock \emph{\bibinfo{journal}{Astron. Nachrichten}}
  \textbf{\bibinfo{volume}{323}}, \bibinfo{pages}{392--394}
  (\bibinfo{year}{2002}).

\bibitem{meunier10}
\bibinfo{author}{{Meunier}, N.}, \bibinfo{author}{{Desort}, M.} \&
  \bibinfo{author}{{Lagrange}, A.~M.}
\newblock \bibinfo{title}{{Using the Sun to estimate Earth-like planets
  detection capabilities . II. Impact of plages}}.
\newblock \emph{\bibinfo{journal}{\aap}} \textbf{\bibinfo{volume}{512}},
  \bibinfo{pages}{A39} (\bibinfo{year}{2010}).

\bibitem{dumusque14}
\bibinfo{author}{{Dumusque}, X.}, \bibinfo{author}{{Boisse}, I.} \&
  \bibinfo{author}{{Santos}, N.~C.}
\newblock \bibinfo{title}{{SOAP 2.0: A Tool to Estimate the Photometric and
  Radial Velocity Variations Induced by Stellar Spots and Plages}}.
\newblock \emph{\bibinfo{journal}{\apj}} \textbf{\bibinfo{volume}{796}}~(2),
  \bibinfo{pages}{132} (\bibinfo{year}{2014}).

\bibitem{queloz01}
\bibinfo{author}{{Queloz}, D.} \emph{et~al.}
\newblock \bibinfo{title}{{No planet for HD 166435}}.
\newblock \emph{\bibinfo{journal}{\aap}} \textbf{\bibinfo{volume}{379}},
  \bibinfo{pages}{279--287} (\bibinfo{year}{2001}).

\bibitem{boisse09}
\bibinfo{author}{{Boisse}, I.} \emph{et~al.}
\newblock \bibinfo{title}{{Stellar activity of planetary host star HD 189
  733}}.
\newblock \emph{\bibinfo{journal}{\aap}} \textbf{\bibinfo{volume}{495}}~(3),
  \bibinfo{pages}{959--966} (\bibinfo{year}{2009}).

\bibitem{dumusque16}
\bibinfo{author}{{Dumusque}, X.}
\newblock \bibinfo{title}{{Radial velocity fitting challenge. I. Simulating the
  data set including realistic stellar radial-velocity signals}}.
\newblock \emph{\bibinfo{journal}{\aap}} \textbf{\bibinfo{volume}{593}},
  \bibinfo{pages}{A5} (\bibinfo{year}{2016}).

\bibitem{simola19}
\bibinfo{author}{{Simola}, U.}, \bibinfo{author}{{Dumusque}, X.} \&
  \bibinfo{author}{{Cisewski-Kehe}, J.}
\newblock \bibinfo{title}{{Measuring precise radial velocities and
  cross-correlation function line-profile variations using a Skew Normal
  density}}.
\newblock \emph{\bibinfo{journal}{\aap}} \textbf{\bibinfo{volume}{622}},
  \bibinfo{pages}{A131} (\bibinfo{year}{2019}).

\bibitem{simola22}
\bibinfo{author}{{Simola}, U.} \emph{et~al.}
\newblock \bibinfo{title}{{Accounting for stellar activity signals in
  radial-velocity data by using change point detection techniques}}.
\newblock \emph{\bibinfo{journal}{\aap}} \textbf{\bibinfo{volume}{664}},
  \bibinfo{pages}{A127} (\bibinfo{year}{2022}).

\bibitem{Bonfanti2023arXiv230210664B}
\bibinfo{author}{{Bonfanti}, A.} \emph{et~al.}
\newblock \bibinfo{title}{{TOI-1055 b: Neptunian planet characterised with
  HARPS, TESS, and CHEOPS}}.
\newblock \emph{\bibinfo{journal}{arXiv e-prints}}
  \bibinfo{pages}{arXiv:2302.10664} (\bibinfo{year}{2023}).

\bibitem{bonfanti20}
\bibinfo{author}{{Bonfanti}, A.} \& \bibinfo{author}{{Gillon}, M.}
\newblock \bibinfo{title}{{MCMCI: A code to fully characterise an exoplanetary
  system}}.
\newblock \emph{\bibinfo{journal}{\aap}} \textbf{\bibinfo{volume}{635}},
  \bibinfo{pages}{A6} (\bibinfo{year}{2020}).

\bibitem{schwarz78}
\bibinfo{author}{{Schwarz}, G.}
\newblock \bibinfo{title}{{Estimating the Dimension of a Model}}.
\newblock \emph{\bibinfo{journal}{Ann. Stat.}}
  \textbf{\bibinfo{volume}{6}}~(2), \bibinfo{pages}{461--464}
  (\bibinfo{year}{1978}) .

\bibitem{gelman1992}
\bibinfo{author}{{Gelman}, A.} \& \bibinfo{author}{{Rubin}, D.~B.}
\newblock \bibinfo{title}{{Inference from Iterative Simulation Using Multiple
  Sequences}}.
\newblock \emph{\bibinfo{journal}{Stat. Sci.}}
  \textbf{\bibinfo{volume}{7}}, \bibinfo{pages}{457--472}
  (\bibinfo{year}{1992}).

\bibitem{Rajpaul2015}
\bibinfo{author}{{Rajpaul}, V.}, \bibinfo{author}{{Aigrain}, S.},
  \bibinfo{author}{{Osborne}, M.~A.}, \bibinfo{author}{{Reece}, S.} \&
  \bibinfo{author}{{Roberts}, S.}
\newblock \bibinfo{title}{{A Gaussian process framework for modelling stellar
  activity signals in radial velocity data}}.
\newblock \emph{\bibinfo{journal}{\mnras}} \textbf{\bibinfo{volume}{452}}~(3),
  \bibinfo{pages}{2269--2291} (\bibinfo{year}{2015}).

\bibitem{pyaneti2}
\bibinfo{author}{{Barrag{\'a}n}, O.}, \bibinfo{author}{{Aigrain}, S.},
  \bibinfo{author}{{Rajpaul}, V.~M.} \& \bibinfo{author}{{Zicher}, N.}
\newblock \bibinfo{title}{{PYANETI - II. A multidimensional Gaussian process
  approach to analysing spectroscopic time-series}}.
\newblock \emph{\bibinfo{journal}{\mnras}} \textbf{\bibinfo{volume}{509}}~(1),
  \bibinfo{pages}{866--883} (\bibinfo{year}{2022}).
  .

\bibitem{Barragan2022}
\bibinfo{author}{{Barrag{\'a}n}, O.} \emph{et~al.}
\newblock \bibinfo{title}{{The young HD 73583 (TOI-560) planetary system: two
  10-M$_{{\ensuremath{\oplus}}}$ mini-Neptunes transiting a 500-Myr-old,
  bright, and active K dwarf}}.
\newblock \emph{\bibinfo{journal}{\mnras}} \textbf{\bibinfo{volume}{514}}~(2),
  \bibinfo{pages}{1606--1627} (\bibinfo{year}{2022}).

\bibitem{Zicher2022}
\bibinfo{author}{{Zicher}, N.} \emph{et~al.}
\newblock \bibinfo{title}{{One year of AU Mic with HARPS - I. Measuring the
  masses of the two transiting planets}}.
\newblock \emph{\bibinfo{journal}{\mnras}} \textbf{\bibinfo{volume}{512}}~(2),
  \bibinfo{pages}{3060--3078} (\bibinfo{year}{2022}).

\bibitem{pyaneti}
\bibinfo{author}{{Barrag{\'a}n}, O.}, \bibinfo{author}{{Gandolfi}, D.} \&
  \bibinfo{author}{{Antoniciello}, G.}
\newblock \bibinfo{title}{{PYANETI: a fast and powerful software suite for
  multiplanet radial velocity and transit fitting}}.
\newblock \emph{\bibinfo{journal}{\mnras}} \textbf{\bibinfo{volume}{482}},
  \bibinfo{pages}{1017--1030} (\bibinfo{year}{2019}).

\bibitem{Cale21}
\bibinfo{author}{{Cale}, B.~L.} \emph{et~al.}
\newblock \bibinfo{title}{{Diving Beneath the Sea of Stellar Activity:
  Chromatic Radial Velocities of the Young AU Mic Planetary System}}.
\newblock \emph{\bibinfo{journal}{\aj}} \textbf{\bibinfo{volume}{162}}~(6),
  \bibinfo{pages}{295} (\bibinfo{year}{2021}).

\bibitem{Blunt23}
\bibinfo{author}{{Blunt}, S.} \emph{et~al.}
\newblock \bibinfo{title}{{Overfitting Affects the Reliability of Radial
  Velocity Mass Estimates of the V1298 Tau Planets}}.
\newblock \emph{\bibinfo{journal}{arXiv e-prints}}
  \bibinfo{pages}{arXiv:2306.08145} (\bibinfo{year}{2023}).

\bibitem{Dorn2015}
\bibinfo{author}{{Dorn}, C.} \emph{et~al.}
\newblock \bibinfo{title}{{Can we constrain the interior structure of rocky
  exoplanets from mass and radius measurements?}}
\newblock \emph{\bibinfo{journal}{Astron. Astrophys.}}
  \textbf{\bibinfo{volume}{577}}, \bibinfo{pages}{A83} (\bibinfo{year}{2015}).

\bibitem{Dorn2017}
\bibinfo{author}{{Dorn}, C.} \emph{et~al.}
\newblock \bibinfo{title}{{A generalized Bayesian inference method for
  constraining the interiors of super Earths and sub-Neptunes}}.
\newblock \emph{\bibinfo{journal}{Astron. Astrophys.}}
  \textbf{\bibinfo{volume}{597}}, \bibinfo{pages}{A37} (\bibinfo{year}{2017}).

\bibitem{Haldemann2020}
\bibinfo{author}{{Haldemann}, J.}, \bibinfo{author}{{Alibert}, Y.},
  \bibinfo{author}{{Mordasini}, C.} \& \bibinfo{author}{{Benz}, W.}
\newblock \bibinfo{title}{{AQUA: a collection of H$_{2}$O equations of state
  for planetary models}}.
\newblock \emph{\bibinfo{journal}{Astron. Astrophys.}}
  \textbf{\bibinfo{volume}{643}}, \bibinfo{pages}{A105} (\bibinfo{year}{2020}).

\bibitem{Hakim2018}
\bibinfo{author}{{Hakim}, K.} \emph{et~al.}
\newblock \bibinfo{title}{{A new ab initio equation of state of hcp-Fe and its
  implication on the interior structure and mass-radius relations of rocky
  super-Earths}}.
\newblock \emph{\bibinfo{journal}{\icarus}} \textbf{\bibinfo{volume}{313}},
  \bibinfo{pages}{61--78} (\bibinfo{year}{2018}).

\bibitem{Sotin2007}
\bibinfo{author}{{Sotin}, C.}, \bibinfo{author}{{Grasset}, O.} \&
  \bibinfo{author}{{Mocquet}, A.}
\newblock \bibinfo{title}{{Mass radius curve for extrasolar Earth-like planets
  and ocean planets}}.
\newblock \emph{\bibinfo{journal}{\icarus}} \textbf{\bibinfo{volume}{191}}~(1),
  \bibinfo{pages}{337--351} (\bibinfo{year}{2007}).

\bibitem{Lopez2014}
\bibinfo{author}{{Lopez}, E.~D.} \& \bibinfo{author}{{Fortney}, J.~J.}
\newblock \bibinfo{title}{{Understanding the Mass-Radius Relation for
  Sub-neptunes: Radius as a Proxy for Composition}}.
\newblock \emph{\bibinfo{journal}{Astrophys. J.}}
  \textbf{\bibinfo{volume}{792}}~(1), \bibinfo{pages}{1}
  (\bibinfo{year}{2014}).

\bibitem{Thiabaud2014}
\bibinfo{author}{{Thiabaud}, A.} \emph{et~al.}
\newblock \bibinfo{title}{{From stellar nebula to planets: The refractory
  components}}.
\newblock \emph{\bibinfo{journal}{Astron. Astrophys.}}
  \textbf{\bibinfo{volume}{562}}, \bibinfo{pages}{A27} (\bibinfo{year}{2014}).

\bibitem{Marboeuf2014}
\bibinfo{author}{{Marboeuf}, U.}, \bibinfo{author}{{Thiabaud}, A.},
  \bibinfo{author}{{Alibert}, Y.}, \bibinfo{author}{{Cabral}, N.} \&
  \bibinfo{author}{{Benz}, W.}
\newblock \bibinfo{title}{{From planetesimals to planets: volatile molecules}}.
\newblock \emph{\bibinfo{journal}{Astron. Astrophys.}}
  \textbf{\bibinfo{volume}{570}}, \bibinfo{pages}{A36} (\bibinfo{year}{2014}).

\bibitem{Venturini2020A&A...643L...1V}
\bibinfo{author}{{Venturini}, J.}, \bibinfo{author}{{Guilera}, O.~M.},
  \bibinfo{author}{{Haldemann}, J.}, \bibinfo{author}{{Ronco}, M.~P.} \&
  \bibinfo{author}{{Mordasini}, C.}
\newblock \bibinfo{title}{{The nature of the radius valley. Hints from
  formation and evolution models}}.
\newblock \emph{\bibinfo{journal}{\aap}} \textbf{\bibinfo{volume}{643}},
  \bibinfo{pages}{L1} (\bibinfo{year}{2020}).

\bibitem{Emsenhuber2021A&A...656A..70E}
\bibinfo{author}{{Emsenhuber}, A.} \emph{et~al.}
\newblock \bibinfo{title}{{The New Generation Planetary Population Synthesis
  (NGPPS). II. Planetary population of solar-like stars and overview of
  statistical results}}.
\newblock \emph{\bibinfo{journal}{\aap}} \textbf{\bibinfo{volume}{656}},
  \bibinfo{pages}{A70} (\bibinfo{year}{2021}).

\bibitem{Izidoro2022ApJ...939L..19I}
\bibinfo{author}{{Izidoro}, A.} \emph{et~al.}
\newblock \bibinfo{title}{{The Exoplanet Radius Valley from Gas-driven Planet
  Migration and Breaking of Resonant Chains}}.
\newblock \emph{\bibinfo{journal}{\apjl}} \textbf{\bibinfo{volume}{939}}~(2),
  \bibinfo{pages}{L19} (\bibinfo{year}{2022}).

\bibitem{Hu2021ApJ...921L...8H}
\bibinfo{author}{{Hu}, R.} \emph{et~al.}
\newblock \bibinfo{title}{{Unveiling Shrouded Oceans on Temperate sub-Neptunes
  via Transit Signatures of Solubility Equilibria versus Gas Thermochemistry}}.
\newblock \emph{\bibinfo{journal}{\apjl}} \textbf{\bibinfo{volume}{921}}~(1),
  \bibinfo{pages}{L8} (\bibinfo{year}{2021}).

\bibitem{Tsai2021ApJ...922L..27T}
\bibinfo{author}{{Tsai}, S.-M.} \emph{et~al.}
\newblock \bibinfo{title}{{Inferring Shallow Surfaces on Sub-Neptune Exoplanets
  with JWST}}.
\newblock \emph{\bibinfo{journal}{\apjl}} \textbf{\bibinfo{volume}{922}}~(2),
  \bibinfo{pages}{L27} (\bibinfo{year}{2021}).

\bibitem{numpy}
\bibinfo{author}{{Harris}, C.~R.} \emph{et~al.}
\newblock \bibinfo{title}{{Array programming with NumPy}}.
\newblock \emph{\bibinfo{journal}{\nat}} \textbf{\bibinfo{volume}{585}}~(7825),
  \bibinfo{pages}{357--362} (\bibinfo{year}{2020}).

\bibitem{matplotlib}
\bibinfo{author}{{Hunter}, J.~D.}
\newblock \bibinfo{title}{{Matplotlib: A 2D Graphics Environment}}.
\newblock \emph{\bibinfo{journal}{Computing in Science and Engineering}}
  \textbf{\bibinfo{volume}{9}}~(3), \bibinfo{pages}{90--95}
  (\bibinfo{year}{2007}).

\bibitem{astropy}
\bibinfo{author}{{Astropy Collaboration}} \emph{et~al.}
\newblock \bibinfo{title}{{The Astropy Project: Sustaining and Growing a
  Community-oriented Open-source Project and the Latest Major Release (v5.0) of
  the Core Package}}.
\newblock \emph{\bibinfo{journal}{\apj}} \textbf{\bibinfo{volume}{935}}~(2),
  \bibinfo{pages}{167} (\bibinfo{year}{2022}).

\bibitem{emcee}
\bibinfo{author}{{Foreman-Mackey}, D.}, \bibinfo{author}{{Hogg}, D.~W.},
  \bibinfo{author}{{Lang}, D.} \& \bibinfo{author}{{Goodman}, J.}
\newblock \bibinfo{title}{{emcee: The MCMC Hammer}}.
\newblock \emph{\bibinfo{journal}{\pasp}} \textbf{\bibinfo{volume}{125}}~(925),
  \bibinfo{pages}{306} (\bibinfo{year}{2013}).

\bibitem{2021AJ....162..114M}
\bibinfo{author}{{MacDonald}, M.~G.}, \bibinfo{author}{{Shakespeare}, C.~J.} \&
  \bibinfo{author}{{Ragozzine}, D.}
\newblock \bibinfo{title}{{A Five-Planet Resonant Chain: Reevaluation of the
  Kepler-80 System}}.
\newblock \emph{\bibinfo{journal}{\aj}} \textbf{\bibinfo{volume}{162}}~(3),
  \bibinfo{pages}{114} (\bibinfo{year}{2021}).

\bibitem{1918AnHar..91....1C}
\bibinfo{author}{{Cannon}, A.~J.} \& \bibinfo{author}{{Pickering}, E.~C.}
\newblock \bibinfo{title}{{The Henry Draper catalogue 0h, 1h, 2h, and 3h}}.
\newblock \emph{\bibinfo{journal}{Annals of Harvard College Observatory}}
  \textbf{\bibinfo{volume}{91}}, \bibinfo{pages}{1--290} (\bibinfo{year}{1918}).

\bibitem{GaiaEDR3}
\bibinfo{author}{{Gaia Collaboration}} \emph{et~al.}
\newblock \bibinfo{title}{{Gaia Early Data Release 3. Summary of the contents
  and survey properties}}.
\newblock \emph{\bibinfo{journal}{\aap}} \textbf{\bibinfo{volume}{649}},
  \bibinfo{pages}{A1} (\bibinfo{year}{2021}).

\bibitem{1997JApA...18..161Y}
\bibinfo{author}{{Yoss}, K.~M.} \& \bibinfo{author}{{Griffin}, R.~F.}
\newblock \bibinfo{title}{{Radial Velocities and DDO, BV Photometry of Henry
  Draper G5-M Stars Near the North Galactic Pole}}.
\newblock \emph{\bibinfo{journal}{Journal of Astrophysics and Astronomy}}
  \textbf{\bibinfo{volume}{18}}, \bibinfo{pages}{161} (\bibinfo{year}{1997}).

\bibitem{2MASS}
\bibinfo{author}{{Skrutskie}, M.~F.} \emph{et~al.}
\newblock \bibinfo{title}{{The Two Micron All Sky Survey (2MASS)}}.
\newblock \emph{\bibinfo{journal}{\aj}} \textbf{\bibinfo{volume}{131}},
  \bibinfo{pages}{1163--1183} (\bibinfo{year}{2006}).

\bibitem{Delisle2017}
\bibinfo{author}{{Delisle}, J.~B.}
\newblock \bibinfo{title}{{Analytical model of multi-planetary resonant chains
  and constraints on migration scenarios}}.
\newblock \emph{\bibinfo{journal}{\aap}} \textbf{\bibinfo{volume}{605}},
  \bibinfo{pages}{A96} (\bibinfo{year}{2017}).

\end{thebibliography}


\newpage\bigskip
\textbf{Data Availability}

The TESS observations used in this study are publicly available at the Mikulski Archive for Space Telescopes (\url{https://archive.stsci.edu/missions-and-data/tess}).
The CHEOPS observations used in this study are available at the CHEOPS mission archive (\url{https://cheops-archive.astro.unige.ch/archive_browser/}).
The ground-based photometry and high-resolution imaging observations are uploaded to ExoFOP (\url{https://exofop.ipac.caltech.edu/tess/target.php?id=347332255}) and are publicly available. 
CARMENES and HARPS-N reduced spectra, together with the derived CCF-based radial velocities and spectral indicators are available in Zenodo (\url{https://doi.org/10.5281/zenodo.8211589}).
All reduced transit photometry and radial velocity measurements used in this work are also provided in Zenodo (\url{https://doi.org/10.5281/zenodo.8211589}).

\newpage\bigskip
\textbf{Code Availability} 

We used the following publicly available codes, resources and Python packages to reduce, analyze and interpret our observations of HD~110067: numpy \cite{numpy}, matplotlib \cite{matplotlib}, astropy \cite{astropy}, lightkurve \cite{lightkurve}, PIPE \cite{Szabo2021A&A...654A.159S,Morris2021A&A...653A.173M}, AstroImageJ \cite{Collins:2017}, raccoon \cite{Lafarga2020}, serval \cite{SERVAL}, ARES \cite{Sousa2007,Sousa2015}, MOOG \cite{Sneden1973}, ZASPE \cite{ZASPE}, emcee \cite{emcee}, CLES \cite{Scuflaire2008}, exoplanet \cite{exoplanet:exoplanet}, MonoTools \cite{MonoTools}, pymc3 \cite{exoplanet:pymc3}, ArviZ \cite{arviz}, GLS \cite{GLS}, MCMCI \cite{bonfanti20}, and pyaneti \cite{pyaneti,pyaneti2}. We can share the code used in the data reduction or data analysis on request.

\newpage\bigskip
\textbf{Acknowledgements}

We acknowledge the use of public TESS data from pipelines at the TESS Science Office and at the TESS Science Processing Operations Center. Resources supporting this work were provided by the NASA High-End Computing (HEC) Program through the NASA Advanced Supercomputing (NAS) Division at Ames Research Center for the production of the SPOC data products.
CHEOPS is an ESA mission in partnership with Switzerland with important contributions to the payload and the ground segment from Austria, Belgium, France, Germany, Hungary, Italy, Portugal, Spain, Sweden, and the United Kingdom. The CHEOPS Consortium would like to gratefully acknowledge the support received by all the agencies, offices, universities, and industries involved. Their flexibility and willingness to explore new approaches were essential to the success of this mission. 
CARMENES acknowledges financial support from the Agencia Estatal de Investigación of the Ministerio de Ciencia e Innovación MCIN/AEI/10.13039/501100011033 and the ERDF “A way of making Europe” through projects PID2019-107061GB-C61, PID2019-107061GB-C66, PID2021-125627OB-C31, and PID2021-125627OB-C32, from the Centre of Excellence “Severo Ochoa'' award to the Instituto de Astrofísica de Canarias (CEX2019-000920-S), from the Centre of Excellence “María de Maeztu” award to the Institut de Ciències de l’Espai (CEX2020-001058-M), and from the Generalitat de Catalunya/CERCA programme. 
Based on observations made with the Italian Telescopio Nazionale Galileo (TNG) operated on the island of La Palma by the Fundación Galileo Galilei of the INAF (Istituto Nazionale di Astrofisica) at the Spanish Observatorio del Roque de los Muchachos of the Instituto de Astrofisica de Canarias.
This article is based on observations made with the MuSCAT2 instrument, developed by ABC, at Telescopio Carlos Sánchez operated on the island of Tenerife by the IAC in the Spanish Observatorio del Teide. 
This paper is based on observations made with the MuSCAT3 instrument, developed by the Astrobiology Center and under financial supports by JSPS KAKENHI (JP18H05439) and JST PRESTO (JPMJPR1775), at Faulkes Telescope North on Maui, HI, operated by the Las Cumbres Observatory.
Tierras is supported by grants from the John Templeton Foundation and the Harvard Origins of Life Initiative. The opinions expressed in this publication are those of the authors and do not necessarily reflect the views of the John Templeton Foundation.
The NGTS facility is operated by the consortium institutes with support from the UK Science and Technology Facilities Council (STFC) under projects ST/M001962/1 and ST/S002642/1. 
Some of the observations presented in this paper were carried out at the Observatorio Astronómico Nacional on the Sierra de San Pedro Mártir (OAN-SPM), Baja California, México. 
This work makes use of observations from the Las Cumbres Observatory global telescope network.
Some of the observations in this paper made use of the High-Resolution Imaging instrument Alopeke and were obtained under Gemini LLP Proposal Number GN-S-2021A-LP-105. Alopeke was funded by the NASA Exoplanet Exploration Program and built at the NASA Ames Research Center by Steve B. Howell, Nic Scott, Elliott P. Horch, and Emmett Quigley. Alopeke was mounted on the Gemini North telescope of the international Gemini Observatory, a program of NSF OIR Lab, which is managed by the Association of Universities for Research in Astronomy (AURA) under a cooperative agreement with the National Science Foundation. on behalf of the Gemini partnership: the National Science Foundation (United States), National Research Council (Canada), Agencia Nacional de Investigación y Desarrollo (Chile), Ministerio de Ciencia, Tecnología e Innovación (Argentina), Ministério da Ciência, Tecnologia, Inovações e Comunicações (Brazil), and Korea Astronomy and Space Science Institute (Republic of Korea). 
This work was supported by the KESPRINT collaboration, an international consortium devoted to the characterization and research of exoplanets discovered with space-based missions.
R.Lu. thanks Prof. Daniel Fabrycky for helpful discussions regarding the orbital dynamics of the HD~110067 system.
R.Lu. acknowledges funding from University of La Laguna through the Margarita Salas Fellowship from the Spanish Ministry of Universities ref. UNI/551/2021-May 26, and under the EU Next Generation funds. 
This work has been carried out within the framework of the NCCR PlanetS supported by the Swiss National Science Foundation under grants 51NF40\_182901 and 51NF40\_205606. 
A.C.Ca. and T.G.Wi. acknowledge support from STFC consolidated grant numbers ST/R000824/1 and ST/V000861/1, and UKSA grant number ST/R003203/1. 
O.Ba. acknowledges that has received funding from the European Research Council (ERC) under the European Union’s Horizon 2020 research and innovation programme (Grant agreement No. 865624)
M.Le. acknowledges support of the Swiss National Science Foundation under grant number PCEFP2\_194576. 
P.F.L.Ma. acknowledges support from STFC research grant number ST/M001040/1. 
Y.Al. acknowledges support from the Swiss National Science Foundation (SNSF) under grant 200020\_192038. 
D.Ga. gratefully acknowledges financial support from the CRT foundation under Grant No. 2018.2323 ``Gaseous or rocky? Unveiling the nature of small worlds''. 
J.A.Eg. acknowledges support from the Swiss National Science Foundation (SNSF) under grant 200020\_192038. 
G.No. thanks for the research funding from the Ministry of Education and Science programme the Excellence Initiative - Research University conducted at the Centre of Excellence in Astrophysics and Astrochemistry of the Nicolaus Copernicus University in Torun, Poland. 
D.Ra. was supported by NASA under award number NNA16BD14C for NASA Academic Mission Services.
M.La. acknowledges funding from a UKRI Future Leader Fellowship, grant number MR/S035214/1. 
V.Ad. is supported by FCT through national funds by the following grants UIDB/04434/2020, UIDP/04434/2020, and 2022.06962.PTDC. 
P.J.Am. acknowledges financial support from the grants CEX2021-001131-S and PID2019-109522GB-C52, both funded by MCIN/AEI/ 10.13039/501100011033 and by ``ERDF: A way of making Europe''.
S.C.C.Ba. acknowledges support from FCT through FCT contracts nr. IF/01312/2014/CP1215/CT0004. 
X.Bo., S.Ch., D.Ga., M.Fr. and J.La. acknowledge their role as ESA-appointed CHEOPS science team members. 
L.Bo., G.Br., V.Na., I.Pa., G.Pi., R.Ra., G.Sc., V.Si., and T.Zi. acknowledge support from CHEOPS ASI-INAF agreement n. 2019-29-HH.0. 
A.Br. was supported by the SNSA. 
Contributions at the Mullard Space Science Laboratory by E.M.Br. were supported by STFC through the consolidated grant ST/W001136/1. 
S.C.Ga. acknowledges support from UNAM PAPIIT-IG101321. 
D.Ch. and J.G-M. thank the staff at the F. L. Whipple Observatory for their assistance in the refurbishment and maintenance of the 1.3-m telescope.  
W.D.Co. acknowledges support from NASA grant 80NSSC23K0429. This is University of Texas Center for Planetary Systems Habitability Contribution 0063. 
K.A.Co. acknowledges support from the TESS mission via subaward s3449 from MIT. 
H.J.De. acknowledges support from the Spanish Research Agency of the Ministry of Science and Innovation (AEI-MICINN) under grant PID2019-107061GB-C66, DOI: 10.13039/501100011033. 
This project was supported by the CNES. 
The Belgian participation to CHEOPS has been supported by the Belgian Federal Science Policy Office (BELSPO) in the framework of the PRODEX Program, and by the University of Liège through an ARC grant for Concerted Research Actions financed by the Wallonia-Brussels Federation. 
L.De. is an F.R.S.-FNRS Postdoctoral Researcher. 
This work was supported by FCT - Fundação para a Ciência e a Tecnologia through national funds and by FEDER through COMPETE2020 - Programa Operacional Competitividade e Internacionalizacão by these grants: UID/FIS/04434/2019, UIDB/04434/2020, UIDP/04434/2020, PTDC/FIS-AST/32113/2017 \& POCI-01-0145-FEDER- 032113, PTDC/FIS-AST/28953/2017 \& POCI-01-0145-FEDER-028953, PTDC/FIS-AST/28987/2017 \& POCI-01-0145-FEDER-028987, O.D.S.De. is supported in the form of work contract (DL 57/2016/CP1364/CT0004) funded by national funds through FCT. 
B.-O.De. acknowledges support from the Swiss State Secretariat for Education, Research and Innovation (SERI) under contract number MB22.00046. 
This project has received funding from the European Research Council (ERC) under the European Union’s Horizon 2020 research and innovation programme (project {\sc Four Aces} grant agreement No 724427). It has also been carried out in the frame of the National Centre for Competence in Research PlanetS supported by the Swiss National Science Foundation (SNSF). D.Eh. acknowledges financial support from the Swiss National Science Foundation for project 200021\_200726. 
E.E-Bo. acknowledges financial support from the European Union and the State Agency of Investigation of the Spanish Ministry of Science and Innovation (MICINN) under the grant PRE2020-093107 of the Pre-Doc Program for the Training of Doctors (FPI-SO) through FSE funds. 
M.Fr. gratefully acknowledges the support of the Swedish National Space Agency (DNR 65/19, 174/18). 
J.G-M. acknowledges support by the National Science Foundation through a Graduate Research Fellowship under grant No. DGE1745303 and by the Ford Foundation through a Ford Foundation Predoctoral Fellowship, administered by the National Academies of Sciences, Engineering, and Medicine. 
The contributions at the University of Warwick by S.Gi. have been supported by STFC through consolidated grants ST/L000733/1 and ST/P000495/1. 
M.Gi. is F.R.S.-FNRS Research Director. 
Y.G.M.Ch. acknowledges support from UNAM PAPIIT-IG101321. 
E.Go. acknowledge the support by the Thueringer Ministerium füër Wirtschaft, Wissenschaft und Digitale Gesellschaft. 
M.N.Gu. is the ESA CHEOPS Project Scientist and Mission Representative, and as such also responsible for the Guest Observers (GO) Programme. M.N.Gu. does not relay proprietary information between the GO and Guaranteed Time Observation (GTO) Programmes, and does not decide on the definition and target selection of the GTO Programme. 
A.P.Ha. acknowledges support by DFG grant HA 3279/12-1 within the DFG Schwerpunkt SPP 1992. 
Ch.He. acknowledges support from the European Union H2020-MSCA-ITN-2019 under Grant Agreement no. 860470 (CHAMELEON). 
S.Ho. gratefully acknowledges CNES funding through the grant 837319. 
This work is partly supported by JST CREST Grant Number JPMJCR1761. 
K.G.Is. is the ESA CHEOPS Project Scientist and is responsible for the ESA CHEOPS Guest Observers Programme. She does not participate in, or contribute to, the definition of the Guaranteed Time Programme of the CHEOPS mission through which observations described in this paper have been taken, nor to any aspect of target selection for the programme. 
J.Ko. gratefully acknowledges the support of the Swedish National Space Agency (SNSA, DNR 2020-00104) and of the Swedish Research Council (VR: Etableringsbidrag 2017-04945). 
K.W.F.La. was supported by Deutsche Forschungsgemeinschaft grants RA714/14-1 within the DFG Schwerpunkt SPP 1992, Exploring the Diversity of Extrasolar Planets. 
This work was granted access to the HPC resources of MesoPSL financed by the Region Ile de France and the project Equip@Meso (reference ANR-10-EQPX-29-01) of the programme Investissements d'Avenir supervised by the Agence Nationale pour la Recherche. 
A.L-E. acknowledges support from the CNES (Centre national d’études spatiales, France). 
This work is partly supported by Astrobiology Center SATELLITE Research project AB022006. 
This work is partly supported by JSPS KAKENHI Grant Number JP18H05439 and JST CREST Grant Number JPMJCR1761.  
H.L.M.Os. acknowledges funding support by STFC through a PhD studentship. 
H.Pa. acknowledges the support by the Spanish Ministry of Science and Innovation with the Ramon y Cajal fellowship number RYC2021-031798-I. 
This work was also partially supported by a grant from the Simons Foundation (PI Queloz, grant number 327127). 
S.N.Qu. acknowledges support from the TESS mission via subaward s3449 from MIT. S.N.Qu. acknowledges support from the TESS GI Program under award 80NSSC21K1056 (G03268). 
L.Sa. acknowledges support from UNAM PAPIIT project IN110122. 
N.C.Sa. acknowledges funding by the European Union (ERC, FIERCE, 101052347). Views and opinions expressed are however those of the author(s) only and do not necessarily reflect those of the European Union or the European Research Council. Neither the European Union nor the granting authority can be held responsible for them. 
N.Sc. acknowledges support from the Swiss National Science Foundation (PP00P2-163967 and PP00P2-190080) and NASA under award number 80GSFC21M0002. 
S.G.So. acknowledge support from FCT through FCT contract nr. CEECIND/00826/2018 and POPH/FSE (EC). 
Gy.M.Sz. acknowledges the support of the Hungarian National Research, Development and Innovation Office (NKFIH) grant K-125015, a PRODEX Experiment Agreement No. 4000137122, the Lend\"ulet LP2018-7/2021 grant of the Hungarian Academy of Science and the support of the city of Szombathely. 
A.Tu. acknowledges funding support from the STFC via a PhD studentship. 
V.V.Ey. acknowledges support by STFC through the consolidated grant ST/W001136/1. 
V.V.Gr. is an F.R.S-FNRS Research Associate. 
J.Ve. acknowledges support from the Swiss National Science Foundation (SNSF) under grant PZ00P2\_208945. 
N.A.Wa. acknowledges UKSA grant ST/R004838/1. 
N.Wa. is partly supported by JSPS KAKENHI Grant Number JP21K20376. 

\newpage\clearpage
\textbf{Author information} \\

\textbf{Authors and Affiliations} \\

\textbf{Author Contributions} 
R.Lu, H.P.Os, A.Le., E.Pa., A.Bo., O.Ba., and T.G.Wi. conceived the project and contributed notably to the writing of this manuscript. R.Lu. and H.P.Os. led the analysis of the photometric data. A.Le. led the dynamical analysis of the system and developed the method with J.-B.De. to predict the orbits of the planets based on their resonant state within the chain. R.Lu., A.Bo. and O.Ba. led the analysis of the radial velocity data and the stellar activity mitigation. T.G.Wi. led the stellar characterization with the help of V.Ad., S.G.So., A.Bo., V.V.Gr., S.Sa., and W.D.Co. 
Y.Al. and J.A.Eg. led the analysis of the internal structures, while L.Fo. and A.Bo. performed the atmospheric evolution simulations. 
D.Ra., J.D.Tw., and J.M.Je. improved the TESS data reduction to recover the missing cadences affected by reflected light and high background.
R.Lu., E.Pa., and G.No. planned and obtained the time for the observations with CARMENES and HARPS-N. CARMENES observations were made possible by M.La., J.C.Mo., P.J.Am., A.Qu., and I.Ri. HARPS-N observations were made possible by I.Ca., J.O-M., F.Mu., H.J.De., J.Ko., D.Ga., J.H.Li., W.D.Co., E.W.Gu., V.V.Ey., H.L.M.Os., S.Re., E.Go., F.Da., and K.W.F.La. 
High-resolution imaging observations from Palomar and Gemini North were made possible by A.W.Bo., D.R.Ci, I.J.M.Cr., S.B.Ho., E.Ma., and J.E.Sc. 
Ground-based photometric observations to catch the transit of planet f were made possible by the MuSCAT2 (R.Lu., E.Pa., N.Na., J.H.Li., K.Ik., E.E-B., J.O-M., N.Wa., F.Mu., G.No., A.Fu., H.Pa., M.Mo., T.Ka., J.P.Le., and T.Ko.), LCO (T.G.Wi., R.Lu., H.P.Os., E.Pa., A.Le., A.Tu., M.J.Ho., Y.Al., and D.Ga.), NGTS (H.P.Os., S.Gi., D.Ba., D.R.An., M.Mo., A.M.S.Sm., E.M.Br., and S.Ud.), Tierras (J.G-M. and D.Ch.), SAINT-EX (N.Sc., Y.G-M-C., L.Sa., S.C-G., and B.-O.De.), and MuSCAT3 (N.Na., J.H.Li., K.Ik., N.Wa., A.Fu., M.Mo., T.Ka., J.P.Le., and T.Ko.) instruments. 
The remaining authors provided key contributions to the development of the TESS and CHEOPS mission. All authors read and commented on the manuscript, and helped with its revision. \\

\textbf{Corresponding author} Correspondence to Rafael Luque (rluque@uchicago.edu). \\

\bigskip\textbf{Competing interests} The authors declare no competing interests.

\newpage\clearpage
\bigskip\textbf{Extended Data Figures and Tables}
\newpage

\setcounter{page}{1}
\setcounter{figure}{0}
\setcounter{table}{0}
\setcounter{section}{1}
\renewcommand{\figurename}{Extended Data Fig.}
\renewcommand{\tablename}{Extended Data Table}

\begin{figure}[t]
    \centering
    \includegraphics[width=0.99\columnwidth]{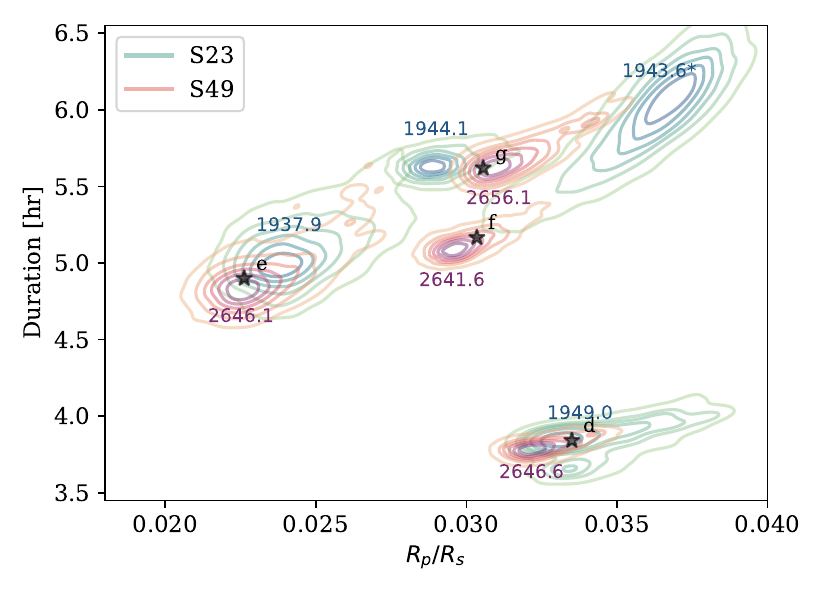}
    \caption{\textbf{Transit duration versus transit depth for all unassigned transits in TESS data}. TESS Sector 23 and Sector 49 are shown as different colors. The numbers above each transit denote the mid-transit time in TJD. Contours represent percentile levels, the innermost one corresponding to the 50th percentile and the outermost to the 99th percentile by increments of 10\%. The transit of planet f in PLD photometry is marked with $^{*}$ to indicate that its properties are heavily affected by pre-transit systematic noise.}
    \label{fig:ind_transits}
\end{figure}


\begin{figure}[t]
  \centering
  \includegraphics[width = .99\linewidth]{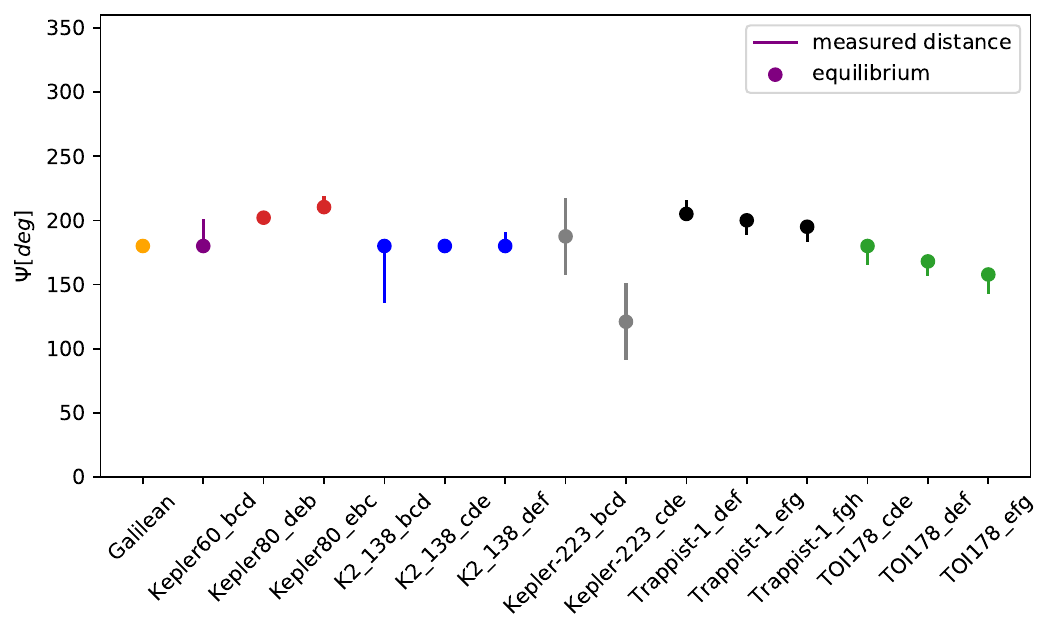}
  \caption{\textbf{Generalized three-body Laplace angles for known systems in resonant chains.} Included are the Galilean satellites, Kepler-60 \cite{Gozdziewski2016MNRAS.455L.104G,RIVERS3}, Kepler-80 \cite{2021AJ....162..114M}, K2-138 \cite{Lopez2019}, Kepler-223 \cite{Mills2016}, TRAPPIST-1 \cite{Agol2021}, and TOI-178 \citep{Leleu2021A&A...649A..26L}. Measurements belonging to the same system are marked with the same color. The line marks the observed distance to the theorized equilibrium (marked with a circle). The distances are estimated at 0th order in eccentricity \cite{Mills2016,SiFa2021}. For most systems, a single estimation of the generalized Laplace angle is made, while \cite{Mills2016} made an estimation for each Kepler quarter.}
  \label{fig:Laplace_amp}
\end{figure}


\begin{figure}[t]
  \centering
  \includegraphics[width = .99\linewidth]{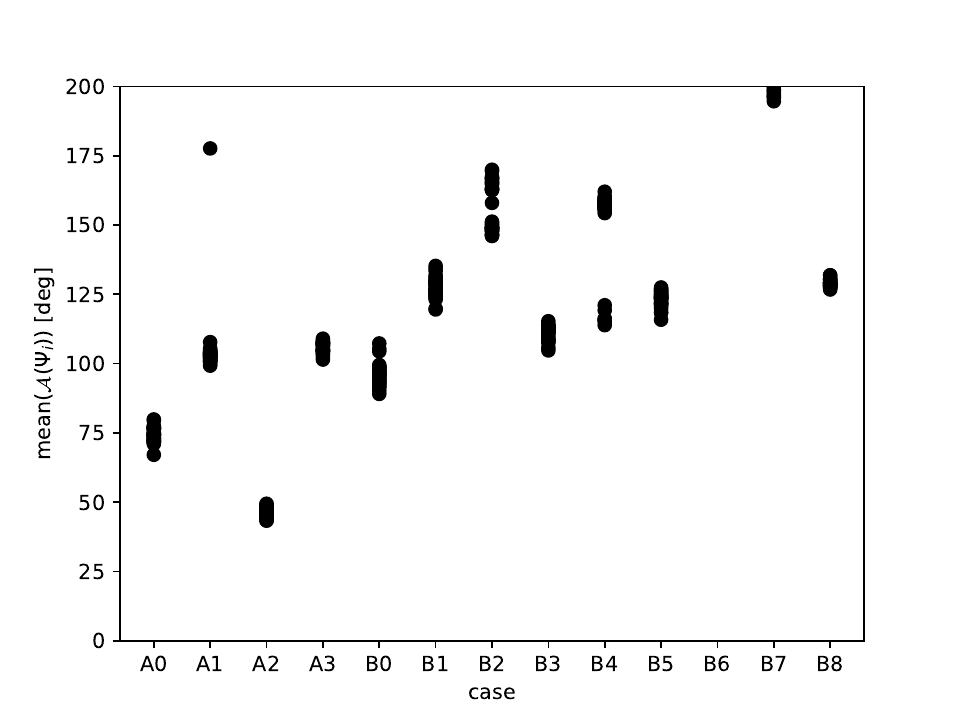}
  \caption{\textbf{Observed distance from the equilibrium for all the simulated scenarios in which planets f and g continue the resonant chain}. The y-axis is converted to the mean peak-to-peak amplitude from the generalized three-body Laplace angle using the following expression: mean($\mathcal{{A}}(\Psi_{i}))=C/4$. Case A2 remains the one that has the potential to be the closest to an equilibrium.}
  \label{fig:Cminimize}
\end{figure}


\begin{figure}[t]
    \centering
    \includegraphics[width=.99\linewidth]{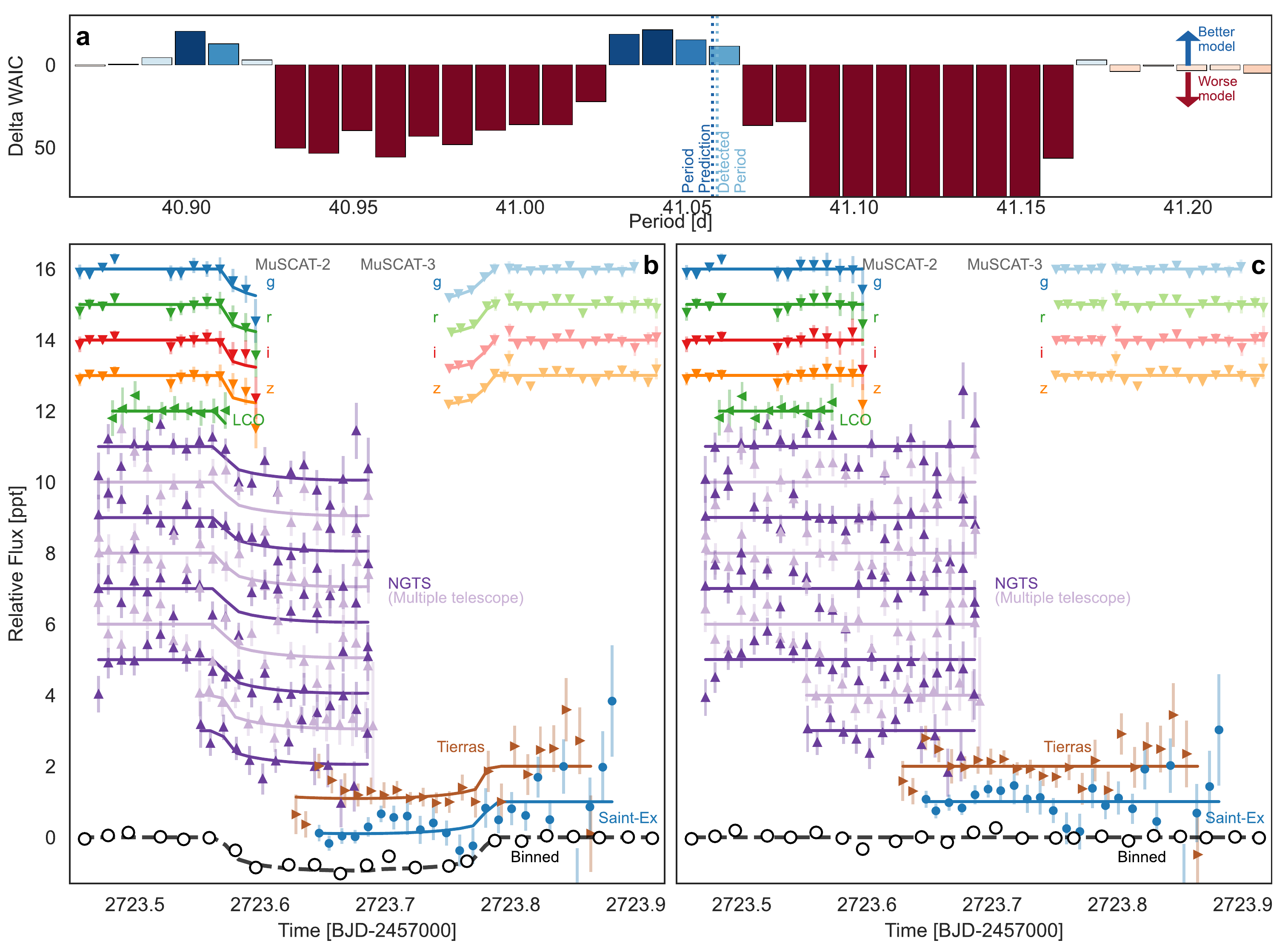}
    \caption{\textbf{Results from the ground-based campaign to detect HD~110067~f}. \textbf{a,} $\Delta$WAIC for each of the constrained period bins when compared to a transit-free model. \textbf{b,c,} Best-fit decorrelated photometry with (\textbf{b}) and without (\textbf{c}) a transit model. Each light curve from each telescope has been offset for clarity. Error bars represent 1$\sigma$ uncertainties. }
    \label{fig:groundcampaign}
\end{figure}


\begin{figure}[t]
    \centering    
    \includegraphics[width=.99\linewidth]{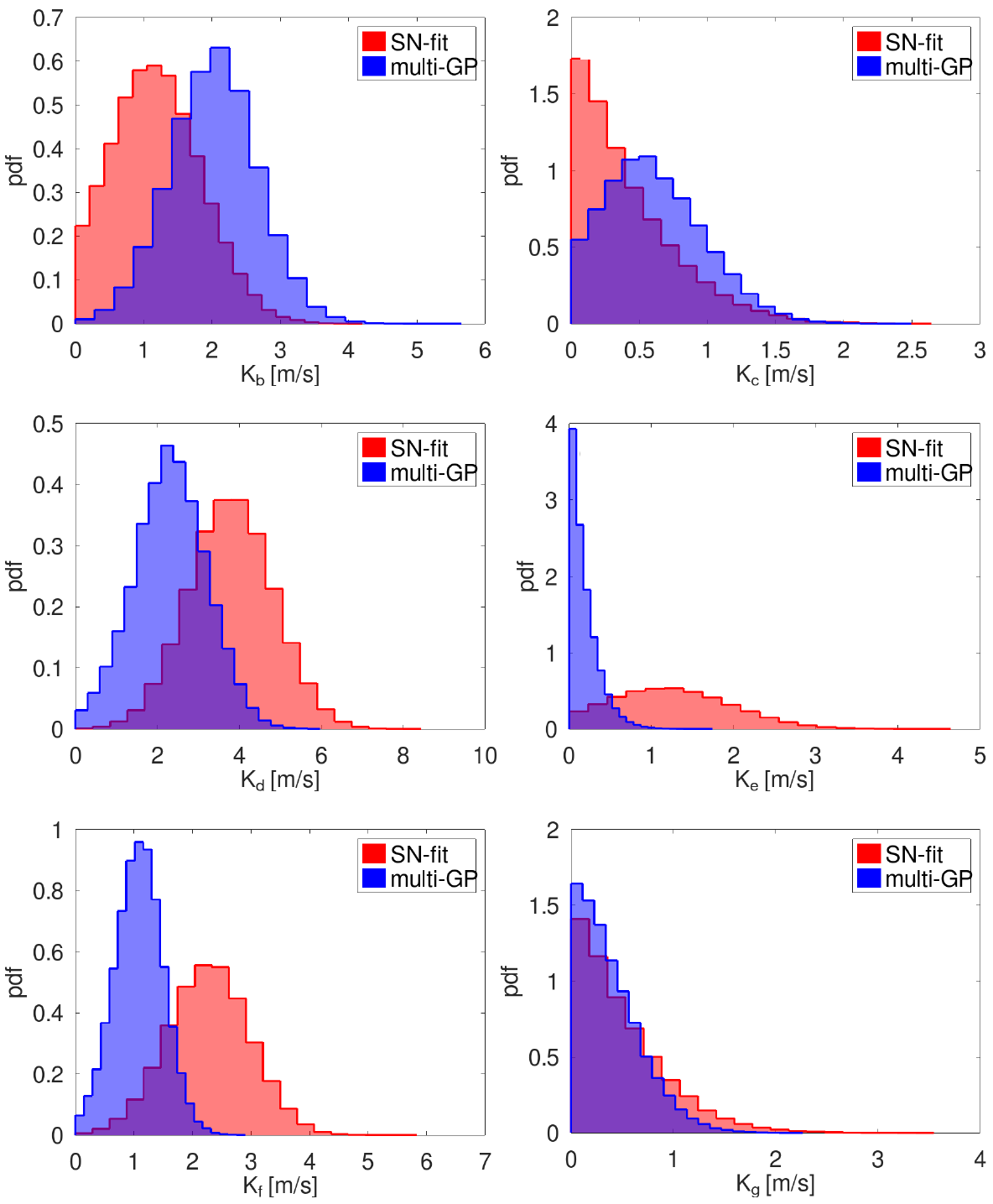}
    \caption{\textbf{Results from the two radial velocity analyses to measure the mass of each of the planets in the HD~110067 system}. Each histogram represents the posterior density function (pdf) of the radial velocity semi-amplitudes as inferred from \textsc{Method I} (red) and \textsc{Method II} (blue). The area underneath each histogram is normalized to unity.}
    \label{fig:Kcomparison}
\end{figure}


\begin{figure}[t]
    \centering
    \includegraphics[width=0.99\columnwidth]{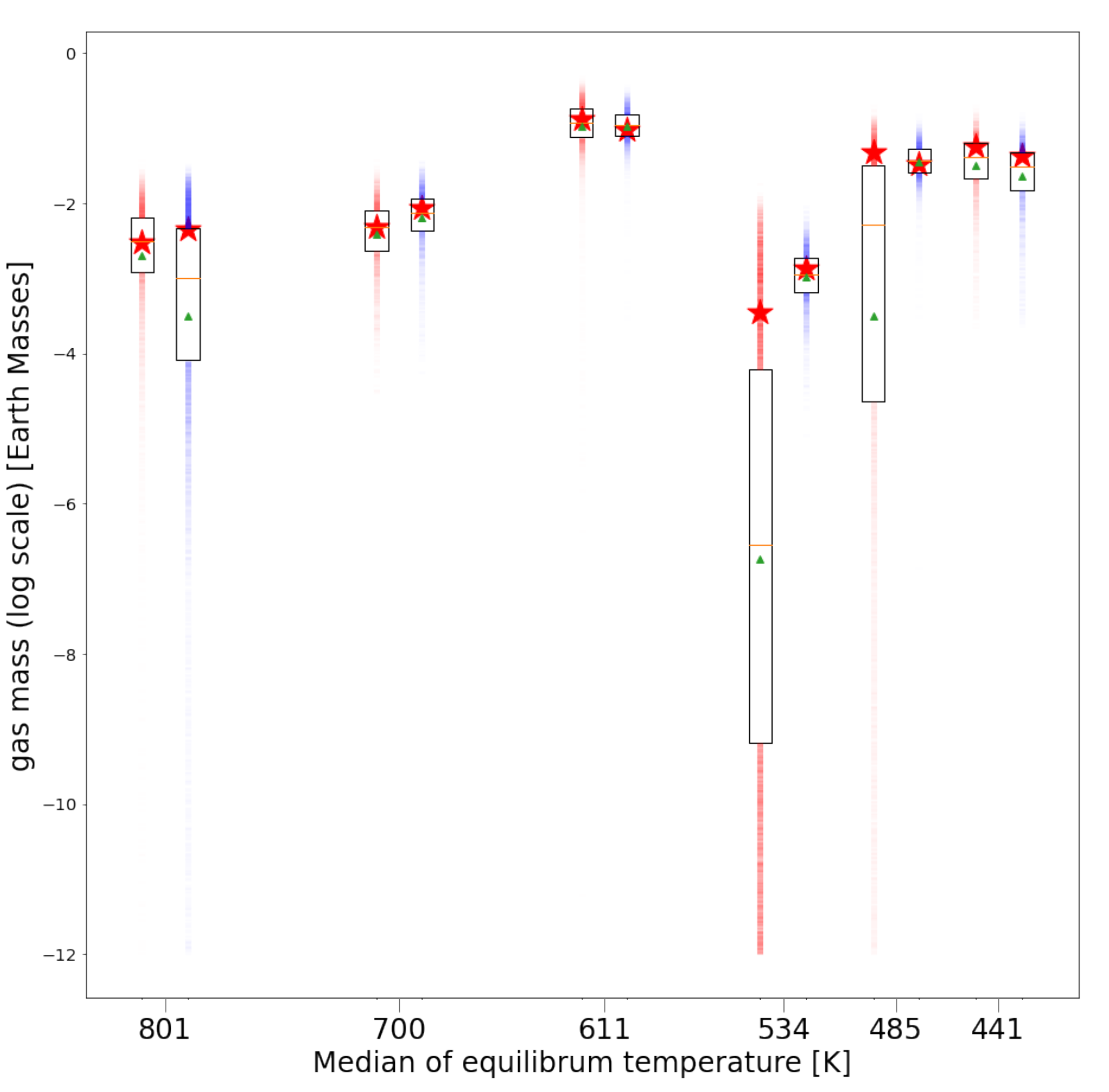}
    \caption{\textbf{Gas mass fraction of the HD~110067 planets as a function of their equilibrium temperature}. We infer two values per planet by assuming the different planetary masses from our \textsc{Method I} (red) and \textsc{Method II} (blue) radial velocity analyses. The boxes, orange lines, green triangles, and red stars represent respectively the 25\% and 75\% percentiles, medians, means, and modes of the posterior distributions. The opacity of the vertical lines is proportional to the posterior distribution. \label{fig:gasfrac_box}}
\end{figure}

\clearpage\newpage

\begin{table}[t]
\caption{\textbf{CHEOPS observing log.} Filler observations aim to catch transits serendipitously in between time-critical observations with higher priority. Boldface notes indicate that a transit event was detected in the data. }\label{tab:cheops_dat}
\footnotesize{
\centering
\begin{tabular}{cccccr}
\hline
\hline
\noalign{\smallskip}
Start (TJD) & Length [hr] & Archive filekey & Av. eff. (\%) & RMS (ppm) & Notes\\
\noalign{\smallskip}
\hline
\noalign{\smallskip}
$ 2681.3341 $ & $ 9.93 $ & PR110048\_TG025701\_V0200 & $ 79 $ & $ 202 $ & e, 35.4d alias \\
$ 2683.1380 $ & $ 10.36 $ & PR110048\_TG025801\_V0200 & $ 68 $ & $ 212 $ & e, 37.3d alias \\
$ 2685.2296 $ & $ 10.4 $ & PR110048\_TG025901\_V0200 & $ 68 $ & $ 196 $ & e, 39.3d alias \\
$ 2690.0127 $ & $ 9.88 $ & PR110048\_TG026601\_V0200 & $ 71 $ & $ 183 $ & d, 21.8d alias \\
$ 2692.9367 $ & $ 4.84 $ & PR100031\_TG052101\_V0200 & $ 75 $ & $ 187 $ & filler \\
$ 2693.3133 $ & $ 5.27 $ & PR100031\_TG052102\_V0200 & $ 60 $ & $ 217 $ & filler \\
$ 2694.6805 $ & $ 8.21 $ & PR100031\_TG052001\_V0200 & $ 73 $ & $ 212 $ & \textbf{Planet b} \\
$ 2702.5843 $ & $ 9.26 $ & PR110048\_TG026701\_V0200 & $ 72 $ & $ 220 $ & d, 22.65d alias \\
$ 2703.0623 $ & $ 10.36 $ & PR110048\_TG026101\_V0200 & $ 61 $ & $ 217 $ & d, 18.86d alias \\
$ 2704.0525 $ & $ 9.34 $ & PR110048\_TG026801\_V0200 & $ 62 $ & $ 213 $ & d, 23.38d alias \\
$ 2704.7545 $ & $ 3.2 $ & PR110031\_TG052201\_V0200 & $ 71 $ & $ 251 $ & filler \\
$ 2706.1956 $ & $ 9.38 $ & PR110048\_TG026401\_V0200 & $ 57 $ & $ 256 $ & d, 19.93d alias\\
$ 2707.9583 $ & $ 9.44 $ & PR110048\_TG028101\_V0200 & $ 66 $ & $ 275 $ & \textbf{Planet d}, 20.5d alias \\
$ 2712.8885 $ & $ 8.37 $ & PR100031\_TG052002\_V0200 & $ 61 $ & $ 240 $ & \textbf{Planet b}\\
$ 2713.9877 $ & $ 8.76 $ & PR110048\_TG028001\_V0200 & $ 60 $ & $ 247 $ & d, 22.5d alias \\
$ 2714.4689 $ & $ 4.19 $ & PR100031\_TG052103\_V0200 & $ 60 $ & $ 258 $ & filler \\
$ 2714.7439 $ & $ 10.08 $ & PR110048\_TG028501\_V0200 & $ 61 $ & $ 249 $ & d, 24.4d alias\\
$ 2715.1871 $ & $ 3.06 $ & PR100031\_TG052202\_V0200 & $ 52 $ & $ 222 $ & filler \\
$ 2717.2854 $ & $ 9.86 $ & PR110048\_TG028401\_V0200 & $ 56 $ & $ 243 $ & g, $2 \times P_e$\\
\noalign{\smallskip}
\hline
\end{tabular}
}
\end{table}


\begin{table}[t]
\centering
\caption{\sep \textbf{Ground-based photometric campaign observing log.} }\label{tab:planetf_log}
{\scriptsize
\begin{tabular}{llccr}
\hline
\hline
\noalign{\smallskip}
Facility & Filter & Start (TJD) & End (TJD) & Notes \\
\noalign{\smallskip}
\hline
\noalign{\smallskip}
MuSCAT-2    & $g$,$r$,$i$,$z_s$   & 2723.4534 & 2723.5991 & Interrupted at 2723.488 for 54\,min \\
LCO         & $z_p$      & 2723.4797 & 2723.5756 & No in-transit data at expected ingress\\
NGTS        & custom     & 2723.4684 & 2723.6904 & From 9 telescopes, 2 started late \\
Tierras     & custom    & 2723.6260 & 2723.8681 & Affected by cirrus \\
SAINT-EX    & zcut      & 2723.6446 & 2723.8826 & Affected by cirrus \\
MuSCAT-3    & $g$,$r$,$i$,$z_s$   & 2723.7479 & 2723.9177 & Interrupted at 2723.795 for 8\,min \\
\noalign{\smallskip}
\hline
\noalign{\smallskip}
\textbf{Transit} & --   & 2723.5869 & 2723.7987 & Expected in/egress for reference \\
\noalign{\smallskip}
\hline
\end{tabular}
}
\end{table}


\begin{table}[t]
\centering
\footnotesize
\caption{\textbf{Stellar parameters of HD~110067.} Error bars represent 1$\sigma$ uncertainties.} \label{tab:star}
\begin{tabular}{lcr}
\hline\hline
\noalign{\smallskip}
Parameter                               & Value                 & Reference \\ 
\hline
\noalign{\smallskip}
\multicolumn{3}{c}{\it Name and identifiers}\\
\noalign{\smallskip}
Name                            & HD~110067                     & {\cite{1918AnHar..91....1C}}      \\
TIC                             & 347332255                     & {\cite{Stassun2018AJ....156..102S}}      \\  
TOI                             & 1835                          & \cite{2021ApJS..254...39G}      \\  
\noalign{\smallskip}
\multicolumn{3}{c}{\it Coordinates and spectral type}\\
\noalign{\smallskip}
$\alpha$                                & 12:39:21.503       & \cite{GaiaEDR3}     \\
$\delta$                                & +20:01:40.03       & \cite{GaiaEDR3}     \\
Epoch                                   & 2000.0             & \cite{GaiaEDR3}     \\
Spectral type                           & K0.0\,V            & \cite{1997JApA...18..161Y}  \\
\noalign{\smallskip}
\multicolumn{3}{c}{\it Magnitudes}\\
\noalign{\smallskip}
$B$ [mag]                               & $9.203\pm0.03$        & \cite{1997JApA...18..161Y}       \\
$V$ [mag]                               & $8.419\pm0.002$       & \cite{1997JApA...18..161Y}       \\
$G$ [mag]                               & $8.17208\pm0.00028$   & \cite{GaiaEDR3}   \\
$J$ [mag]                               & $6.952\pm0.023$       & \cite{2MASS}       \\
$H$ [mag]                               & $6.561\pm0.017$       & \cite{2MASS}       \\
$K_s$ [mag]                             & $6.492\pm0.018$       & \cite{2MASS}       \\
\noalign{\smallskip}
\multicolumn{3}{c}{\it Parallax and kinematics}\\
\noalign{\smallskip}
$\pi$ [mas]                             & $31.037\pm0.022$     & \cite{GaiaEDR3}             \\
$d$ [pc]                                & $32.220\pm0.023$     & \cite{GaiaEDR3}             \\
$\mu_{\alpha}\cos\delta$ [$\mathrm{mas\,yr^{-1}}$]  & $-81.96 \pm 0.08$     & \cite{GaiaEDR3}          \\
$\mu_{\delta}$ [$\mathrm{mas\,yr^{-1}}$]            & $-104.59 \pm 0.04$    & \cite{GaiaEDR3}          \\
$U$ [$\mathrm{km\,s^{-1}}]$             &  +7.50$\pm$0.01       & This work      \\
$V$ [$\mathrm{km\,s^{-1}}]$             & -13.56$\pm$0.03       & This work      \\
$W$ [$\mathrm{km\,s^{-1}}]$             & -4.06$\pm$0.20        & This work      \\
\noalign{\smallskip}
\multicolumn{3}{c}{\it Photospheric parameters}\\
\noalign{\smallskip}
$T_{\mathrm{eff}}$ [K]                      & $5266 \pm 64$         & This work   \\
$\log g$                                    & $4.54 \pm 0.03$       & This work   \\
{[Fe/H]}                                    & $-0.20 \pm 0.04$      & This work   \\
{[Mg/H]}                                    & $-0.21 \pm 0.06$      & This work   \\
{[Si/H]}                                    & $-0.19 \pm 0.03$      & This work   \\
$v \sin i_\star$ [$\mathrm{km\,s^{-1}}$]    & $2.5 \pm 1.0$         & This work   \\
\noalign{\smallskip}
\multicolumn{3}{c}{\it Physical parameters}\\
\noalign{\smallskip}
$M$ [$M_{\odot}$]                       & $0.798\pm0.042$       & This work       \\
$R$ [$R_{\odot}$]                       & $0.788 \pm 0.008$     & This work       \\
Age [Gyr]                               & $8.1 \pm 4.0$         & This work       \\
\noalign{\smallskip}
\hline
\end{tabular}
\end{table}


\begin{table}[t]
\centering
\begin{footnotesize} 
\caption{\textbf{Distance of the estimated generalized three-body Laplace angle $\Psi_{e=0}$ to the closest equilibrium for all period ratios that are not excluded by available observations}. Case A assumes that the mono-transit at 2641.5778\,TJD belongs to the 5th planet, and 2656.0944\,TJD belongs to the 6th planet. Case B assumes the opposite. The flag column $^*=1$ indicates that the position of the equilibria varies with the masses of the planets, in which case the equilibrium is recomputed using \cite{Delisle2017}, with masses computed using a mass-radius relation for sub-Neptunes \cite{Otegi2020}. For $^*=0$, the equilibrium is 180\,degrees.} 
\label{tab:casesAB} 
\centering 
\begin{tabular}{lllllllllll} 
\hline \hline
  \noalign{\smallskip}
Case & $P_f/P_e$ & $P_g/P_f$ & $P_f$ & $P_g$ &  $\Delta \Psi_{bcd}$ & $\Delta \Psi_{cde}$ & $\Delta \Psi_{def}$ & $\Delta \Psi_{efg}$ & $^*$  \\ 
 &  &  & (days) & (days) &  (deg) & (deg) & (deg) & (deg) &   \\ 
  \noalign{\smallskip}
\hline
  \noalign{\smallskip}
A0&$4/3$&$2/1$&$41.0575$&$82.1150$&$44.17$&$18.55$&$7.67$&$95.50$&$1$ \\
A1&$4/3$&$3/2$&$41.0575$&$61.5862$&$44.17$&$36.13$&$88.05$&$79.21$&$1$ \\
A2&$4/3$&$4/3$&$41.0575$&$54.7433$&$44.17$&$18.06$&$4.89$&$9.81$&$1$ \\
A3&$4/3$&$6/5$&$41.0575$&$49.2690$&$44.17$&$37.57$&$80.14$&$95.33$&$1$ \\
  \noalign{\smallskip}
\hline 
  \noalign{\smallskip}
B0&$2/1$&$2/1$&$61.5862$&$123.172$&$44.17$&$10.00$&$43.60$&$21.79$&$0$ \\
B1&$2/1$&$3/2$&$61.5862$&$92.3793$&$44.17$&$10.00$&$43.60$&$106.64$&$0$ \\
B2&$2/1$&$4/3$&$61.5862$&$82.1150$&$44.17$&$10.00$&$43.60$&$168.50$&$0$ \\
B3&$2/1$&$6/5$&$61.5862$&$73.9035$&$44.17$&$10.00$&$43.60$&$1.21$&$0$ \\
B4&$3/2$&$2/1$&$46.1897$&$92.3793$&$44.17$&$10.00$&$73.33$&$167.00$&$0$ \\
B5&$3/2$&$6/5$&$46.1897$&$55.4276$&$44.17$&$10.00$&$73.33$&$100.43$&$0$ \\
B6&$4/3$&$2/1$&$41.0575$&$82.1150$&$44.17$&$38.56$&$136.29$&$72.06$&$1$ \\
B7&$4/3$&$4/3$&$41.0575$&$54.7433$&$44.17$&$38.07$&$133.51$&$177.37$&$1$ \\
B8&$5/4$&$5/4$&$38.4914$&$48.1142$&$44.17$&$10.00$&$52.81$&$110.80$&$0$ \\
  \noalign{\smallskip}
\hline 
\end{tabular} 
\end{footnotesize} 
\end{table}

\clearpage\newpage

\bigskip\textbf{Supplementary Information} 
Supplementary Information is available for this paper. 

\bigskip\textbf{Rights and permissions} Reprints and permissions information is available at www.nature.com/reprints.

\end{document}